\pdfoutput=1
\documentclass[11pt,twoside,a4paper,cmspaper,final,collab]{cms-tdr}
\def\svnVersion{9a5916b}\def\svnDate{2026/03/03}\def\cmsCernNoTag{CERN-EP-2024-308}\def\cmsCernDate{\today}\def\cmsMessage{Published in Nature as \href{http://dx.doi.org/10.1038/s41586-026-10168-5}{\doi{10.1038/s41586-026-10168-5}.}}
\begin{document}\cmsNoteHeader{SMP-23-002}

\newcommand{\ud}{\mathrm{d}}
\newcommand{\sigmaUL}{\ensuremath{\sigma^{\mathrm{U}{+}\mathrm{L}}}\xspace}
\newcommand{\alphaS}{\ensuremath{\alpha_{s}}\xspace}

\newcommand{\photospp}{\textsc{Photos++}\xspace}
\newcommand{\MiNNLO}{\textsc{MiNNLO}$_\text{PS}$\xspace}
\newcommand{\horace}{\textsc{Horace}\xspace}
\newcommand{\pMiNNLO}{\POWHEG~\MiNNLO\xspace}
\newcommand{\winhac}{\textsc{winhac}\xspace}
\newcommand{\Matrix}{\textsc{matrix}\xspace}
\newcommand{\radish}{\textsc{RadISH}\xspace}
\newcommand{\matrixradish}{\ensuremath{\Matrix{+}\radish}\xspace}
\newcommand{\mcfm}{\textsc{mcfm}\xspace}
\newcommand{\dyturbo}{\textsc{DYTurbo}\xspace}
\newcommand{\scetlib}{\textsc{SCETlib}\xspace}
\newcommand{\nnlojet}{\text{NNLOJET}\xspace}
\newcommand{\scetlibdyturbo}{\ensuremath{\scetlib{+}\dyturbo}\xspace}
\newcommand{\scetlibnnlojet}{\ensuremath{\scetlib{+}\nnlojet}\xspace}
\newcommand{\scetlibminnlo}{\ensuremath{\scetlib{+}\textsc{MiNNLO}_{\text{PS}}}\xspace}
\newcommand{\NtLL}{\ensuremath{\mathrm{N}^{3}\mathrm{LL}}\xspace}
\newcommand{\NfLL}{\ensuremath{\mathrm{N}^{4}\mathrm{LL}}\xspace}
\newcommand{\NtpzLL}{\ensuremath{\mathrm{N}^{3{+}0}\mathrm{LL}}\xspace}
\newcommand{\NmpkLL}{\ensuremath{\mathrm{N}^{m{+}k}\mathrm{LL}}\xspace}
\newcommand{\NtpoLL}{\ensuremath{\mathrm{N}^{3{+}1}\mathrm{LL}}\xspace}
\newcommand{\NfpzLL}{\ensuremath{\mathrm{N}^{4{+}0}\mathrm{LL}}\xspace}
\newcommand{\NtLLpNNLO}{\ensuremath{\mathrm{N}^{3}\mathrm{LL}{+}\mathrm{NNLO}}\xspace}
\newcommand{\NtpzLLpNNLO}{\ensuremath{\mathrm{N}^{3+0}\mathrm{LL}{+}\mathrm{NNLO}}\xspace}
\newcommand{\NtpoLLpNNLO}{\ensuremath{\mathrm{N}^{3+1}\mathrm{LL}{+}\mathrm{NNLO}}\xspace}
\newcommand{\NtpzLLpNtLO}{\ensuremath{\mathrm{N}^{3+0}\mathrm{LL}{+}\mathrm{N}^{3}\mathrm{LO}}\xspace}
\newcommand{\NfpzLLpNNLO}{\ensuremath{\mathrm{N}^{4+0}\mathrm{LL}{+}\mathrm{NNLO}}\xspace}
\newcommand{\NtpoLLpNtLO}{\ensuremath{\mathrm{N}^{3+1}\mathrm{LL}{+}\mathrm{N}^{3}\mathrm{LO}}\xspace}
\newcommand{\NfpzLLpNtLO}{\ensuremath{\mathrm{N}^{4+0}\mathrm{LL}{+}\mathrm{N}^{3}\mathrm{LO}}\xspace}
\newcommand{\NtLO}{\ensuremath{\mathrm{N}^{3}\mathrm{LO}}\xspace}
\newcommand{\NfLLpNtLO}{\ensuremath{\mathrm{N}^{4}\mathrm{LL}{+}\mathrm{N^{3}LO}}\xspace}
\newcommand{\NLOpHO}{\ensuremath{\mathrm{NLO}{+}\mathrm{HO}}\xspace}

\newcommand{\utmu}{\ensuremath{u_{\mathrm{T}}^{\mu}}\xspace}
\newcommand{\ut}{\ensuremath{u_{\mathrm{T}}}\xspace}
\newcommand{\utvec}{\ensuremath{\vec{u}_{\mathrm{T}}}\xspace}
\newcommand{\mw}{\ensuremath{m_{\PW}}\xspace}
\newcommand{\gw}{\ensuremath{\Gamma_{\PW}}\xspace}
\newcommand{\mtw}{\ensuremath{m_{\mathrm{T}}^{\PW}}\xspace}
\newcommand{\mwp}{\ensuremath{m_{\PWp}}\xspace}
\newcommand{\mwm}{\ensuremath{m_{\PWm}}\xspace}
\newcommand{\mwpm}{\ensuremath{m_{\PWpm}}\xspace}
\newcommand{\mz}{\ensuremath{m_{\PZ}}\xspace}
\newcommand{\mv}{\ensuremath{m_{\PV}}\xspace}
\newcommand{\ptw}{\ensuremath{\pt^{\PW}}\xspace}
\newcommand{\ptz}{\ensuremath{\pt^{\PZ}}\xspace}
\newcommand{\ptv}{\ensuremath{\pt^{\PV}}\xspace}
\newcommand{\yv}{\ensuremath{y^{\PV}}\xspace}
\newcommand{\yw}{\ensuremath{y^{\PW}}\xspace}
\newcommand{\ayw}{\ensuremath{\abs{y^{\PW}}}\xspace}
\newcommand{\ayz}{\ensuremath{\abs{y^{\PZ}}}\xspace}
\newcommand{\met}{\ensuremath{p_{\mathrm{T}}^{\text{miss}}}\xspace}
\newcommand{\mt}{\ensuremath{m_{\mathrm{T}}}\xspace}
\newcommand{\qt}{\ensuremath{q_{\mathrm{T}}}\xspace}

\newcommand{\mll}{\ensuremath{m_{\Pell\Pell}}\xspace}
\newcommand{\mmm}{\ensuremath{m_{\PGm\PGm}}\xspace}
\newcommand{\yll}{\ensuremath{y_{\Pell\Pell}}\xspace}
\newcommand{\ymm}{\ensuremath{y^{\PGm\PGm}}\xspace}
\newcommand{\ptmumu}{\ensuremath{\pt^{\PGm\PGm}}\xspace}
\newcommand{\ptmumuvec}{\ensuremath{\ptvec^{\PGm\PGm}}\xspace}
\newcommand{\absetal}{\ensuremath{\abs{\eta^{\Pell}}}\xspace}

\newcommand{\ptl}{\ensuremath{\pt^{\Pell}}\xspace}
\newcommand{\ptnu}{\ensuremath{\pt^{\PGn}}\xspace}
\newcommand{\ptmu}{\ensuremath{\pt^{\PGm}}\xspace}
\newcommand{\etamu}{\ensuremath{\eta^{\PGm}}\xspace}
\newcommand{\aeta}{\ensuremath{\abs{\eta}}\xspace}
\newcommand{\aetamu}{\ensuremath{\abs{\eta^{\PGm}}}\xspace}
\newcommand{\psera}{\ensuremath{\eta}\xspace}
\newcommand{\etapt}{\ensuremath{(\pt, \eta)}\xspace}
\newcommand{\qmu}{\ensuremath{q^{\PGm}}\xspace}
\newcommand{\etaptqmu}{\ensuremath{(\ptmu, \etamu, \qmu)}\xspace}
\newcommand{\etaptmu}{\ensuremath{(\ptmu, \etamu)}\xspace}
\newcommand{\ptymumu}{\ensuremath{(\ptmumu, \ymm)}\xspace}
\newcommand{\ptut}{\ensuremath{(\utmu\text, \ptmu)}\xspace}
\newcommand{\phimu}{\ensuremath{\phi^{\PGm}}\xspace}

\newcommand{\lpost}{16.8}
\providecommand{\PGgst}{\HepParticle{\PGg}{}{\ast}}
\newcommand{\wj}{\ensuremath{\PW+\text{jets}}\xspace}
\newcommand{\zj}{\ensuremath{\PZ+\text{jets}}\xspace}
\newcommand{\wmn}{\ensuremath{\PW\to\PGm\PGn}\xspace}
\newcommand{\wtn}{\ensuremath{\PW\to\PGt\PGn}\xspace}
\newcommand{\wln}{\ensuremath{\PW\to\Pell\PGn}\xspace}
\newcommand{\zmm}{\ensuremath{\PZ\to\PGm\PGm}\xspace}
\newcommand{\zgmm}{\ensuremath{\PZ/\PGgst\to\PGm\PGm}\xspace}
\newcommand{\jmm}{\ensuremath{\JPsi\to\PGm\PGm}\xspace}
\newcommand{\ztt}{\ensuremath{\PZ\to\PGt\PGt}\xspace}
\newcommand{\zll}{\ensuremath{\PZ\to\Pell\Pell}\xspace}
\newcommand{\umm}{\ensuremath{\PgUa\to\PGm\PGm}\xspace}

\newcommand{\chisq}{\ensuremath{\chi^2}\xspace}
\newcommand{\wlike}{{\PW}\nobreakdashes-like\xspace}
\newcommand{\wlikens}{{\PW}\nobreakdashes-like}
\newcommand{\muR}{\ensuremath{\mu_{\mathrm{R}}}\xspace}
\newcommand{\muF}{\ensuremath{\mu_{\mathrm{F}}}\xspace}
\newcommand{\nvtx}{\ensuremath{n_{\text{vtx}}\xspace}}
\newlength{\cmsFigWidth}
\ifthenelse{\boolean{cms@external}}{
\providecommand{\cmsLeft}{upper\xspace}\providecommand{\cmsRight}{lower\xspace}
\setlength{\cmsFigWidth}{0.5\textwidth}
\newcommand{\insupp}[2]{#1}
\newcommand{\natcap}[1]{\bgroup\bfseries #1\egroup}
\def\appendixname{Methods}
}{
\providecommand{\cmsLeft}{left\xspace}\providecommand{\cmsRight}{right\xspace}
\setlength{\cmsFigWidth}{0.95\textwidth}
\newcommand{\insupp}[2]{#2}
\newcommand{\natcap}[1]{#1}
}

\newenvironment{dataavailability}{\section*{Data availability}}{\par}\newenvironment{codeavailability}{\bmhead{Code availability}}{\par}
\newenvironment{interests}{\section*{Competing interests}}{\par}
\newenvironment{authorcontributions}{\section*{Author contributions}}{\par}

\newboolean{cms@standalone}
\setboolean{cms@standalone}{true}
\providecommand{\FIXME}[1]{\textcolor{red}{FIXME!} #1\xspace}
\newcommand{\suppmat}{Supplementary information\xspace}
\cmsNoteHeader{SMP-23-002}

\title{High-precision measurement of the \texorpdfstring{\PW}{W} boson mass with the CMS experiment}

\date{\today}

\abstract{
In the standard model of particle physics, the masses of the \PW and \PZ bosons, the carriers of the weak interaction, are uniquely related. A precise determination of their masses is important because quantum loops of heavy, undiscovered particles could modify this relationship. Although the \PZ mass is known to the remarkable precision of 22 parts per million (2.0\MeV), the \PW mass is known much less precisely. A global fit to measured electroweak observables predicts the \PW mass with 6\MeV uncertainty~\cite{PDG2024,Haller:2018nnx,PhysRevD106033003}. Reaching a comparable experimental precision would be a sensitive and fundamental test of the standard model, made even more urgent by a recent challenge to the global fit prediction by a measurement from the CDF Collaboration at the Fermilab Tevatron collider~\cite{CDF:2022hxs}. Here we report the measurement of the \PW mass by the CMS Collaboration at the CERN LHC, based on a large data sample of $\PW\to\PGm\PGn$ events collected in 2016 at the proton-proton collision energy of 13\TeV. The measurement exploits a high-granularity maximum likelihood fit to the kinematic properties of muons produced in \PW decays. By combining an accurate determination of experimental effects with marked in situ constraints of theoretical inputs, we reach a precise measurement of the \PW mass, of $80\,360.2 \pm 9.9\MeV$, in agreement with the standard model prediction.}

\hypersetup{%
pdfauthor={CMS Collaboration},%
pdftitle={High-precision measurement of the W boson mass with the CMS experiment},%
pdfsubject={CMS},%
pdfkeywords={CMS, physics, mass, W boson}}

\maketitle 

\label{sec:intro}

Precision measurements of fundamental parameters have played a major part in the development of the 
standard model (SM) of particle physics, which provides an
accurate description of the known elementary particles and their interactions.
Over the span of several decades, they provided increasingly precise 
estimates for the masses of the \PW and \PZ bosons, top quark, and Higgs boson, 
which helped guide the experimental programs aimed at their discoveries. 
With the observation of the Higgs boson at the CERN~\cite{ATLAS:2012yve,CMS:2012qbp,CMS:2013btf}
Large Hadron Collider (LHC)
and the determination of its mass, 
all the parameters in the electroweak (EW) sector of the SM 
are now constrained by experimental measurements. 
Nevertheless, the SM is widely believed to be incomplete, 
given that it does not explain certain fundamental observations, 
such as the asymmetry of matter and antimatter in the universe and the existence of dark matter.
In the SM, the \PW and \PZ boson masses, \mw and \mz, 
are uniquely related to the coupling strengths of the weak and electromagnetic interactions. 
If the measured masses and couplings deviated from the predicted relation, 
it would be a clear sign of physics beyond the SM,
likely due to new particles that, while too heavy to be directly produced at existing accelerators,
interact
via quantum loops with the \PW and \PZ bosons~\cite{heinemeyer2013implications,lopez2014delta}. 

Following the observations of the \PW and \PZ bosons at the CERN $\mathrm{S}\Pp\Pap\mathrm{S}$ collider~\cite{UA1:1983crd,UA2:1983tsx,UA1:1983398, Bagnaia:146503},
\mz was measured
with the exceptional precision 
of 22 parts per million ($\mz = 91\,188.0 \pm 2.0$\MeV~\cite{PDG2024}),
predominantly by the experiments operating at the CERN 
LEP collider through
measurements of resonant \PZ boson production in precise beam energy scans~\cite{ALEPH:2005ab}.
The \mw measurement at LEP~\cite{ALEPH:2013dgf} was based on the direct reconstruction of
pair-produced \PW bosons, whose production rate 
in electron-positron collisions is several orders of magnitude lower than the \PZ boson production rate. 
Consequently, the uncertainty in the \mw measurement was an order of magnitude larger than that of \mz.
Subsequent measurements performed at the Fermilab Tevatron~\cite{D0:2012kms}
and the LHC~\cite{ATLAS:2017rzl, LHCb:2021bjt,ATLAS:2024erm}
contributed to
the current experimental average of $\mw = 80\,369.2 \pm 13.3$\MeV~\cite{Amoroso:2023pey,PDG2024}.
The value of \mw
derived from the predicted relationships of EW parameters in the SM 
and independently measured observables, known as a global EW fit, 
$\mw = 80\,353 \pm 6$\MeV~\cite{PDG2024,Haller:2018nnx,PhysRevD106033003},
is significantly more precise.
As such, improving the direct measurement of \mw tests the SM and enhances sensitivity to new physics. 
The experimental combination does not include the most precise single measurement, performed by the CDF Collaboration,
$\mw = 80\,433.5 \pm 9.4$\MeV~\cite{CDF:2022hxs}.
The strong disagreement between this value
and both the SM expectation and the other measurements~\cite{Amoroso:2023pey}
represents a major puzzle in the field of particle physics.
An independent high-precision \mw measurement is, therefore, of the utmost importance.
In this paper we report the results of the first \PW boson mass determination by the CMS Collaboration.
Our measurement is based on the analysis of more than 100~million reconstructed \PW boson decays
selected from a sample of proton-proton ($\Pp\Pp$) collisions ---
one of the largest samples used for measuring \mw.
Together with an accurate determination 
of the experimental effects, this large dataset allows us
to markedly reduce the theoretical and experimental 
uncertainties in our measurement.
This result constitutes a substantial step towards resolving the \PW boson mass puzzle.

\section{Analysis strategy}
\label{sec:strategy}

At hadron colliders, 
jets from the hadronization of the quark-antiquark pair produced in the decay of the \PW boson
cannot be selected and calibrated with sufficient accuracy for a precise \mw measurement. 
Therefore, measurements of \mw rely on the \PW boson decay to a charged lepton $\Pell$ and a neutrino $\PGn$,
\wln, in which the \PW boson cannot be fully 
reconstructed
because neutrinos are not directly measurable in collider detectors.
In the rest frame of the decaying \PW boson,
the mass of the \PW boson is equally shared between the momenta of the neutrino and of the charged lepton.
In the laboratory frame, 
the transverse components of the charged lepton and neutrino momenta (\ptl and \ptnu) 
exhibit characteristic peaks at around $\mw/2$,
though their exact distributions depend on the transverse momentum of the \PW boson itself, \ptw.
Therefore, \mw can be indirectly measured through \ptl or by
the transverse component of the negative vector momentum sum of all measured particles in the event, \ptvecmiss,
an estimator of \ptnu.
The \ptmiss (magnitude of \ptvecmiss) and the transverse mass \mtw,
defined in analogy to the two-body mass as
$\mtw = \sqrt{\smash[b]{2\,\pt^{\ell}\,\ptmiss-\ptvec^{\ell}\cdot\ptvecmiss}}$,
are powerful observables in the Tevatron measurements~\cite{D0:2012kms, CDF:2022hxs}.
However,
their sensitivity to \mw
in LHC measurements~\cite{ATLAS:2017rzl,ATLAS:2024erm,LHCb:2021bjt}
is weakened by the \ptmiss resolution, which degrades in the presence of
a large number of $\Pp\Pp$ collisions in the same or adjacent bunch crossings (pileup).
While such channels can provide important cross-checks, 
these considerations inform our \mw determination strategy, 
which focuses on the kinematic distributions of the charged lepton in \wln events. 

Among the three leptonic decays, we exploit the muon ($\mu$) channel, 
since it offers the best experimental precision with the multipurpose, nearly hermetic CMS detector~\cite{CMS:2008xjf}.
The CMS apparatus~\cite{CMS:2008xjf} is designed to trigger on~\cite{CMS:2020cmk,CMS:2024aqx,CMS:2016ngn} and identify 
electrons, muons, photons, and (charged and neutral) hadrons~\cite{CMS:2020uim,CMS:2018rym,CMS:2014pgm}. 
A global event reconstruction algorithm~\cite{CMS:2017yfk} aims to reconstruct all individual particles in an event, 
combining information provided by the all-silicon inner tracker 
and by the crystal electromagnetic and brass-scintillator hadron calorimeters, 
operating inside a 3.8\unit{T} superconducting solenoid, 
with data from the gas-ionization muon detectors embedded in the flux-return yoke outside the solenoid. 
Charged-particle trajectories (tracks) are built from energy deposits
in each layer of the silicon detector, referred to as ``hits.''
Muon tracks typically have at least 12 hits, each of which is measured  with an 
accuracy of ${\approx}15\mum$ in the bending plane.
The muon momentum is derived from the curvature of the corresponding track,
with a typical resolution for \ptmu = 40\GeV of $\approx$1\% ($\approx$4\%)
in the central (forward) region of the detector.

Our measurement relies on a deep understanding of 
both the experimental and theoretical sources of systematic uncertainty.
The muon momentum scale (the largest source of uncertainty in the measurement) is calibrated to a few parts per hundred-thousand
by using a sample of dimuon decays of the \JPsi resonance.
Muons from \PgUa meson and \PZ boson decays are used for independent validations.
The predicted \ptmu distribution depends on the theoretical modeling of the \ptw distribution and 
on the parton distribution functions (PDFs), 
which describe the momentum distributions of the quarks and gluons inside the protons.
The PDFs strongly affect the \PW boson polarization and, hence, 
the kinematic distributions of the decay leptons~\cite{Manca_2017}.
To minimize the impact of these uncertainties on our measurement,
we aggregate selected data and simulated \wmn events into a highly-granular three-dimensional distribution
depending on \ptmu, \etamu, and \qmu, 
where $\etamu = -\ln\tan (\theta/2)$ is the muon pseudorapidity,
$\theta$ is the muon polar angle with respect to the beam line,
and \qmu is the muon electric charge.
This distribution is uniformly divided into
48 \etamu bins from $-2.4$ to 2.4, 30 \ptmu bins from 26 to 56\GeV, and two bins in \qmu ($+1$ or $-1$).
The \mw value is extracted from a binned maximum likelihood fit to this distribution,
using template shapes for the signal and background processes.

Our analysis uses state-of-the-art calculations to describe the \PW and \PZ boson production.
The predictions combine an all-order resummation of logarithmically enhanced soft and collinear gluon emissions
at next-to-next-to-next-to-leading logarithmic (\NtLL) accuracy
with next-to-next-to-leading order (NNLO) accuracy in perturbative quantum chromodynamics (QCD)~\cite{Ebert:2020dfc}.
The nonperturbative
motion of the partons inside the proton
is described by a phenomenological model~\cite{Billis:2024dqq}.
We incorporate a new proposal for ``theory nuisance parameters'' (TNPs)~\cite{TNPs} 
to parameterize the impact of unknown perturbative corrections. 
These models, combined with uncertainty profiling~\cite{PDG2024} in the binned maximum likelihood fit
to the \etaptqmu distribution,
allow the in situ determination of the \ptw spectrum with our \wmn data and
reduce its uncertainty to subleading importance in the measurement.
In contrast to previous \mw measurements
at hadron colliders~\cite{CDF:2022hxs,D0:2012kms, LHCb:2021bjt,ATLAS:2024erm},
we do not rely
on measurements of \PZ boson production to modify the predicted \ptw distribution.
As shown in Ref.~\cite{CMS:SMP-18-012}, our procedure also
significantly constrains the PDFs, the second largest source of uncertainty in our \mw measurement.

We have also developed an alternative analysis approach 
where \mw is extracted simultaneously with the 
angular distributions of the muon from the \PW boson decay.
This procedure is based on the general parameterization of the 
production cross section of a spin-1 boson and its decay to leptons
in terms of nine helicity states~\cite{mirkes:Wangular}.
For each bin in the two-dimensional \ptw and \PW boson rapidity (\yw) space,
and separately for the $\PW^+$ and $\PW^-$ bosons,
each helicity component leads to a different \etaptmu distribution.
We perform a differential analysis, 
encoding the variations of the helicity components as alternative templates 
fitted to the \etaptqmu distributions.
While this method, referred to as ``helicity fit", is less sensitive to \mw,
it provides a valuable cross-check of the nominal result 
by relaxing some assumptions about the \PW boson production and, hence, reducing the
dependence of the measurement on theoretical predictions.

We validate the experimental and theoretical inputs of the measurement 
with two \mz determinations.
First, we extract \mz through a maximum likelihood fit to the \zmm dimuon mass distribution.
Then, we perform a ``\wlike measurement" of \mz using only one of the two decay muons, 
mimicking the conditions of the \mw analysis. 
We model the \PZ and \PW boson production and their associated uncertainties with a common parameterization and perturbative accuracy, allowing the \wlike \mz measurement to serve as a robust validation of the predictions and uncertainties relevant for the \mw determination.
The consistency of these results with the precise \mz value obtained at LEP confirms the robustness of our \mw measurement.

The vector-boson ($\PV=\PW$ or \PZ) mass and width are defined in the running-width scheme~\cite{Bardin:1988xt}.
The analysis is conducted following the ``data blinding'' concept~\cite{Klein:2005di}:
it is optimized on simulated event samples and a random offset, between $-500$ and 500\MeV, 
is applied to the \mw and \mz values until all the procedures are established. 
In the following sections we briefly discuss the most important aspects of the \mw measurement, 
with further details given in the Methods section (Section~\ref{sec:Methods}).

\section{Event samples and selection criteria}
\label{sec:samples}

The measurements are made using a sample of $\Pp\Pp$ collisions 
at $\sqrt{s}=13\TeV$
collected in 2016 and corresponding to an integrated luminosity of $\lpost \pm 1.2$\%\fbinv~\cite{CMS:2021xjt}.
The events are preselected by an online trigger algorithm that requires the presence of 
at least one muon with $\ptmu > 24\GeV$, 
isolated from other energy deposits in the detector and 
satisfying quality criteria for tracks reconstructed 
in the silicon tracker and muon detectors~\cite{CMS:2008xjf,CMS:2020cmk,CMS:2016ngn}.
Selected \wmn events have exactly one muon 
with $\aetamu < 2.4$ and $26 < \ptmu < 56$\GeV.
The selected muon must be isolated from other particles and
satisfy selection criteria meant to reduce backgrounds and ensure a high-quality reconstruction.
To suppress backgrounds and enhance the purity of the W boson signal,
we require $\mtw > 40\GeV$; no upper limit on \mtw is imposed.
Machine-learning techniques are used to improve the resolution 
of the reconstructed \ptmiss~\cite{DeepMET},
enhancing the separation between signal and background events.
The simulated \ptmiss distribution is corrected 
using measured \zmm events, 
as discussed in Section~\ref{sec:recoil}. 
A total of 117 million data events are selected by these criteria.

The selection criteria for the dimuon and \wlike \mz measurements 
are designed to be maximally consistent with those of the \mw analysis.
Selected events have exactly two muons satisfying the same criteria, 
but with $\ptmu < 60\GeV$ because the \PZ boson is heavier than the \PW boson.
The two muons must have opposite electric charge and 
a dimuon invariant mass in the $60 < \mmm < 120$\GeV range.
A total of 7.5 million \zmm data events are selected. 
For the \wlike \mz analysis,
only one muon from the \PZ boson decay is considered to form the \etaptqmu templates.
The other muon is treated as a neutrino and excluded from 
the \ptmiss computation~\cite{CMS:2019ctu}.
The nominal \wlike \mz event sample is defined such that 
only positive muons are considered for selection from odd-numbered events and negative muons 
are considered for selection from even-numbered events
An alternative sample is defined by reversing the event number parity and \qmu matching.
Each event is considered only once per sample.
The results obtained from the two samples are not fully independent due to the partial overlap of selected events.
A total of 7.5 million \zmm data events are selected,
of which 7.4 million are selected for the nominal \wlike \mz sample.

Monte Carlo (MC) generators are used to produce large samples of simulated events
that are used to guide the analysis and to assess the consistency of the data with different hypotheses
for the value of \mw.
Simulated \PW and \PZ boson event samples are generated with \MiNNLO~\cite{Monni:2019whf,Monni_2020},
interfaced with \PYTHIA~\cite{Sjostrand:2015}.
The \MiNNLO predictions are scaled by two-dimensional binned corrections in the \PW or \PZ boson \pt and rapidity 
obtained with \scetlib~\cite{Ebert:2020dfc,Billis_2021,Billis:2024dqq},
thereby achieving \NtLLpNNLO
accuracy and improving the description of the data.
The CT18Z PDF set~\cite{Hou:2019efy} at NNLO accuracy is used.
The detector response is simulated using a detailed description of the CMS detector, 
implemented with the \GEANTfour package~\cite{AGOSTINELLI2003250}. 
More details on the data sample, event selection, and simulation are given in Section~\ref{sec:samples-selection-details}.

The average number of pileup interactions in data is 25, with a tail extending up to 44. 
The simulated distributions of the number of pileup interactions 
and the position along the beam line of the $\Pp\Pp$ collision producing the muon
are corrected to match the measured distributions, 
so as to accurately capture their impact on the muon reconstruction efficiency. 
The efficiency predicted by the simulation
is corrected to match that measured in data, as discussed in Section~\ref{sec:eff}. 

The main backgrounds in the selected \wmn data sample result from events 
with nonprompt muons, primarily from decays of heavy-flavor hadrons,
or with prompt muons, from \zmm decays where one muon misses the detector acceptance. 
Smaller backgrounds include \wtn and \ztt events,
with muons from \PGt lepton decays, 
as well as top quark-antiquark pair, single top quark, and diboson production. 
As discussed in Section~\ref{sec:ABCD}, 
the nonprompt-muon background is evaluated with the extended ABCD method~\cite{choi2021improved},
using sideband regions in data.
The uncertainty in the estimated background yields is dominated by the nonprompt-muon
component and contributes 3.2\MeV to the uncertainty in \mw.

\section{High-precision muon momentum calibration}
\label{sec:muon}

Reconstructing and calibrating the muon momentum
requires an exceptionally detailed understanding of 
the features that affect the trajectories
of charged particle tracks. 
In particular, the alignment of the tracking detector components,
the magnetic field throughout the tracking volume, 
and the material distribution, 
which governs the energy loss and multiple scattering
(interactions with electrons or nuclei
that lead to small-angle deflections),
must be precisely determined.
Hits in the silicon tracker are much more important to the track
determination than those in the muon system for
the \ptmu range relevant to this analysis.
Therefore,
we reconstruct the muon momentum exclusively using the silicon pixel and strip detectors,
restricting the volume of the detector where these features must
be accurately controlled.

The muon tracks are reconstructed using algorithms and conditions
specifically developed for this analysis, 
including a magnetic field mapping and a material model with a higher precision 
than those used in the standard CMS reconstruction. 
The alignment procedure~\cite{CMSAlignment2022} used to determine the position and orientation 
of the silicon modules has been extended to include fine-granularity corrections 
for the magnetic field and energy loss.  
The correspondence 
between the measured track curvature and the muon momentum is calibrated 
using a sample of events in which the dimuon invariant mass is
consistent with the well-established mass of the \JPsi resonance~\cite{PDG2024}. 
We extract parameterized corrections in fine bins of \etamu 
and extrapolate across the relevant range of \ptmu 
using a model that takes into account small offsets 
in the magnetic field, alignment, and tracker material 
remaining after the initial correction procedure. 
We validate our results using samples of \umm and \zmm events. 
The uncertainty in the procedure is evaluated 
from the deviations of the \etamu-binned correction parameters from zero 
when applying the corrections derived from \jmm events to \zmm events
and constraining \mz to the value of Ref.~\cite{PDG2024}.

Extrapolating corrections to the muon momentum resolution 
from the relatively low momentum range typical of muons from the \JPsi meson decay
to the higher momentum range of muons from \PZ boson decays
is more challenging than for the momentum
scale calibration, because multiple scattering has a large impact on the muon momentum resolution 
and is highly momentum dependent.
For this reason we correct the
muon momentum resolution in simulation to
match the measured resolution in data using both \jmm and \zmm events. 
The muon momentum calibration contributes 4.8\MeV
to the \mw uncertainty, primarily due to the scale calibration.
Additional details on the muon momentum scale and resolution calibrations are 
given in Section~\ref{sec:muon-reco}.

\section{Theoretical corrections and uncertainties}
\label{sec:theory}

The uncertainties in the predictions for \PZ and \PW boson production include contributions reflecting the 
limited knowledge of the PDFs,
the missing higher-order (HO) perturbative corrections in the QCD and EW interactions,
and the nonperturbative effects. 
The \ptw spectrum cannot be directly measured with high precision given the limited \ptmiss resolution.
Although the \ptz spectrum is measured precisely, using it to 
infer the \ptw spectrum requires estimating theoretical uncertainties in the \ptw/\ptz ratio, 
which depend strongly on the assumed uncertainty correlation~\cite{Bizon:2019zgf}.
Therefore, 
we do not apply corrections derived from the measured \ptmumu spectrum to the \PW boson simulation.
Instead, corrections to the \ptw spectrum come from \wmn data events via the profiling procedure
employed in the maximum likelihood fits used to extract results.
This approach relies on the high accuracy of the theoretical predictions, 
novel techniques to model their uncertainties and correlations across phase space, 
and the large statistical power of the analyzed data sample. 

The \scetlib calculation
parameterizes the dominant sources of uncertainty in the 
\PW and \PZ boson transverse momentum spectra (\ptv) 
due to perturbative and nonperturbative effects.
Perturbative uncertainties are represented by the TNPs of Ref.~\cite{TNPs}.
The TNPs have a true, but unknown, value that is varied according to its expected magnitude. 
The closure of this procedure is demonstrated for \PZ boson production 
in Ref.~\cite{Cridge:2025wwo} and we have independently verified that 
fitting a HO prediction (\eg, at fourth-order logarithmic accuracy) 
using a lower order prediction 
(\eg, at \NtLL with the next-order TNPs) as the fit model yields TNP values consistent with the known values.
The calculations treat the quarks as massless. Possible modifications
due to the true quark masses
are effectively absorbed into the other sources of modeling uncertainty.
We have tested other alternatives for the \ptw modeling, 
at equivalent or higher perturbative orders, 
and confirmed that the variation in \mw is within the uncertainty evaluated 
from our nominal prediction at \NtLLpNNLO accuracy. 

The relative fractions of the \PZ and \PW boson helicity states 
and their uncertainties due to missing HO perturbative corrections 
are evaluated at NNLO in QCD using \MiNNLO;
we have verified their consistency with the fixed-order NNLO QCD predictions 
of \dyturbo~\cite{Camarda:2019zyx} and \mcfm~\cite{Campbell:2019dru}.
Uncertainties due to the PDFs, including their impact on the \PW boson helicity states, are evaluated by 
propagating the Hessian eigenvectors of the CT18Z PDF set~\cite{PDFLHC}. 
Their contribution to the uncertainty in \mw is 4.4\MeV.
We have repeated the \mw measurement using seven alternative PDF sets.
Additional details on these studies, corrections, and uncertainties 
are given in Sections~\ref{sec:ptmodeling}--\ref{sec:addValidation}.

\section{Measurement of the \texorpdfstring{\PZ}{Z} and \texorpdfstring{\PW}{W} boson masses}
\label{sec:results}

The results are obtained from binned maximum likelihood fits in which 
systematic uncertainties are represented by nuisance parameters with Gaussian constraints~\cite{Conway:2011in}.
We allow the systematic uncertainties to be constrained and 
the central values to be pulled, with respect to their initial values,
through the profile likelihood function~\cite{PDG2024} used in the fits.
Common sources of uncertainty are correlated across bins of the distribution. 
The parameter of interest, \ie, the mass of the \PW or \PZ boson (\mv), 
is an unconstrained parameter in the fit.
The effect of different \mv values on the distributions is derived from 
a continuous interpolation around the nominal value in the fit,
set to the world-average experimental value, 
and from variations of \mv by $\pm 100\MeV$, evaluated from the 
full matrix-element-level calculation of the \MiNNLO simulation.
We verified that the fit correctly extracts the simulated \mw value
to within 0.1\MeV accuracy for twenty points within this range.
The construction and minimization of the likelihood is implemented 
using the \textsc{TensorFlow} software package~\cite{TensorFlow}, in which the
use of automatic differentiation~\cite{Baydin:2015tfa} of gradients in the likelihood
minimization allows the \mw and \wlike \mz likelihood fits to be
computationally feasible and numerically stable, 
despite involving approximately 3000 bins and 4000 nuisance parameters.

\subsection{Extraction of the \texorpdfstring{\PZ}{Z} boson mass from the dimuon mass spectrum}

We extract \mz from a binned maximum likelihood fit to the dimuon mass distribution, 
in 25 bins of \mmm and 14 bins of \etamu of the muon with the largest $\abs{\etamu}$.
Compared with the world-average value~\cite{PDG2024},
dominated by measurements at the LEP collider, 
we obtain
\begin{equation}
\mz^{\PGm\PGm} - \mz^{\mathrm{PDG}} = -2.2 \pm 4.8\MeV.
\end{equation}
The largest uncertainties result from
the muon momentum calibration (4.6\MeV)
and from the size of the data sample (1.0\MeV).

\begin{figure}[ht]
\centering
\includegraphics[width=0.45\textwidth]{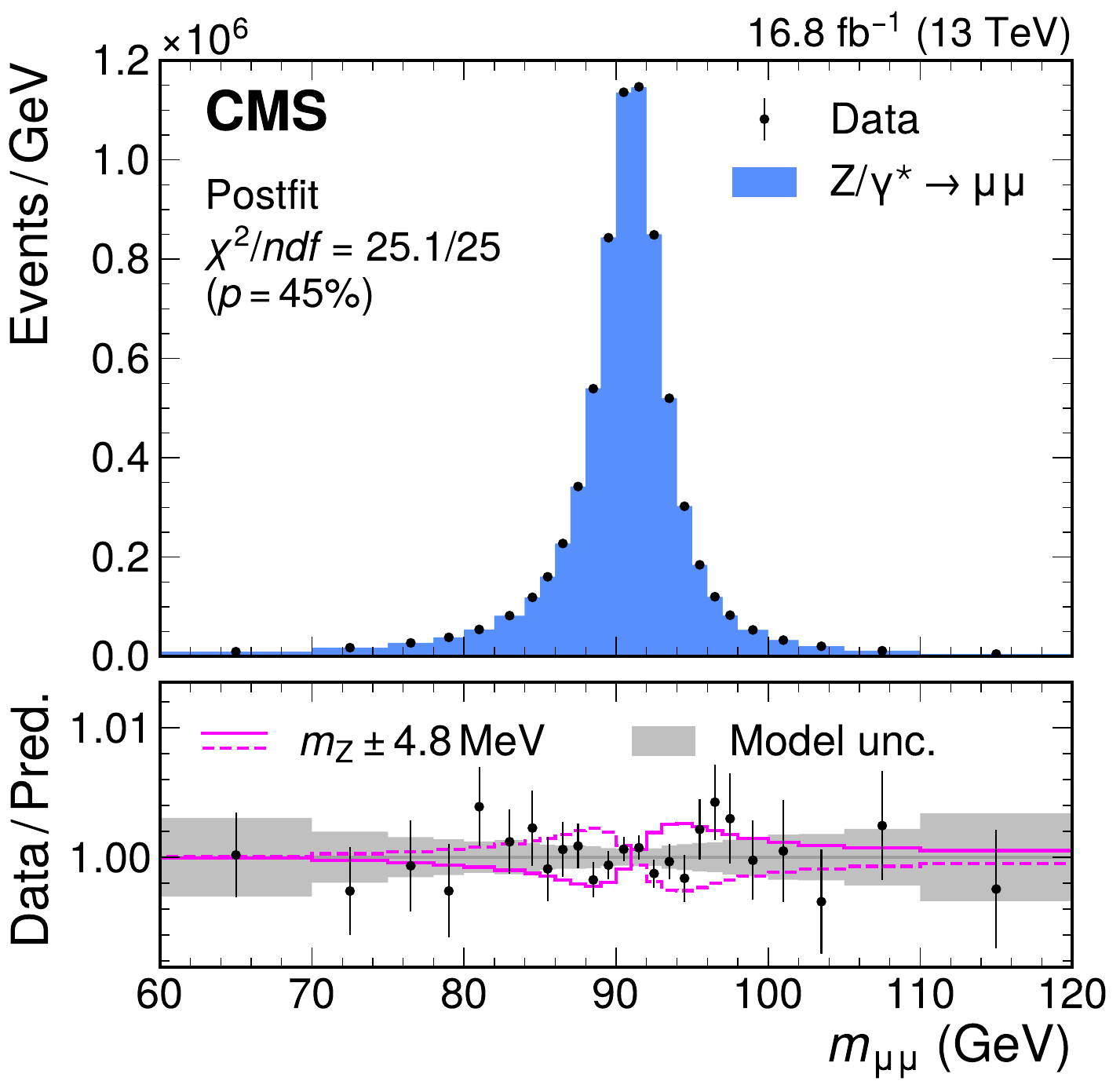}
\caption{\natcap{The \PZ boson mass measurement.}
Measured and simulated \zmm dimuon mass distributions. 
The postfit \zmm distribution is shown in blue. The small contributions
of other processes are included but not visible.
The bottom panel shows the ratio between the number of events in data and the total nominal prediction.
The vertical bars represent the statistical uncertainties in the data.
The total uncertainty in the prediction 
after the uncertainty profiling procedure (gray band) 
and the effect of a $\pm$4.8\MeV variation of \mz (magenta lines) are also shown, 
illustrating the precision of the achieved understanding of the distribution.
}\label{fig:mz}
\end{figure}

Figure~\ref{fig:mz} shows the measured and simulated \zmm dimuon mass distributions, 
with the predictions adjusted to reflect the best fit values of nuisance parameters 
obtained from the maximum likelihood fit (referred to as ``postfit predictions''). 
The excellent consistency of our result with $\mz^{\mathrm{PDG}}$
is a powerful validation of the muon reconstruction, momentum scale calibration, 
and corrections.
Although \zmm events are not used to determine the values of the parameterized muon momentum scale calibration, 
they are used, together with the $\mz^{\mathrm{PDG}}$ value~\cite{PDG2024},
to define the systematic uncertainties,
as described in Section~\ref{sec:muon-reco}.
Therefore, our \mz value is not a measurement independent of the experimental world average.

\subsection{\texorpdfstring{\PW}{W}-like measurement of the \texorpdfstring{\PZ}{Z} boson mass}

The \wlike \mz analysis extracts \mz from a binned maximum likelihood fit 
to the \etaptqmu distribution of the selected muons.
As described in Section~\ref{sec:samples}, two event samples are used. 
The result for the analysis configuration selecting positive muons in odd event-number events,
compared with the experimental \mz average~\cite{PDG2024}, is
\ifthenelse{\boolean{cms@external}}{
\begin{equation*}
  \begin{aligned}
\mz^{\PW\text{-like}} - \mz^{\mathrm{PDG}} &= -6 \pm 7 \stat \pm 12 \syst \\
&= -6 \pm 14\MeV,
\end{aligned}
\end{equation*}
}
{
\begin{equation*}
\mz^{\PW\text{-like}} - \mz^{\mathrm{PDG}} = -6 \pm 7 \stat \pm 12 \syst = -6 \pm 14\MeV,
\end{equation*}
}
showing that $\mz^{\PW\text{-like}}$ agrees with $\mz^{\mathrm{PDG}}$ (and with $\mz^{\PGm\PGm}$).
Both the helicity fit analysis and the analysis using the configuration with the alternative muon charge and event number parity matching 
provide \mz values that agree with the baseline result to within one standard deviation.

We validate the accuracy of the theory modeling and corresponding uncertainties
by measuring the \ptz directly in \zmm events.
Using the \PZ boson production model described in Section~\ref{sec:theory},
we fit the two-dimensional distribution of the dimuon \pt and rapidity \ptymumu to the observed \zmm data.
The consistency of the predictions and their uncertainties with the data 
is assessed with a goodness-of-fit test based on a saturated
model,
in which an unconstrained normalization parameter is introduced for each 
bin of the likelihood~\cite{lindsey}.
We conclude that the model and the data are compatible, 
given the $p$-value of 16\% that is evaluated from the ratio of the nominal and saturated likelihoods.

\begin{figure}[ht]
\centering
\includegraphics[width=0.45\textwidth]{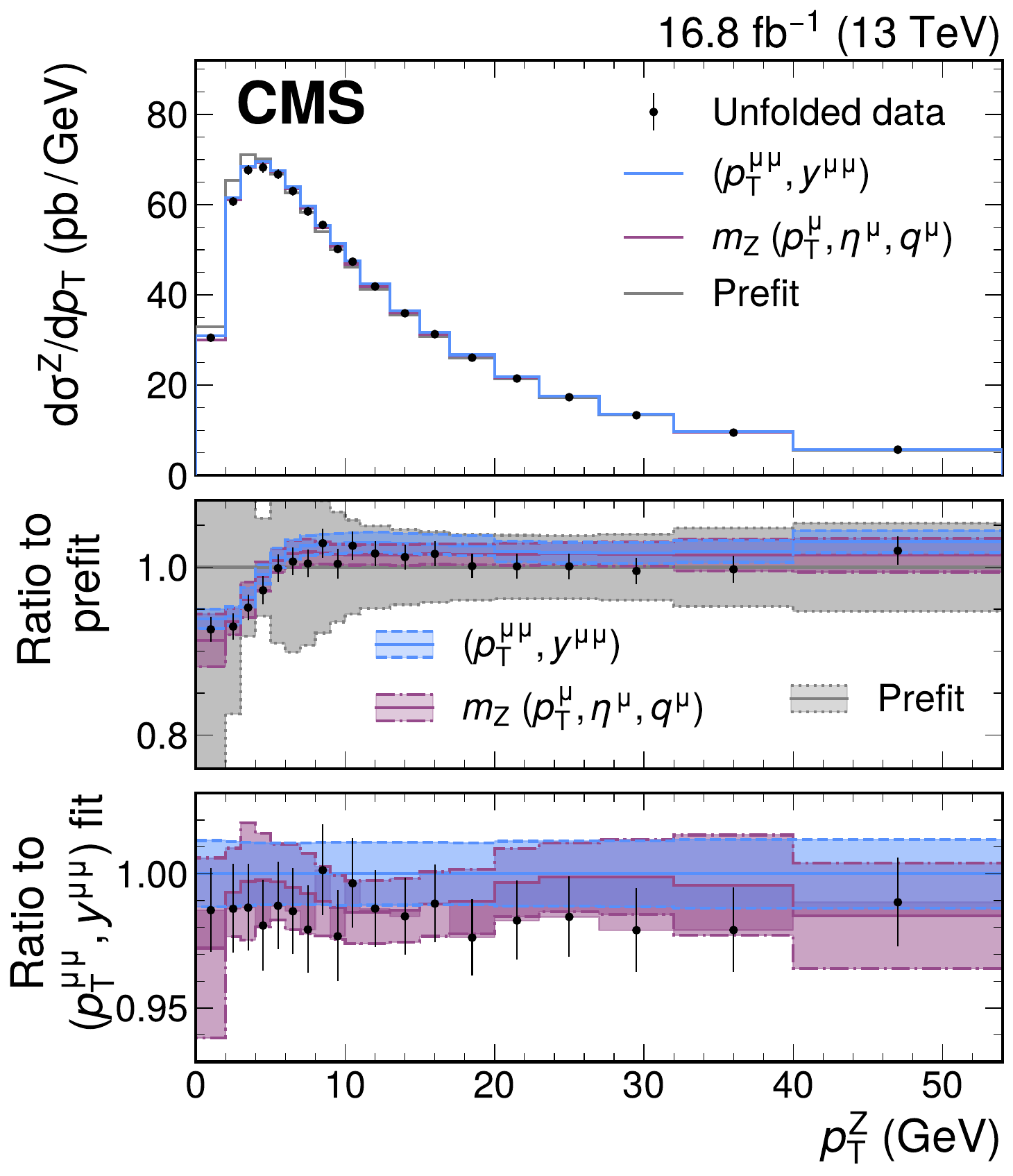}
\caption{\natcap{Validation of the theory model.} 
Unfolded measured \ptz distribution (points) compared with the generator-level \scetlibminnlo predictions
before (prefit, gray) 
and after adjusting the nuisance parameters to the best fit values
obtained from the \wlike \mz fit (magenta) or from the direct fit to the \ptmumu distribution (blue).
The center panel shows
the ratio of the predictions and unfolded data to the prefit prediction.
The uncertainty in the prefit prediction is shown by the shaded gray area.
The bottom panel shows the ratio of the predictions and unfolded data
to the postfit prediction from the
fit to the \ptymumu distribution. The postfit uncertainties in the predictions
are shown in the shaded
magenta and blue bands.
The vertical bars represent the total uncertainty in the unfolded data.}
\label{fig:ptll_gen_postfit}
\end{figure}

The results of the \ptmumu fit are not an input to the \wlike \mz  or \mw measurements.
Rather, we independently determine values for the nuisance parameters 
describing the \ptz modeling from the \wlike \mz measurement
and verify that they are consistent with those from the direct fit to \ptmumu. 
Figure~\ref{fig:ptll_gen_postfit} shows the generator-level \ptz distribution, 
with the predictions adjusted by the nuisance parameter values obtained from
the two independent fits to the data.
To test the accuracy of the adjusted predictions in describing our data, 
we account for effects of the detector response and resolution 
by ``unfolding'' our measurement to the generator level, 
as described in Section~\ref{sec:addValidation}.
The self-consistency of the postfit distributions from the two fits,
as well as their consistency with the data,
confirms the robustness of the predictions and of the uncertainty model,
and demonstrates the ability of the \etaptmu distribution to constrain the \ptv modeling in situ. 
This result supports our treatment of the \PW boson production modeling in the \mw analysis.

Section~\ref{sec:addValidation} gives more details 
on the stability of our \wlike \PZ boson mass measurement 
under different modeling assumptions
and its consistency with the measured \ptmumu distribution.

\subsection{Measurement of the \texorpdfstring{\PW}{W} boson mass}
\label{sec:mW}

Having validated the analysis steps using the \PZ boson data, 
we proceed with the determination of the \PW boson mass.
A fit is performed to the \etaptqmu distribution, 
shown in Fig.~\ref{fig:pteta}, and the observed \mw value is 
\ifthenelse{\boolean{cms@external}}{
  \begin{equation*}
    \begin{aligned}
    \mw &= 80\,360.2 \pm 2.4 \stat \pm 9.6\syst\\
    &= 80\,360.2 \pm 9.9\MeV,
  \end{aligned}
\end{equation*}  
}{
\begin{equation*}
\mw = 80\,360.2 \pm 2.4 \stat \pm 9.6\syst = 80\,360.2 \pm 9.9\MeV,
\end{equation*}
}
in agreement with the EW fit prediction,
$\mw = 80\,353 \pm 6$\MeV~\cite{PDG2024,Haller:2018nnx,PhysRevD106033003},
and with other experimental results, except the latest measurement reported by the CDF Collaboration~\cite{CDF:2022hxs}.
The EW fit prediction is based on relationships
between \mw and other experimental observables,
including the \PZ boson, Higgs boson, and top quark masses, 
the fine-structure constant, and the muon lifetime.
The uncertainty in the prediction is due to
missing HO terms in the perturbative calculation
used to derive the predicted relationship between the experimental inputs
and from uncertainties in the experimental inputs themselves.
The two sources of uncertainty are of comparable size.

\begin{figure}[ht]
\centering
\includegraphics[width=0.45\textwidth]{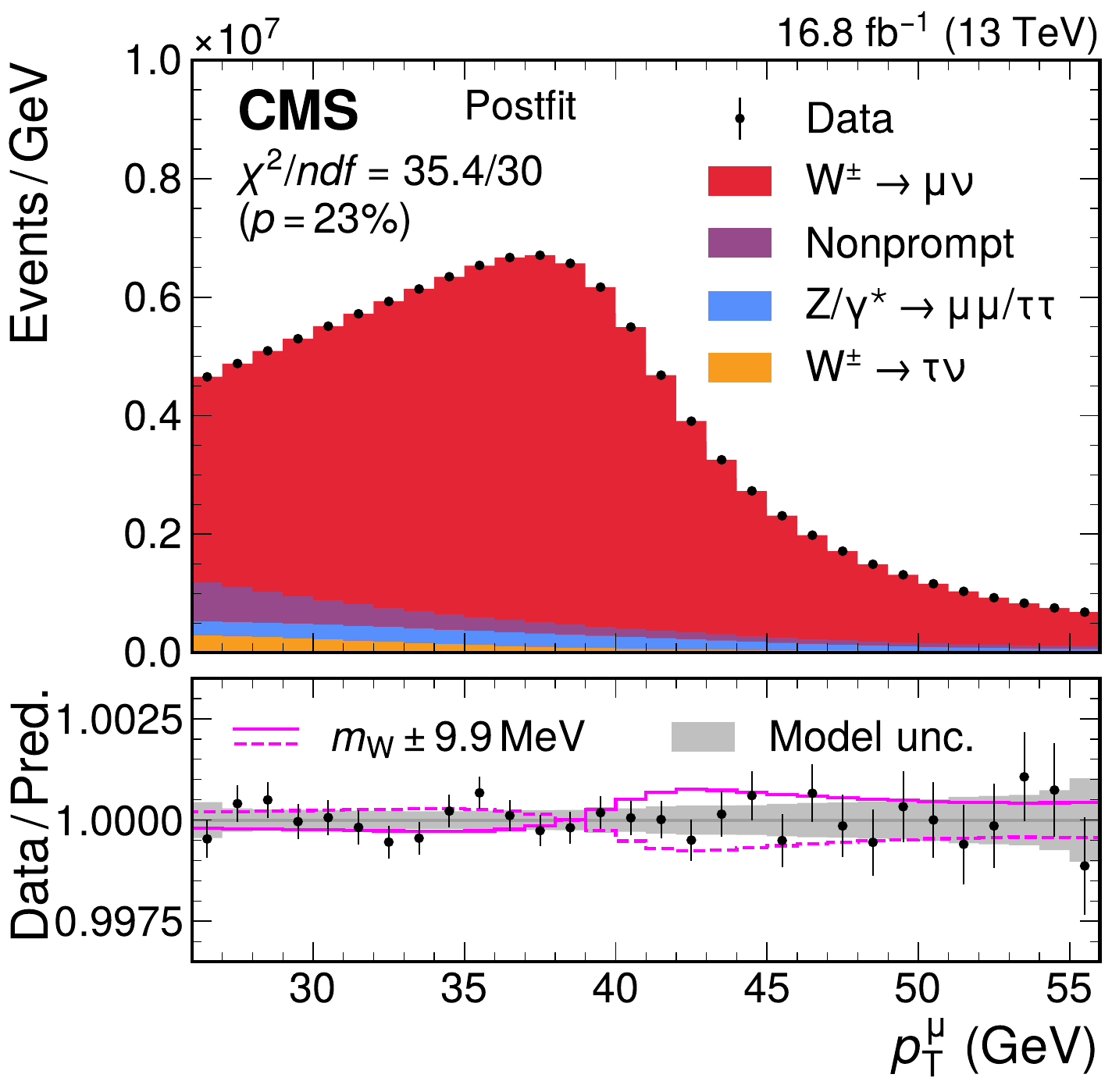}
\caption{\natcap{The \PW boson mass measurement.}
Measured and postfit 
\ptmu distributions, showing the sensitivity to \mw from the
characteristic peak at ${\sim}\mw/2$.
The predicted \wmn contribution, shown in red, 
reflects the measured value of \mw.
The dominant background contributions are shown as 
colored filled histograms.
The bottom panel shows the ratio between the number of events observed in data,
including variations in the predictions,
and the total nominal prediction.
The vertical bars represent the statistical uncertainties in the data.
A shift in the \mw value shifts the peak of the distribution, as illustrated by the
solid and dashed magenta lines, which show an increase or decrease of \mw by 9.9\MeV. 
The total contribution of all theoretical and experimental uncertainties in the predictions, 
after the uncertainty profiling in the maximum likelihood fit, 
is shown by the gray band.}
\label{fig:ptl_mw}
\end{figure}

Figure~\ref{fig:ptl_mw} compares the measured and the postfit predicted \ptmu distributions, 
with \mw adjusted to the observed value. 
The effect on the \ptmu distribution of a 9.9\MeV variation in \mw is shown
to illustrate the degree to which the distribution and its uncertainties
are controlled, enabling the high precision of the measurement.
The main uncertainties in the \mw measurement are due to 
the muon momentum calibration (4.8\MeV) and the PDF uncertainties (4.4\MeV).
Detailed breakdowns of the \mw measurement uncertainty 
are provided in Table~\ref{tab:impacts_all}.
The robustness of the result with respect to the theory model is tested further 
by performing the \mw measurement with the helicity fit, as discussed in Section~\ref{sec:agnostic}.
The result, $80\,360.8 \pm 15.2\MeV$, is consistent with the nominal value.

\section{Discussion}

In this paper we report the first \PW boson mass measurement by the CMS Collaboration
at the CERN LHC. The result is markedly more precise than previous LHC measurements.
The \PW boson mass is extracted from a sample of 
117~million selected \wmn events,
collected in 2016 at the proton-proton collision energy of 13\TeV,
via a highly granular binned maximum likelihood fit to the three-dimensional distribution 
of the muon \ptmu, \etamu, and electric charge.
Novel experimental techniques have been used, 
together with state-of-the-art theoretical models, 
to improve the measurement accuracy.  
The muon momentum calibration, based on \jmm decays, 
as well as the data analysis methods and the treatment of the theory calculations used in the \mw measurement 
have been extensively validated by extracting \mz and \ptz 
both from a direct \zmm dimuon analysis and from a \wlike analysis of the \PZ boson data.

\begin{figure*}[ht]
\centering
\includegraphics[width=0.55\textwidth]{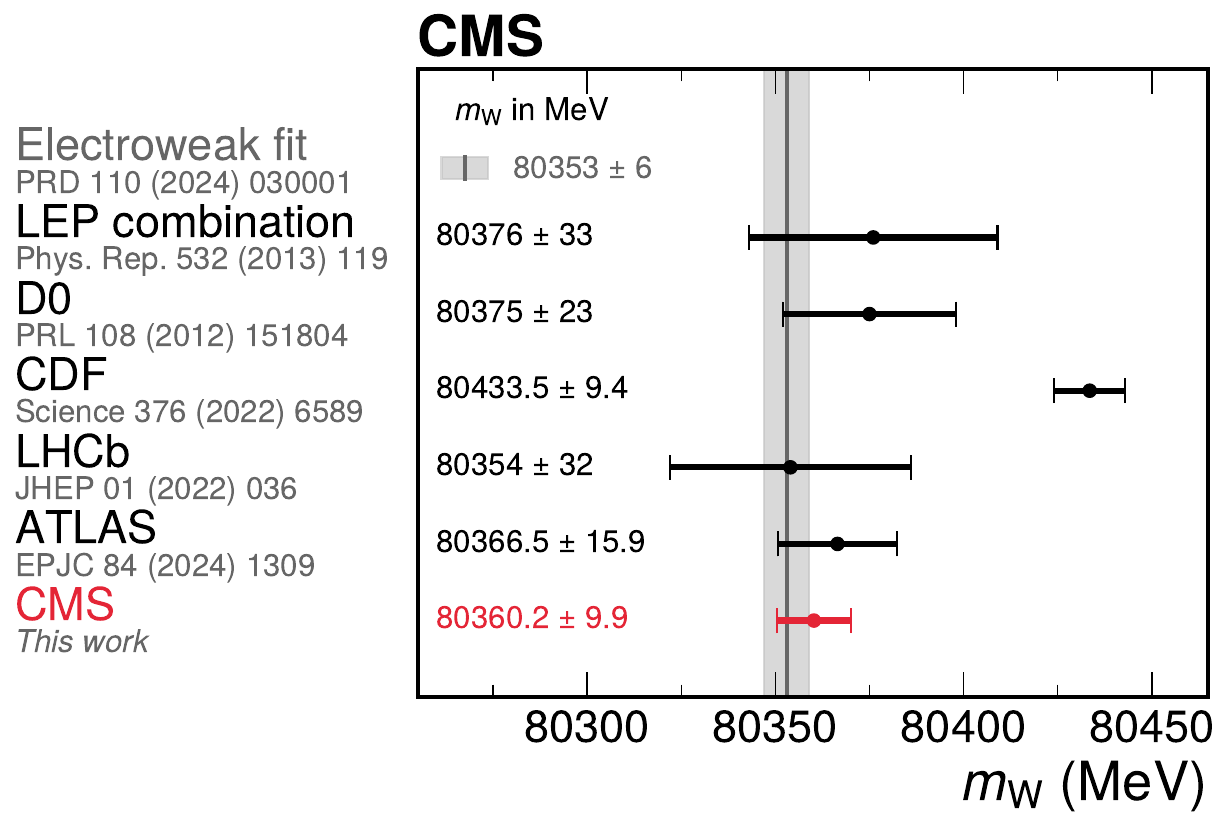}
\caption{\natcap{Comparison with other experiments and the EW fit prediction.}
The \mw measurement from this analysis (in red) is
compared with the combined measurement of experiments at LEP~\cite{ALEPH:2013dgf}, and with the measurements performed by the D0~\cite{D0:2012kms},
CDF~\cite{CDF:2022hxs}, LHCb~\cite{LHCb:2021bjt}, and ATLAS~\cite{ATLAS:2024erm} experiments.
The global EW fit prediction~\cite{PDG2024,Haller:2018nnx,PhysRevD106033003} is represented by the gray vertical band,
with the shaded band showing its uncertainty.}
\label{fig:summary}
\end{figure*}

As shown in Fig.~\ref{fig:summary},
the measured value, $\mw = 80\,360.2 \pm 9.9\MeV$, 
agrees with the standard model expectation from the electroweak fit
and is in disagreement with the measurement reported by the CDF Collaboration.
Our result has similar precision to the CDF Collaboration measurement and 
is significantly more precise than all other measurements.
The dominant sources of uncertainty are
the muon momentum calibration and the parton distribution functions.
Uncertainties in the modeling of \PW boson production are subdominant
due to novel approaches used to parameterize and constrain
the predictions and their corresponding uncertainties in situ with the data.
This result constitutes a significant step 
towards achieving an experimental measurement of \mw with a precision matching that of the EW fit.

\vfill\newpage
\appendix
\numberwithin{figure}{section}
\numberwithin{table}{section}

\section{Methods}
\label{sec:Methods}

\subsection{Event samples and selection criteria}
\label{sec:samples-selection-details}

The dataset used for this analysis, roughly half of the full 2016 sample,
ensures an optimal performance of the CMS detector, especially for the reconstruction
of charged particle tracks~\cite{CMSAlignment2022}.
The data and simulation were processed with the most recent version of the reconstruction software, 
including improvements to particle identification and reconstruction developed for this analysis,
and with the latest detector calibration and description of the operating conditions.

We simulate \PW and \PZ boson production
at NNLO in QCD using the \MiNNLO~Wj and~Zj~\cite{Monni:2019whf,Monni_2020} (rev.~3900) processes
in \POWHEG-\textsc{box-v2}~\cite{Nason:2004rx, Frixione:2007vw, Alioli:2010xd}, 
interfaced with \PYTHIA8.240~\cite{Sjostrand:2015} for the parton shower and hadronization, 
and with \photospp3.61~\cite{Golonka:2005pn,Davidson:2010ew} for final-state photon radiation. 
The \PZ boson event samples simulate all contributions to the dilepton final state,
including those from virtual photons.
We use the CP5 underlying event tune~\cite{CMS:2019csb}, with the hard primordial-\kt parameter set to 2.225\GeV,
obtained from a dedicated optimization using the \ptmumu data of Ref.~\cite{CMS:2019raw}.
The $(G_{\mu}, \mw, \mz)$ and $(G_{\mu}, \sin^2\theta_\text{eff}, \mz)$ 
EW input schemes are used for \PW and \PZ boson production, respectively.
The CT18Z PDF set~\cite{Hou:2019efy} at NNLO accuracy was chosen for the nominal analysis,
before unblinding the result,
given its good description of our \PW and \PZ data
and because the expected shifts in \mw from using other modern PDF sets
are within its uncertainties.
Additional NNLO PDF sets are studied using event-level weights in the \pMiNNLO sample:
NNPDF3.1~\cite{NNPDF:2017mvq}, 
NNPDF4.0~\cite{NNPDF40}, 
CT18~\cite{Hou:2019efy}, 
MSHT20~\cite{Bailey_2021}, and 
PDF4LHC21~\cite{Ball_2022}.
We also consider the MSHT20aN3LO approximate \NtLO PDF set~\cite{Cridge:2023ryv}. 
The \pMiNNLO generator is also used to simulate events 
with \PW or \PZ bosons decaying to \PGt leptons, 
with the same theory corrections on the boson production kinematic distributions 
as those applied to the samples with muonic decays.
To ensure that the MC sample size is not a significant source of uncertainty in the measurement,
simulated samples of more than 4~billion (400~million) \PW (\PZ) boson production events have been produced.
The EW production of lepton pairs or of a \PW boson 
in association with a quark through photon-photon or photon-quark scattering 
is simulated at LO using \PYTHIA8.240~\cite{Bertone:2017bme}.
Top quark and diboson production are simulated at NLO QCD accuracy
using \MGvATNLO v2.6.5~\cite{Alwall:2014hca} and \POWHEG-\textsc{box-v2}~\cite{Melia:2011tj}, respectively,  
interfaced with \PYTHIA8.240 for the parton shower and hadronization.   
Quarkonia production is simulated using \PYTHIA8 
interfaced with \photospp v3.61 for final-state photon radiation.  
Single-muon events have been simulated for additional 
validation of the muon reconstruction and calibration.

While the muon system is not used for the muon momentum evaluation, it is crucial for triggering and identification.
The selected muons must have a reconstructed track in both the silicon tracker and the muon detectors, 
with a consistent track fit for hits in both detector subsystems, 
and pass additional quality criteria to ensure a high purity of the selected events. 
We use the ``medium'' identification working point~\cite{CMS:2020uim}, 
whose efficiency is better than 98\% for signal muons.
The muons must have a transverse impact parameter smaller than 500\mum with respect to the beam line
and be isolated from hadronic activity in the detector. 
The muon isolation is defined as the pileup-corrected ratio between \ptmu and
the sum of the \pt of all other reconstructed physics objects within a cone centered around the muon~\cite{CMS:2018rym}. 
The isolation of selected muons must be smaller than 15\%,
using a cone of radius $\Delta R = \sqrt{\smash[b]{(\Delta\phi)^2 + (\Delta\eta)^2}} = 0.4$,
where $\Delta\phi$ and $\Delta\eta$ are, respectively, the distance in the $\phi$ and $\eta$ 
coordinates between the muon and the physics objects considered in the sum.
Only charged particles 
within 2\unit{mm} of the muon track along the beam axis are considered in the isolation sum. 
The distance is evaluated between the points of closest approach to the beam line for each track.
The same criteria are used to select charged particles used in the \ptmiss calculation.
Our definition differs from the standard CMS approach, where
charged particles in the isolation and \ptmiss sums are defined with respect to the vertex
that maximizes the sum of $\pt^2$ of the associated physics objects~\cite{CMS-TDR-15-02}. 
This change of definition is needed to minimize the rate at which the wrong vertex is chosen,
which is negligible in \zmm events but, with the standard CMS algorithm,
ranges from 1 to 5\% for \wmn events, depending on \ptw. 
Indeed, to ensure the validity of the isolation and \ptmiss corrections 
measured with \zmm events and applied to \wmn events
(as described in Sections~\ref{sec:eff} and~\ref{sec:recoil})
it is important to make sure that there are no differences in their dependence on the vertex selection.

Muons used in both the \mz and \mw analyses are selected by the same trigger, 
requiring the presence of at least one muon with $\ptmu > 24\GeV$, to guarantee 
maximal consistency in terms of 
event selection and efficiency corrections.
Events with electrons (or additional muons) with $\pt > 10$\,(15)\GeV 
satisfying looser identification criteria are rejected~\cite{CMS:2018rym, CMS:2020uim}.
In the \mw analysis, the selected muon must have $26 < \ptmu < 56$\GeV.
The upper threshold is increased to 60\GeV for the \wlike \mz measurement.
These thresholds restrict the selected events to the \ptmu range where the 
trigger and reconstruction efficiencies are measured most accurately.
The selected muon must be geometrically matched to the object that triggered the event, within a cone of radius $\Delta R = 0.3$.
In the \wlike analysis, where two muons are reconstructed,
the matching is only required for the muon used to form the \etaptmu template.
This choice avoids the need to evaluate correlations 
in the triggering efficiency 
in events where both muons satisfy the trigger requirements.
For consistency with the W boson selection,
\wlike events must satisfy $\mt > 45$\GeV ($\sim\!\mv/2$). In this case, \mt is calculated 
from the selected muon and the \ptvecmiss value obtained by excluding
the other muon from the vector sum.

Events are rejected if they contain electrons with $\pt >10$\GeV 
satisfying the identification criteria of the veto working point
(which has 95\% efficiency for genuine electrons~\cite{CMS:2018rym})
or additional muons of $\pt >15$\GeV matching the loose criteria
(with an efficiency above 99\% for real muons~\cite{CMS:2020uim}).
The electron veto rejects the residual contribution of events from top quark and boson pair production, 
and from \ztt decays with one $\tau$ lepton decaying to a muon and the other to an electron. 
The electron veto efficiency has a negligible impact on the analysis. 
The muon veto efficiency and the corresponding uncertainties are discussed in Section~\ref{sec:eff}.

The single muon selection efficiency is 85\%,
evaluated from simulated \wmn and \zmm events.
The fraction of \wmn events in the selected data sample is 89\%.
The signal purity of the selected dimuon sample is larger than 99.5\%, 
given the stronger suppression of the backgrounds due to the double muon selection and invariant mass requirement.
While the \wlike \mz analysis provides a stringent test of the analysis strategy
in an almost background-free environment, 
the significant background from nonprompt muons in the \mw analysis must
be validated by other means,
as discussed in Section~\ref{sec:ABCD}.

\subsection{Efficiency corrections}
\label{sec:eff}

The \mw measurement is based on a fit to the measured \etaptqmu distribution
using simulated templates for the signal and most background processes. 
Therefore, it is important that the simulation can accurately reproduce 
the efficiency of the event selection in the \etaptmu bins used in the analysis.
Corrections to the simulated muon efficiencies are determined from data 
with the tag-and-probe (T\&P) method~\cite{CMS-PAPERS-EWK-10-002}, 
using events from the same \zmm sample that we use in the analysis, 
except that we apply a looser event selection.

The efficiencies are measured differentially in \etaptmu for different 
stages of the muon selection, factorized as: reconstruction of a standalone track in the muon chambers; matching of a standalone muon with a track in the tracker to form a global muon candidate (tracking); impact parameter and identification quality criteria of the global muon track; trigger selection; muon isolation.
The efficiencies are evaluated in the measured and simulated event samples, 
for each of the five sequential stages,
and their ratios are used as scale factors (SFs) to reweight the simulated events.
Efficiencies are determined from the fraction of selected events where the probe muon passes the selection whose efficiency is being evaluated. 
Background events with at least one nonprompt muon are subtracted when computing the efficiency in data.
These background contributions are estimated by fitting the sum of a signal and a background model to the observed \mmm distribution. 
The \zmm contribution is modeled by a simulated template from the \MiNNLO sample, convolved with a Gaussian shape to account for differences in the momentum scale and resolution between data and simulation. An alternative signal model, defined by the convolution of a Breit--Wigner distribution and a resolution function that has a Gaussian core and asymmetric exponential tails, is used to assess the systematic uncertainty. The background component is modeled using an exponential function, except for the reconstruction and tracking steps in the failing probe samples, for which the background fraction is large and its shape at low \mmm is sculpted by the \ptmu selection. For these steps, the background model is an exponential decay distribution that transitions to an error function for $\mmm < \mz$ to capture threshold effects. Third- or fourth-order polynomials are tested as alternative background shapes.

Misalignment or other effects in the reconstruction of tracks in the muon chambers, 
which are used for triggering and identification purposes, 
can result in charge dependent biases in the measured efficiencies. 
To properly account for them, efficiencies are measured separately for each muon charge
except for the isolation step, for which the charge asymmetry is found to be negligible. 
The largest asymmetry is in the trigger SFs, rising to 5\% in the most forward region of the detector and for $\ptmu < 35\GeV$.
The muon isolation is sensitive to the 
vector sum of the momenta of charged and neutral hadrons in the event, referred to as the hadronic recoil (\utvec). 
The angular distance between the muon and \utvec is different between \wmn and \zmm events, 
leading to a bias in the muon isolation efficiency measured using \PZ boson decays. 
For a given \ptz value,
the bias is larger for low \ptmu, 
when the muon is more likely produced in the direction opposite to that of the \PZ boson \pt
and in the vicinity of the recoil.
The trigger efficiencies are also affected 
because of the isolation requirement applied at the trigger level. 
To account for this effect, 
the trigger and isolation efficiencies are measured triple-differentially 
in the muon \etaptmu and 
in the projection of \utvec along the \ptmu direction, \utmu.
The corrections are applied to \wmn events using the \PW boson
recoil, after correcting its distribution as described in Section~\ref{sec:recoil}.

The statistical uncertainty in the SFs originates from 
the limited sample of measured and simulated \zmm events in the T\&P estimate, 
whereas systematic uncertainties stem from the modeling of the \zmm mass distributions
with signal and background components 
when extracting the efficiencies in the measured event sample. 
We evaluate these systematic uncertainties by 
repeating the efficiency measurements in the data sample after varying the signal or background models, 
taking the difference with respect to the nominal efficiency as the uncertainty. 

To mitigate the effects of statistical fluctuations and discrete bin edges, 
the SFs are smoothed as a function of \ptmu using a polynomial interpolation.
Third-order polynomials (second-order for the tracking step) properly model the \ptmu dependence of the binned SFs, as determined with statistical tests.
Trigger and isolation SFs are smoothed using two-dimensional polynomials in the \ptut space, 
with third (second) order in the \ptmu (\utmu) direction.
The smoothing acts independently on each \etamu bin, and no smoothing is performed versus \etamu because physical boundaries in the detector 
might produce genuine discontinuities in the \etamu dependence of the efficiency.
Instead, a smooth dependence on \ptmu is expected in the momentum range of interest.
The smoothing simplifies the treatment of the SF statistical uncertainties in the analysis fit
and also leads to reduced uncertainties in \mw
by imposing that the measured SFs are correlated across the \ptmu or \ptut bins.
We have verified with pseudo-data tests that, within the SF uncertainties, neither the smoothing procedure nor the chosen polynomial order induces a bias in the measured value of \mw.

Statistical uncertainties in the SFs are implemented in the likelihood as 2784 nuisance parameters, defined from the independent variations of the smoothing fit parameters according to the eigenvectors of their covariance matrix for each of the 48 \etamu bins. 
The number of \ptmu variations is determined by the order of the smoothing polynomial used for each step. 
These nuisance parameters are uncorrelated versus \etamu and \qmu, and modify the \ptmu spectrum in a continuous way.

Systematic uncertainties in the SFs are estimated as the difference, after the \ptmu smoothing, 
between the nominal and alternative SFs resulting from the variation of the signal or background models in the T\&P mass fits.
These are correlated across the \etaptqmu bins, because the same signal and background models are employed in all T\&P fits.
However, we also implement additional uncertainties uncorrelated among \etamu bins, 
resulting in 49 nuisance parameters assigned to each efficiency step
to account for the change in the T\&P signal model. Similar uncertainties are implemented for the reconstruction and tracking SFs to reflect the change in the background model.
In total, the systematic uncertainty in the SFs is encoded in 343 nuisance parameters, correlated between muon charges.
The statistical and systematic components of the SF uncertainties, after the smoothing,
have a similar contribution to the uncertainty in \mw, 
and their combined effect is 3.0\MeV.

Dedicated SFs and uncertainties are derived for the muon veto selection employed in the single-muon analysis. 
These SFs are used to correct the simulated yields of the \PZ boson background process 
in events where the second prompt muon falls inside the \etaptmu acceptance window
but fails the reconstruction or identification criteria of the veto.
This component of the \zmm background is characterized by a \ptmu distribution for the selected muon similar to that of \PW boson decays, but with the peak located at higher values of \ptmu. Moreover, because of the high efficiency of the veto selection, close to unity in many \etaptmu bins, small efficiency variations between
data and simulation can result in relatively large corrections for the probability to fail the veto. Therefore, even though this background component constitutes a small fraction of the total \zmm background, its shape and normalization must be accurately controlled to avoid a bias in the measured \mw. The veto SFs are determined and smoothed with the same technique as for other SFs, but are applied to simulated events as a function of the \etaptmu of the second generator-level muon
(evaluated after final-state radiation), 
taken as a proxy for the nonreconstructed muon. 

The veto SFs are measured for $\ptmu > 15 \GeV$, split by \qmu and factorized as three independent terms accounting for muon reconstruction, tracking, and loose identification.
They differ from the nominal analysis SFs because of a slight tuning of the matching criteria between the inner and outer muon track to cope with the lower \ptmu threshold.
Uncertainties in veto SFs are encoded in 581 nuisance parameters, affecting only the \zmm background. The statistical uncertainty derives from varying the parameters of the \ptmu smoothing polynomials, independently in each \etamu bin and \qmu. Systematic uncertainties, related to the \mmm modeling in the T\&P fits, are implemented following the same \etamu granularity and correlation scheme as the standard SFs.

The contribution of the veto SFs to the uncertainty in \mw is smaller than 0.5\MeV, 
reflecting the fact that the \PZ boson background sensitive to these SFs is strongly suppressed by the veto.
The nominal muon veto restricts the selection to ``global muons,''
which have a high-quality track in both the tracker and muon detectors.
An alternative definition has also been tested, 
with the muon inner track not required to be 
matched to a track reconstructed in the muon detectors. 
This looser selection has higher efficiency for prompt muons and, therefore, 
provides better rejection of the \zmm background, 
at the cost of larger systematic uncertainties in the measured veto SFs 
because of the combination of different categories of reconstructed muons.
Tests using pseudo-data generated with either veto selection have been carried out, 
showing that the measured \mw values agree within less than 0.1\MeV between the two veto selections
and that the residual bias in \mw is covered by the veto SF uncertainties. 

Further corrections and corresponding uncertainties are applied to the simulated events. 
Partial mistiming of signals in the muon detectors led to the incorrect assignment of the triggered event 
to the previous proton bunch crossing for a small fraction of events~\cite{CMS:2020cmk}. 
This is known as ``prefiring", and caused a reduction in the trigger efficiency.
A correction for this effect is determined in bins 
of \etamu and \ptmu~\cite{CMS:2021yvr}.
The correction increases with \etamu and varies between 0.5\% and 2\%. 
A similar issue originating from the prefiring of the electromagnetic calorimeter triggers
affects the analysis through hadronic jets containing photons or electrons not rejected by the veto. 
The total contribution of the prefiring to the uncertainty in \mw is about 0.7\MeV.

\begin{figure}[ht]
\centering
\includegraphics[width=0.45\textwidth]{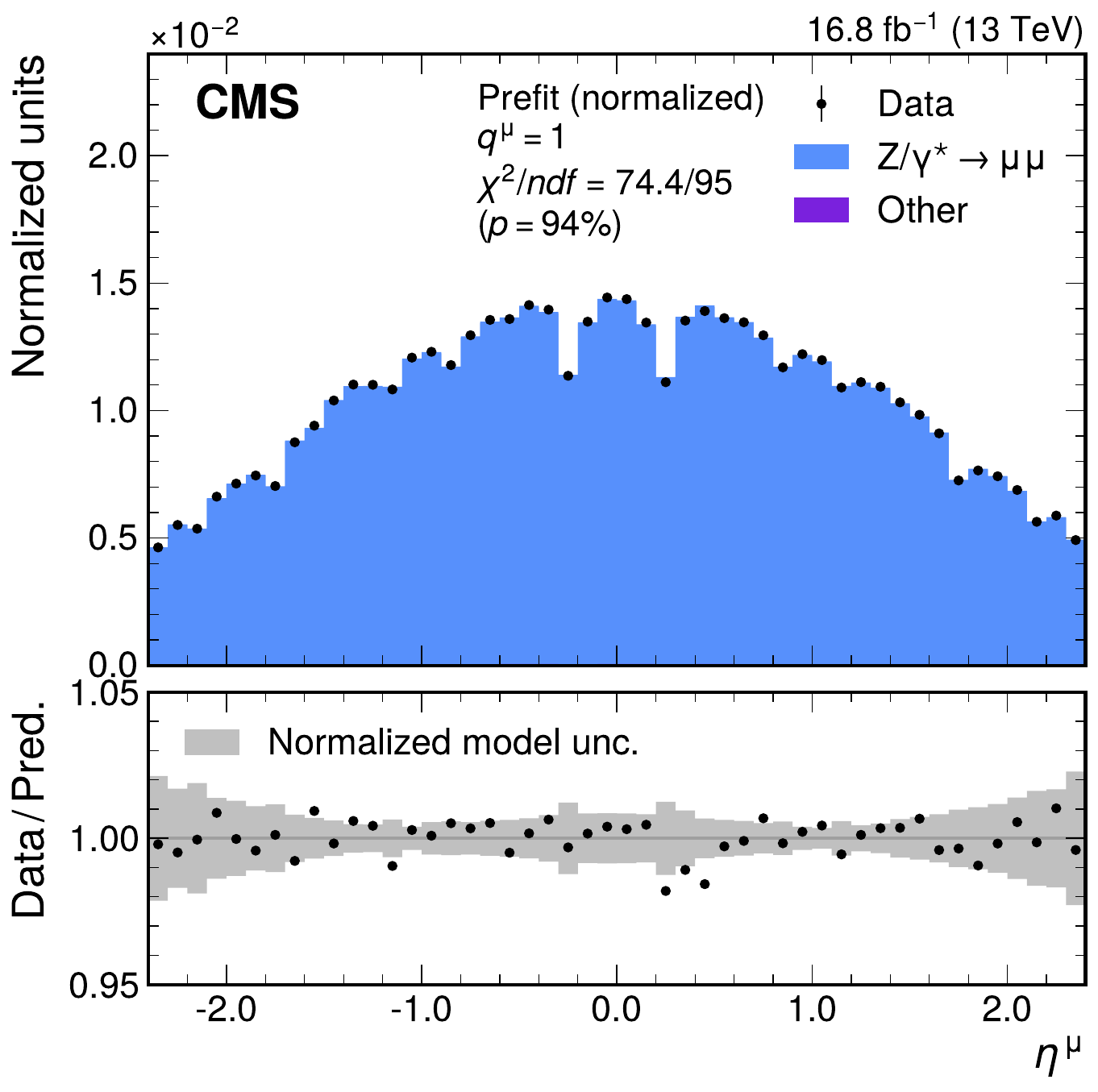}
\includegraphics[width=0.45\textwidth]{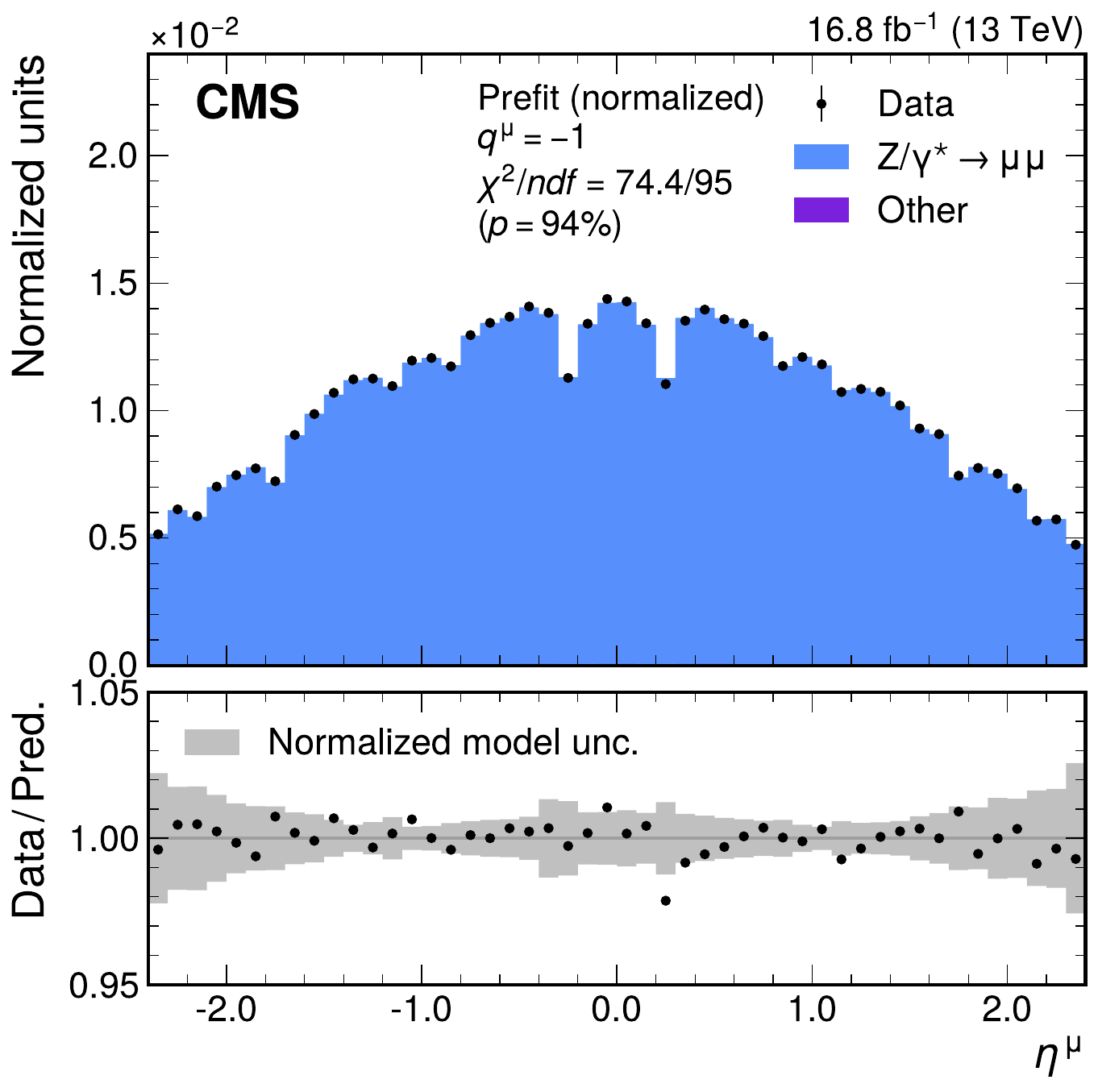}
\caption{\natcap{Measured and predicted \etamu distributions in \zmm events with the \wlike \PZ boson selection 
for positively (left) and negatively (right) charged muons.}
The normalization of the simulated spectrum is scaled to the measured distribution
to better illustrate the level of agreement between the two. 
The shaded bands correspond to the total (statistical and systematic) uncertainty on the normalized distributions.
The normalization of the prediction is scaled by about 1\%, 
which is less than the luminosity and total cross section uncertainties.
The vertical bars represent the statistical uncertainties in the data. 
The bottom panel shows the ratio of the number of events observed in data
and of normalized variations in the predictions
to that of the total nominal prediction.}
\label{fig:muon_pt_Wlike_1Deta}
\end{figure}

The quality of the experimental corrections applied to the simulation is validated using \zmm events, 
which offer a pure sample of prompt muons, 
comparing the predicted distribution of the selected muon \etamu with the measured one, for each muon charge, 
as shown in Fig.~\ref{fig:muon_pt_Wlike_1Deta}. 
The agreement between measured and simulated data is within 2\% in all bins, 
and the difference is covered by the uncertainty.

\subsection{Hadronic recoil calibration}
\label{sec:recoil}
 
To further improve the modeling of \met and \mt in the simulation, 
hadronic recoil corrections are derived using measured \zmm events,
by exploiting the relation between \utvec and the 
\ptmumu vector, $\utvec = -\ptmumuvec$.
The components of \utvec parallel and perpendicular to \ptmumuvec are
modeled independently with spline-based parameterizations as functions of \ptmumu,
for both data and simulated events.
The parameterization yields
two-dimensional probability distribution functions, expressed in terms of 
\ptmumu and the \utvec component.
Although the resulting parameters are not directly physical, they provide a smooth and flexible description of the 
\utvec magnitude (\ut) over the full \ptmumu range.

\begin{figure}[ht!]
\centering
\includegraphics[width=0.45\textwidth]{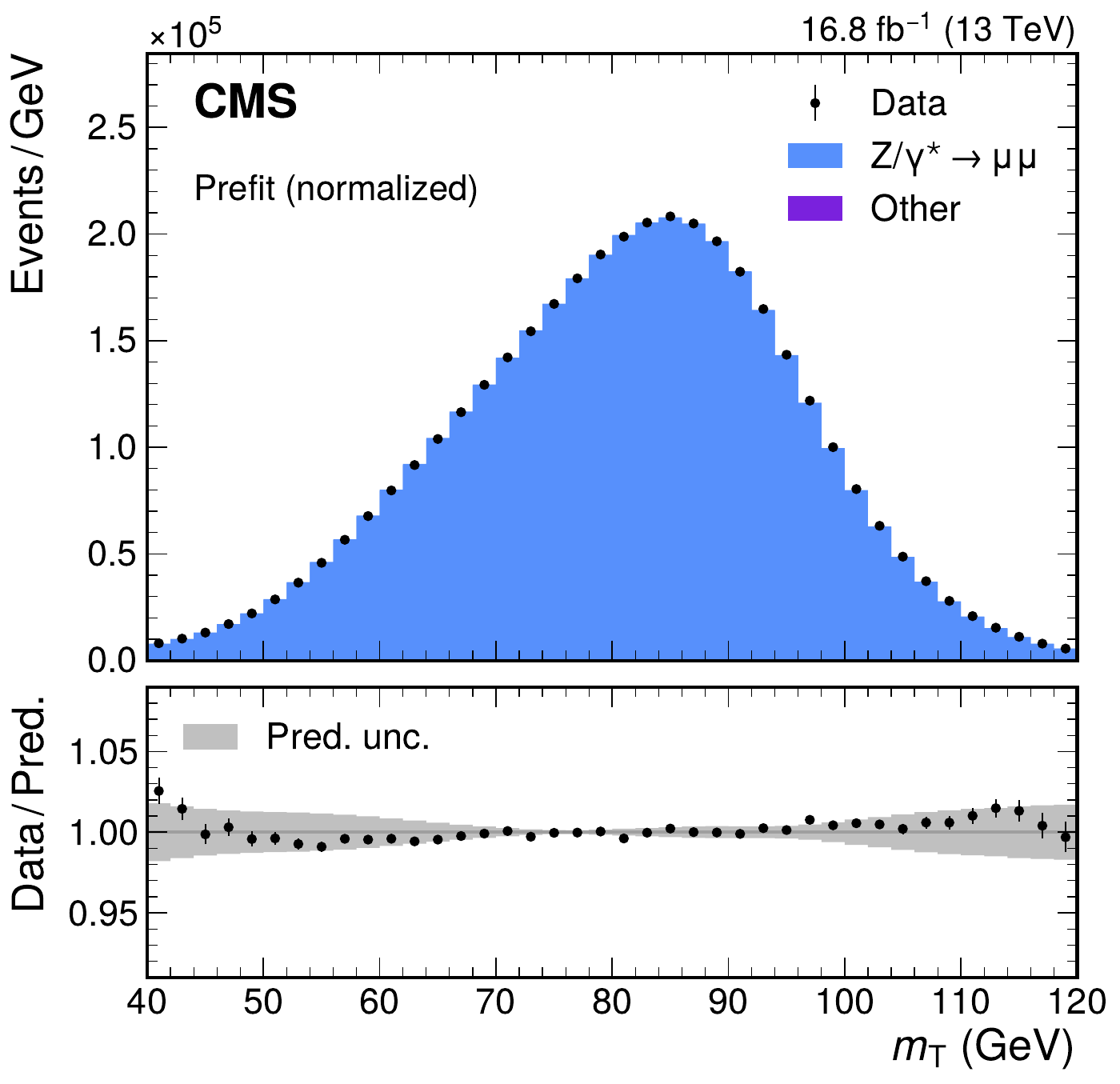}
\includegraphics[width=0.45\textwidth]{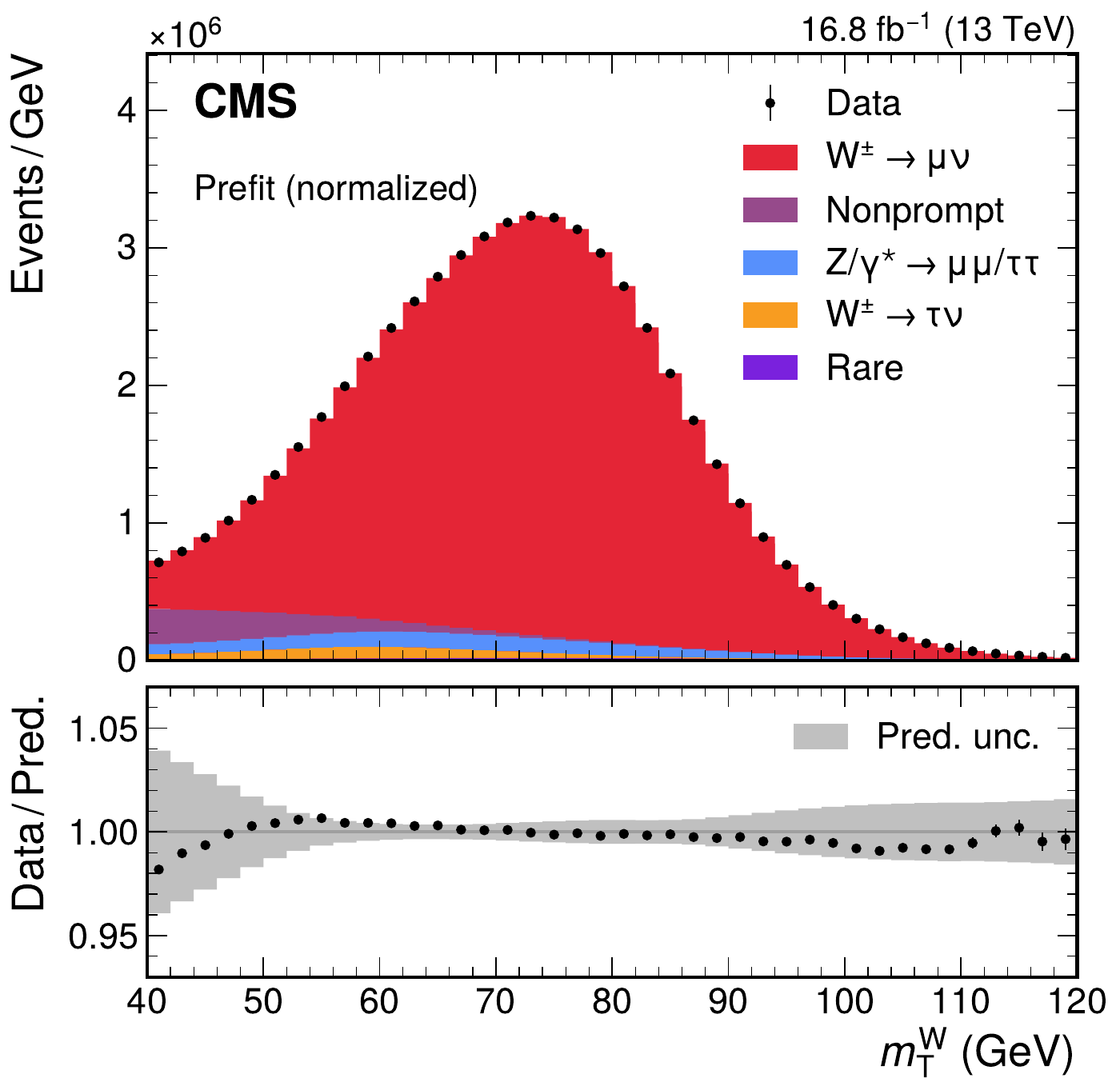}
\caption{\natcap{Measured and predicted \mt distributions in \zmm and \wmn events, 
after calibrating \ut.}
Given that \mtw is an input variable, the extended ABCD method does
not give suitable predictions for the \mtw distribution. Therefore,
the nonprompt background contribution to \PW boson production 
is estimated from the QCD multijet simulation.
The predictions are those prior to the fit to data.
The total uncertainties (statistical and systematic) are represented by the gray band 
and the normalization of the simulated spectrum is scaled to the measured distribution 
to better illustrate their agreement.
The vertical bars represent the statistical uncertainties in the data.
The bottom panel shows the ratio of the number of events observed in data
and of variations in the predictions
to that of the total nominal prediction.}
\label{fig:mt_recoil1}
\end{figure}

Subsequently, for each simulated event, a new value of \ptmiss is computed using an inverse cumulative distribution function transformation of the function mapping the simulated templates to data.
The corrections, derived from \zmm events and parameterized in \ptmumu, are applied to simulated \wmn events as a function of the \ptw,
where the \ptw is built from the reconstructed muon and the generator-level neutrino.
Figure~\ref{fig:mt_recoil1} shows the \ut-corrected transverse mass distribution for \zmm and \wmn events.
\insupp{Figure 18 in the \suppmat}{Figure~\ref{fig:recoil_calib}} shows the impact of the \ut correction on the parallel and perpendicular components of \utvec for \zmm events.
Apart from a slight disagreement in normalization between the measured and simulated distributions 
(unrelated to \ut and accounted for by other uncertainties), 
the scale and resolution of the corrected \ut in simulation match those of the data at the subpercent level.

The uncertainty in the corrections is evaluated from the statistical uncertainty of the fits 
that parameterize the correction of the simulation to the data. 
Their impact on \mw is assessed by varying the correction parameters 
according to the eigenvectors of the fit covariance matrix.
We have verified that their contribution to the uncertainty in \mw is below 0.3\MeV.
Because these variations are computationally expensive to evaluate, and 
their contribution to the \mw uncertainty is negligible,
they are not included in the nominal fit configuration.

\subsection{Nonprompt-muon background determination}
\label{sec:ABCD}

The nonprompt-muon background consists primarily of events 
where muons originate from decays of heavy-flavor hadrons
Despite the large suppression applied by the muon selection criteria, 
a significant contribution from this background remains in the \PW boson selection.
We evaluate it using data from sideband regions defined by 
inverting the \mt selection, the muon isolation requirement, or both. 
To account for correlations between the isolation and the \mt sideband regions,
the ``extended ABCD method'' proposed in Ref.~\cite{choi2021improved} is used.
In this method, 
the low-\mt sideband region is divided into two regions with $\mt < 20\GeV$ or $20 < \mt < 40\GeV$, 
each one further split into events passing or failing the muon isolation criterion,
such that the signal region is complemented by five sideband regions of isolation and \mt, 
compared with the typical three of the classic ABCD method.
The extended ABCD method accounts for a linearly varying isolation efficiency as a function of \mt, exploiting the two low-\mt regions to extrapolate the expected isolation efficiency to the third high-\mt region, contrary to the standard ABCD approach where a constant efficiency is assumed across the entire \mt space. 

In each sideband region, the nonprompt-muon component is evaluated by subtracting from the data 
the contribution of processes with prompt muons, estimated from simulation. 
For each bin of \etaptqmu, the two low-\mt regions are used to obtain a 
transfer factor that is applied to events that satisfy the \mt selection of the signal region but fail 
the muon isolation requirement, to obtain an estimate of the nonprompt-muon background in the signal region.

To reduce the impact on \mw of the statistical fluctuations in the nonprompt-muon background template, the \ptmu distribution from the ABCD prediction is smoothed using an exponential of a third-order polynomial. 
The minimum order of the polynomial to correctly describe the \ptmu shape is determined based on statistical tests.
The statistical uncertainties of the data are accounted for by propagating the uncertainties 
in the smoothing function parameters through the analysis.
This procedure results in 384 independent variations reflecting the four coefficients of the smoothing polynomials,
the two muon charges, and the 48 \etamu bins. 
These uncertainties change both the shape and normalization of the \ptmu distribution in each \etamu bin.

The prompt-muon contamination in the sideband regions is modeled with simulated events, 
with all the corrections applied as for the signal region, 
including the appropriate combination of SFs for events that 
fail the isolation requirement in the nonisolated sideband regions.
All experimental and theoretical systematic uncertainties in the prompt-muon contamination 
are propagated to the sideband regions by repeating 
the subtraction of the prompt component and the determination of the smoothing parameters 
in the sideband regions for each variation. 
In this way, uncertainties stemming from experimental or theoretical sources 
are also assigned to the nonprompt-muon background, 
and the correct correlation structure between prompt- and nonprompt-muon events 
is consistently taken into account in the uncertainty model. 

The extended ABCD method is validated using both simulated nonprompt-muon events from QCD multijet production and a control sample of data enriched in events with nonprompt muons.
Simulated background events permit a stringent test of the internal consistency of the method with no signal contamination. However, they might not accurately describe the background processes in data and a complementary data-driven check with the control region is necessary.
The control region selects nonprompt muons matched to secondary vertices, which originate from the decay of heavy flavor hadrons and appear displaced from the beam line in the transverse plane. Other selection criteria are the same as in the signal region.
The signal contamination in this sample is below 2\%.

We observe that the extended ABCD method overestimates the nonprompt-muon yields 
in the data control region. 
This overestimate is also seen 
in simulated QCD multijet events when comparing the prediction of the extended ABCD method applied to the simulation to the direct prediction in the high-\mt and low-isolation region.
This discrepancy originates from the nonlinear correlation between the nonprompt-muon isolation efficiency and \mt. To account for this effect, a correction is applied to the predicted yields. This correction has a stable value of 85\% across different bins of \etaptmu and is used to scale down the overall normalization of the prediction. This value is consistent between data and simulation in the control region and, using simulated events, is also confirmed by testing the method directly in the signal region.
A 5\% uncertainty is assigned to the predicted background normalization, which covers the largest differences observed when validating the extended ABCD region in the data control region and in simulation.

\begin{figure*}[htp]
\centering
\includegraphics[width=0.48\textwidth]{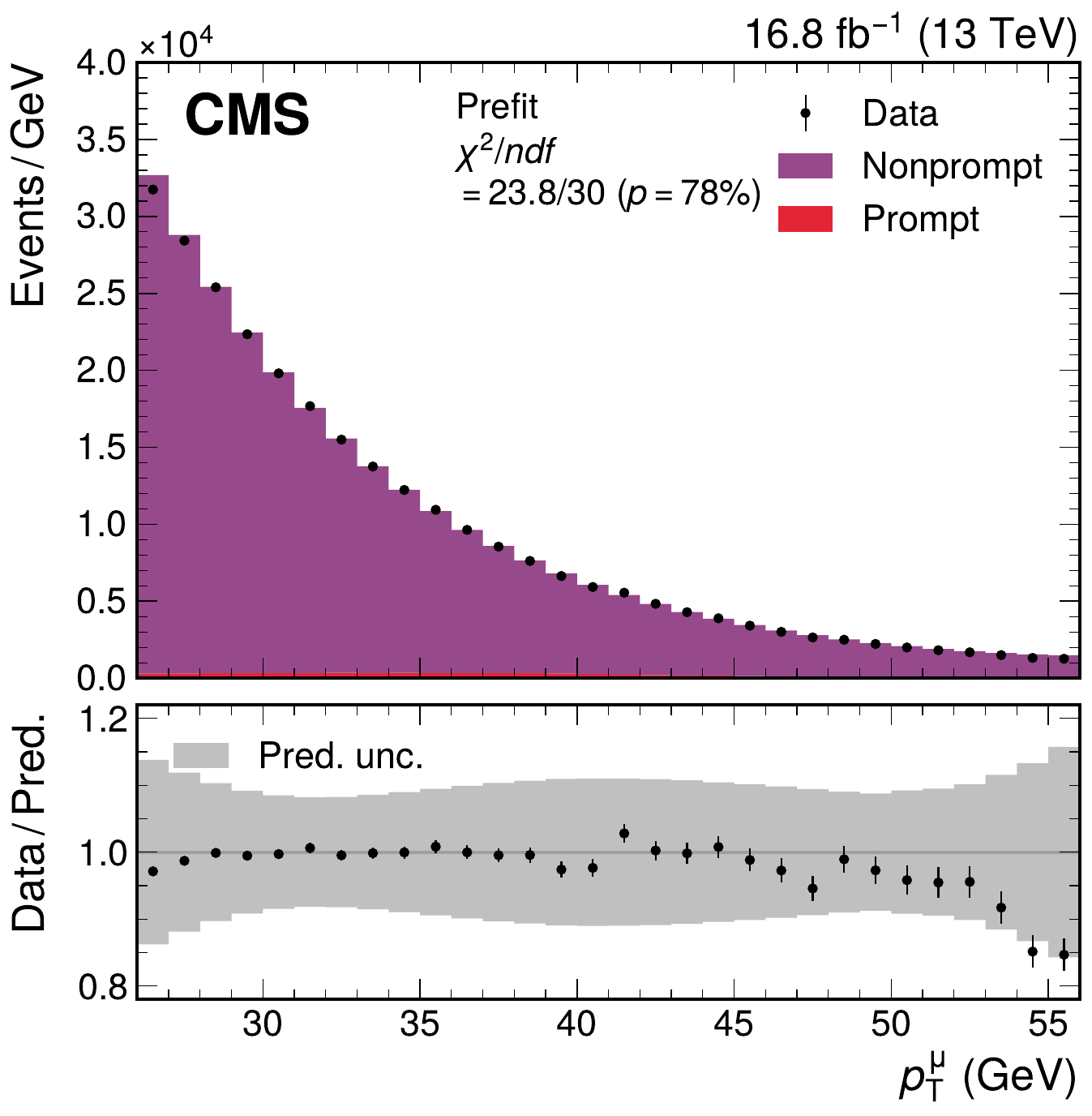}
\includegraphics[width=0.48\textwidth]{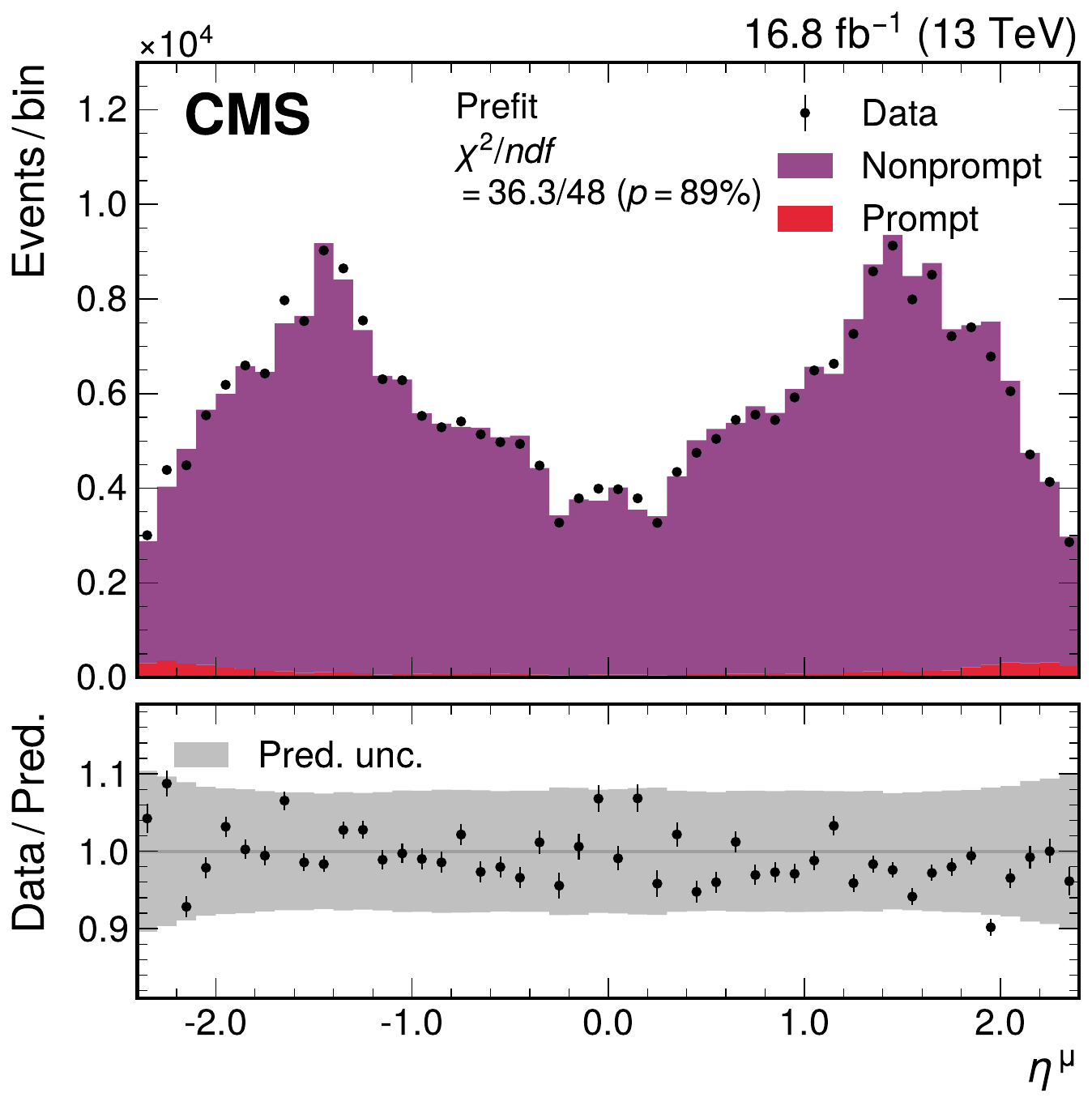}
\includegraphics[width=0.48\textwidth]{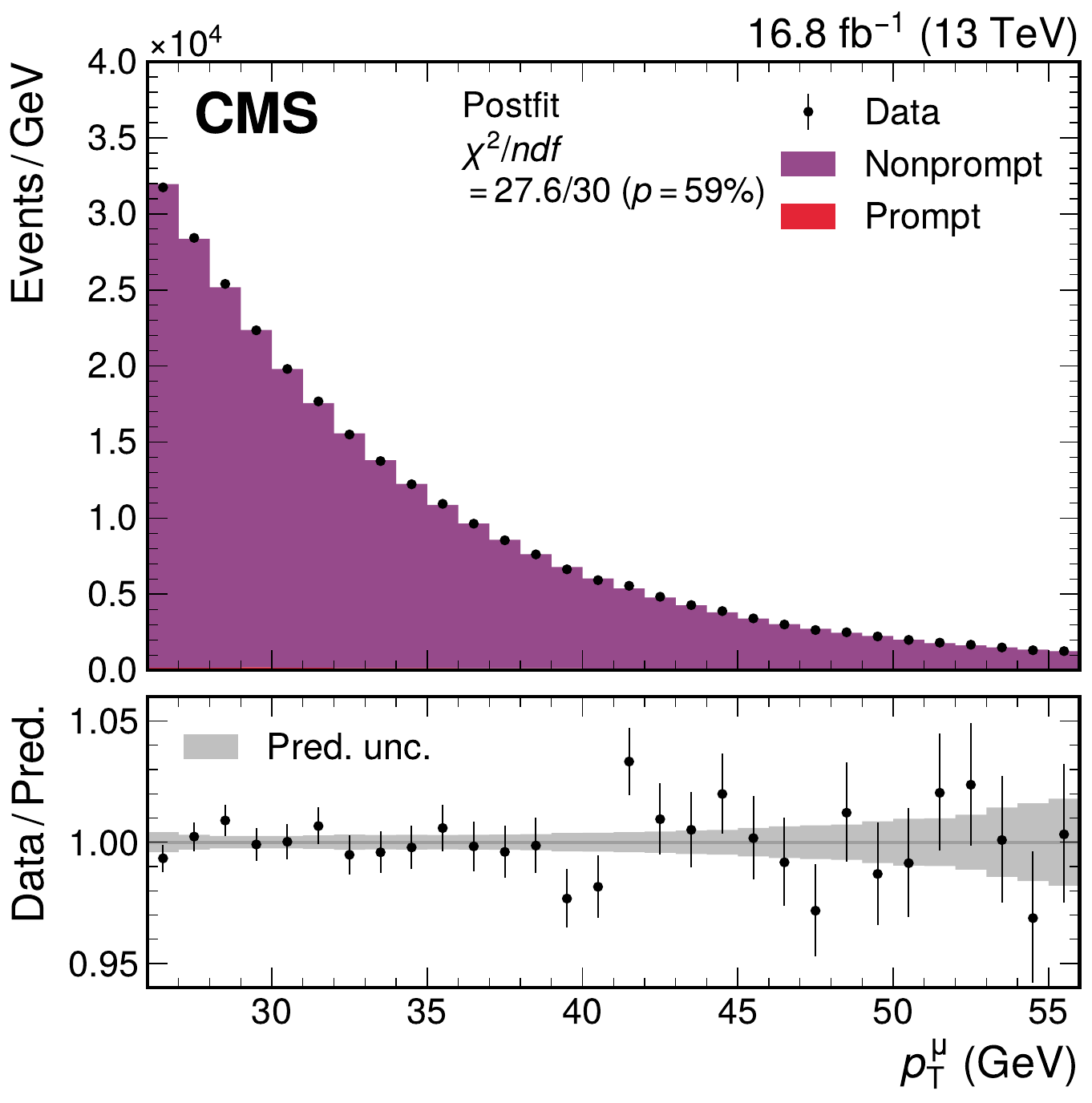}
\includegraphics[width=0.48\textwidth]{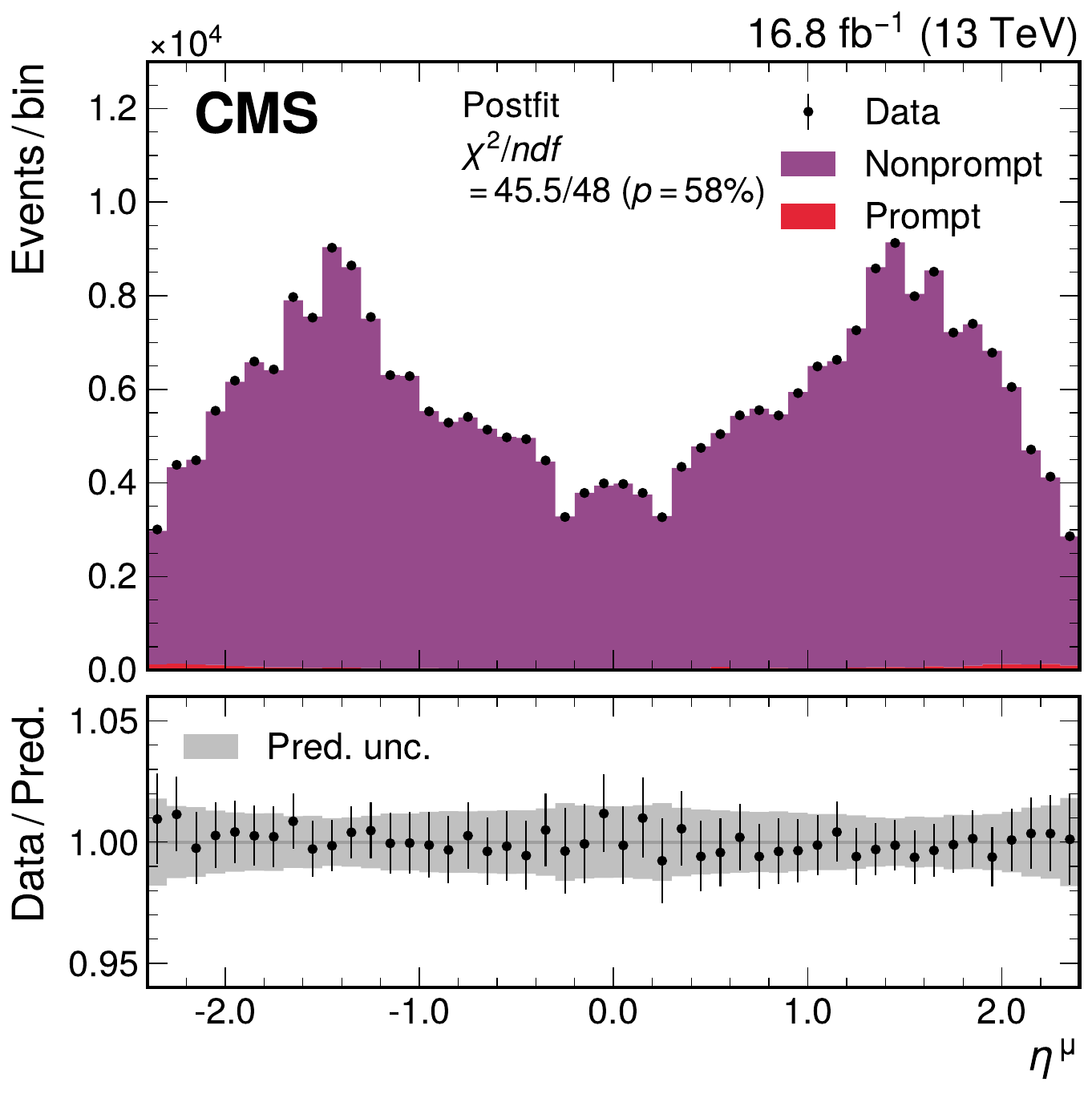} 
\caption{\natcap{The observed data and the prediction of the extended ABCD method
before (top) and after (bottom) the maximum likelihood fit, 
for the \ptmu (left) and \etamu (right) distributions, 
in a control region enriched in events with nonprompt muons 
matched to displaced secondary vertices.} 
Small contributions from events with a prompt muon, 
evaluated using simulated samples, are shown by the red histogram. 
The total uncertainties (statistical and systematic) are represented by the gray bands.
The vertical bars represent the statistical uncertainties in the data.
The bottom panel shows the ratio of the number of events observed in data
and of variations in the predictions
to that of the total nominal prediction.}
\label{fig:QCDBackground_SVCR_prefit}
\end{figure*} 

We test the background uncertainty model by performing maximum likelihood fits to the \etaptqmu distribution in the control region with secondary vertices, both with simulated and data events.
To cover residual shape effects, two additional nuisance parameters are assigned to vary the linear and quadratic coefficients of the smoothing polynomial, fully correlated across \etamu and \qmu bins. The prefit agreement between the nonprompt-muon background predicted by the extended ABCD method and the data in the control region, after correcting the normalization, is shown in Fig.~\ref{fig:QCDBackground_SVCR_prefit}. 
The uncertainty model covers the discrepancies in the shape and normalization, as confirmed by the postfit distributions in Fig.~\ref{fig:QCDBackground_SVCR_prefit} and the goodness-of-fit values.
Finally, we perform a test with biased pseudo-data directly in the signal region. 
In this test, the pseudo-data differ from the nominal prediction by the difference between the extended ABCD prediction and the direct prediction in high-\mt and low-isolation region, 
evaluated from the QCD multijet simulation. 
The shift in \mw from the fit to the biased pseudo-data is within the nonprompt-muon background uncertainty.

The total uncertainty in \mw from the nonprompt-muon background is 3.2\MeV.
It includes the normalization uncertainty (0.2\MeV), the two systematic variations of the coefficients (2.5\MeV), 
and the statistical uncertainty of the smoothing function coefficients (1.9\MeV).
Additional uncertainties result from variations of the predicted nonprompt-muon background due to
experimental and theoretical effects, 
which modify the prompt-muon contamination that is subtracted from data in the sideband regions.
They are accounted for as part of the corresponding 
experimental and theoretical uncertainties in \mw from the respective sources.
We verified that adopting an alternative smoothing algorithm shifts the observed \mw
by less than the associated uncertainty.

\subsection{Muon momentum calibration}
\label{sec:muon-reco}

The muon tracks are first reconstructed using a standard pattern recognition 
and Kalman filter track fit~\cite{CMS:2014pgm}.
To improve the accuracy of the track parameter determination, 
the tracks are then refitted using a continuous variable helix (CVH) fit, 
a global $\chi^2$ fit that 
extends the generalized broken-line fit~\cite{BLOBEL20111760, BLOBEL200614} 
to incorporate continuous energy loss and multiple scattering from finite material elements. 
The detailed material model of the CMS detector used for our simulation 
is based on the initial design of the tracker material and support structures as well as 
in situ measurements using collision data, such as Ref.~\cite{CMS:2018wqs}.
This model is incorporated into the track fit 
using the {\GEANTfour}e propagator~\cite{AGOSTINELLI2003250,1610988,ALLISON2016186}.  
To model the magnetic field, we use a parameterization of the detailed 
three-dimensional solenoidal field map~\cite{Klyukhin:2022jmc} rather than the less accurate, 
but computationally faster, finite-element model used in the standard reconstruction.  
The starting point for the alignment corresponds to what is used in the standard CMS
reconstruction~\cite{CMSAlignment2014, CMSAlignment2022}.  
As compared with the standard track fit, additional quality criteria are used to select pixel hits, 
and a refined parameterization of the local hit position is used 
for the trapezoidal strip modules in the endcaps of the strip detector. 
To ensure an accurate modeling of track hit positions in the simulation, 
the numerical precision of the helix-surface intersection in \GEANTfour 
has been increased with respect to the standard CMS simulation.
The Kalman filter track fit is only used to associate hits to a given track. 
The CVH fit is applied to data and simulation, and used to determine all track parameters,
including the track momentum.

Although the models used to describe the magnetic field, the material distribution, 
and the detector alignment 
are the most accurate currently available, a few sources of potential biases remain. 
The magnetic field model is based on measurements made 
in the ground-level assembly hall rather than in the cavern
and does not account for differences in the field induced by material in the detector and surroundings. 
The simulation geometry underlying the material model 
might not provide a perfect description of the real detector
and there are inaccuracies in the Gaussian model used to incorporate material effects. 
Finally, the alignment is affected by small residual biases in the alignment procedure
and by so-called ``weak modes"
(misalignment patterns, including global translations, twists, and radial expansions, that bias the parameter extraction
from the track but do not impact the overall $\chi^2$ of the track fits~\cite{CMSAlignment2014}).
To correct for these biases, 
we developed a generalized correction procedure that extends the standard alignment procedure.
The alignment degrees of freedom are parameterized by the three translation 
and three rotation degrees of freedom per 
tracker module, 
albeit without extra parameters for module deformation or residual time dependence. 
The parameterization is extended with additional parameters to correct 
the $z$~component of the magnetic field and the energy loss from material in the vicinity of each module.
The correction parameters are derived from a sample of \jmm decays using the CVH fit, 
imposing the additional constraints that the muons are produced from a common vertex 
and that the muon pair has a mass consistent with that of the \JPsi meson.  
These are needed to constrain weak modes in the alignment, magnetic field, and energy loss parameters.

The correction procedure is effective in absorbing local biases 
in the magnetic field, energy loss, and alignment, but remains subject to weak modes, 
as well as to residual biases resulting from limitations in the Gaussian \JPsi meson mass constraint.
Convolution effects from the finite detector resolution 
and for final-state radiation are only accounted for in an approximate manner, 
and background contributions are not considered.
To correct for these potential biases, 
residual corrections are derived from fits to the \jmm dimuon mass distribution, in two steps.
In the first step, we extract correction factors in fine bins over a four-dimensional space 
constructed from the \ptmu and \etamu of the two muons, 
in order to adjust the muon momentum scale in data to that of the simulation.
In these fits, 
the signal model is based on templates from simulation, 
convolved with a Gaussian whose mean and standard deviation ($\sigma$)
account for the residual scale and resolution difference.
The combinatorial background is represented by an exponential function.
In the second step, the muon momentum scale correction factors are translated into correction 
parameters for each individual muon. The conversion between the four-dimensional corrections and 
the per-muon correction parameters is performed with a $\chi^2$ minimization. 
The residual corrections to the muon momenta are binned in \etamu and parameterized as a function of \qmu and the curvature,
$k \equiv 1/\pt$, as
\begin{equation}
\frac{\delta k}{k} = A_{i\eta} -\epsilon_{i\eta} \, k + q \, M_{i\eta}/k.
\label{eq:calibmodel}
\end{equation}
The $i\eta$ subscript indicates the corresponding $\eta$ bin of the correction parameters.
The parameters are independent for the 48~$\eta$ bins of width~0.1 and are integrated over the $\phi$ coordinate.
The $A_{i\eta}$ term corresponds to a small adjustment of the magnetic field. 
The $\epsilon_{i\eta}$ term is the first one in a Taylor series expansion for the effect 
of mismodeling the energy loss between the interaction point and the first hit measurement. 
The $M_{i\eta}$ term expresses the bias in the track sagitta resulting from a misalignment of the tracker 
in the plane transverse to the magnetic field. 
The expression captures the leading behavior once local biases in the magnetic field, material, 
and alignment are corrected. 
In the presence of sufficiently large local biases, additional terms would appear with a more complicated functional form.
Using MC simulation, we have validated that residual biases are well described by this functional form,
after performing the track refit and applying the generalized global corrections.
The corrections are then applied by shifting the reconstructed curvature of the measured muons.
Illustrative fits to the dimuon mass distribution in \jmm events are shown in \insupp{Fig.~17 in the \suppmat}{Fig.~\ref{fig:jpsiMass_fits}}, 
and the parameters of Eq.~(\ref{eq:calibmodel})
extracted from the fits to data are shown in \insupp{Fig.~18 in the \suppmat}{Fig.~\ref{fig:calib_corr_factors}}.

To avoid extrapolating the muon momentum resolution corrections 
from the relatively low momentum values typical of muons from \JPsi decays, 
we calibrate the muon momentum resolution using both \jmm and \zmm events.
The resolution corrections are derived from fits to the \jmm and \zmm dimuon mass distributions, 
binned in the \ptmu and \etamu of the positively and negatively charged muon as for the scale corrections, and
after correcting the momentum scale using the calibration parameters previously extracted from the \JPsi sample.   
The resolution is parameterized as a function of the curvature as
\begin{equation}
\frac{\sigma_k^2}{k^2} = a_{i\eta}^2 + \frac{c_{i\eta}^2}{k^2} + \frac{b_{i\eta}^2}{1 + d_{i\eta}^2 k^2},
\label{eq:resolutionmodel}
\end{equation}
where the parameters $a_{i\eta}$, $c_{i\eta}$, $b_{i\eta}$, and $d_{i\eta}$ parameterize the contributions 
to the curvature resolution from multiple scattering, hit resolution, and the correlations between them induced by the track fit. 
These parameters are computed in 24~$\eta$ bins of width 0.2 for each of the four terms, separately for data and simulation,
and are applied by smearing the reconstructed curvature of the simulated muons 
through a Gaussian distribution with the width corresponding to the difference in quadrature between data and simulation.

\begin{figure*}[ht]
\centering
\includegraphics[width=0.48\textwidth]{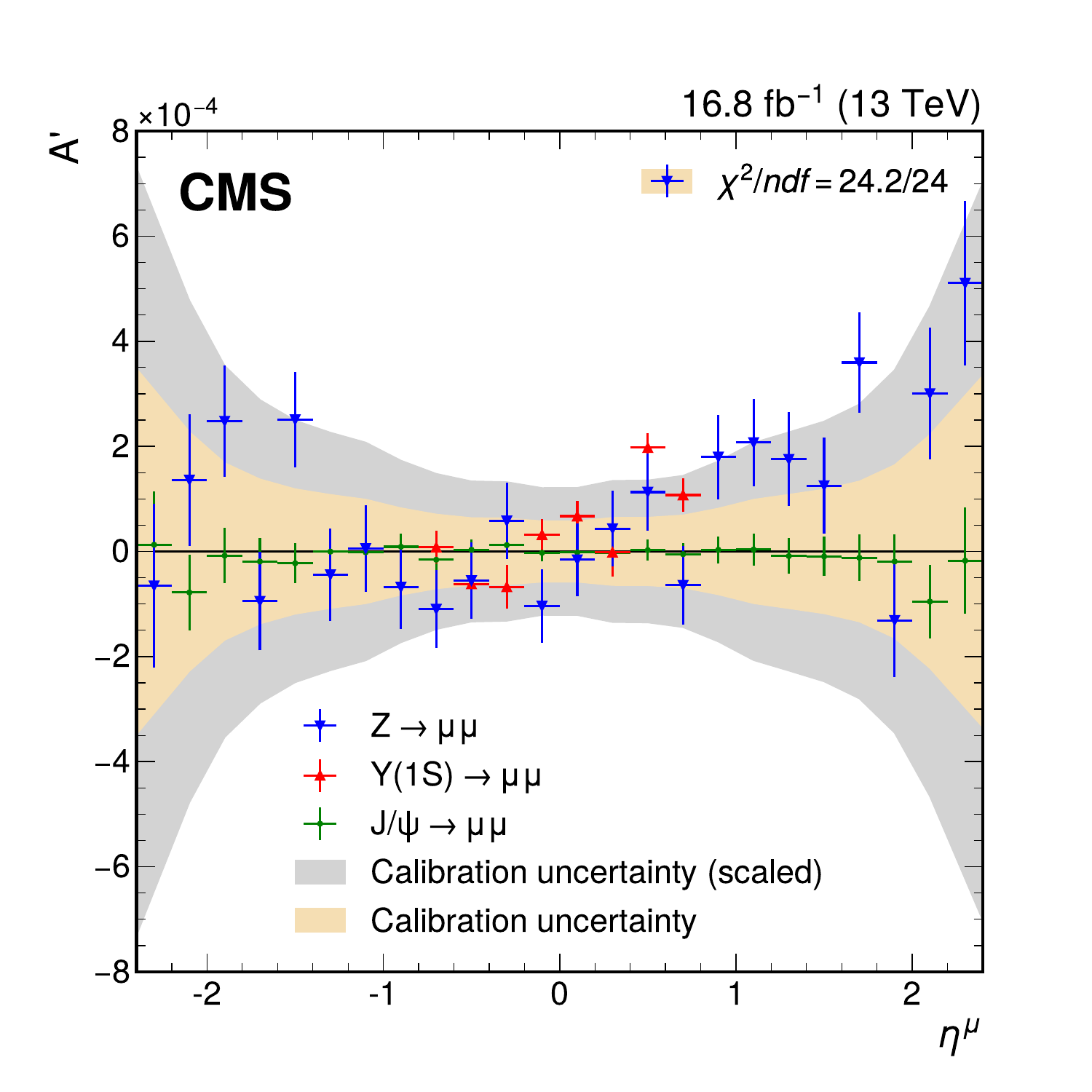}
\includegraphics[width=0.48\textwidth]{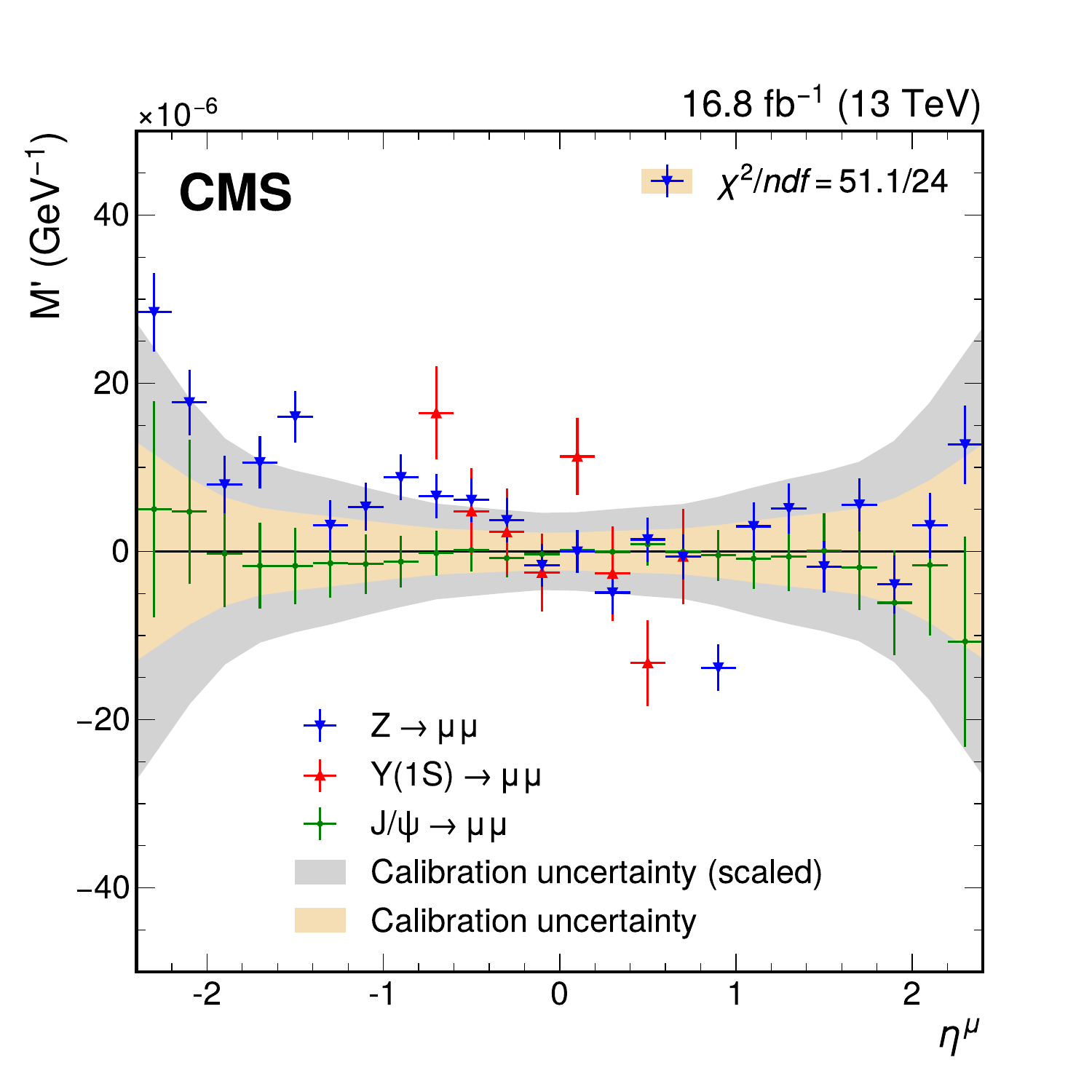}
\caption{\natcap{Charge-independent ($A^{\prime}$, left) and charge-dependent ($M^{\prime}$, right) residual scale differences 
between data and simulation, 
after applying the corrections derived from the  \jmm event sample.} 
The scale differences are evaluated separately in \jmm, \umm, and \zmm events to assess the 
consistency of the muon momentum scale calibration in the different event samples.
The charge-independent comparison probes a magnetic-field-like difference, 
whereas the charge-dependent comparison reflects a misalignment-like term.
The points with error bars represent the scale parameters and statistical uncertainties 
associated with the closure test performed with \jmm (green), \umm (red), and \zmm (blue) events.
The yellow band represents the corresponding statistical uncertainty 
in the calibration parameters derived from the \JPsi calibration sample. 
The filled gray band shows this uncertainty scaled by a factor of~2.1, as described in the text.  
The $\chi^2$ values correspond to the compatibility of the scale parameters with zero for the closure test 
performed with \zmm events.
The statistical uncertainties for these parameters, as well as for the calibration parameters 
derived from the \JPsi sample (without the 2.1 uncertainty scaling factor applied), are taken into account.
The calibration parameter uncertainties are fully uncorrelated from the \PZ and \PgUa closure test uncertainties, 
but very strongly correlated with the \JPsi closure uncertainties, since they use the same data.}
\label{fig:zclosure}
\end{figure*}

The calibration is validated using \jmm, \umm, and \zmm events,
by computing the residual muon momentum scale difference between the measured and simulated distributions,
after applying all corrections, following the same two-step procedure used to derive the calibration factors.
The residual scale differences between the event samples
are obtained in 24 bins of~\etamu
following the parameterization of Eq.~(\ref{eq:calibmodel}) with $\epsilon_{i\eta} = 0$.
The resulting closure parameters, corresponding to
a charge-independent magnetic-field-like residual ($A^{\prime}$) and 
a charge-dependent alignment-like residual ($M^{\prime}$),
are shown in Fig.~\ref{fig:zclosure}.
The $\chi^2$ compatibility test for the \jmm calibration applied to \zmm events demonstrates 
that there is consistency within the statistical uncertainties for the charge-independent residuals. 
A small inconsistency for the charge-dependent residuals is seen, indicating a systematic uncertainty source.
Given the momentum range relevant for \wmn events, 
the magnetic field and alignment effects are dominant with respect to those reflecting energy loss.
The \umm events are only used
to validate the calibration in the central region of the detector, where the dimuon mass resolution allows us 
to select a high purity sample of muons from the \PgUa meson decay.
The small deviations from zero in the \jmm
events are due to the larger \etamu bin sizes used for the validation step
and from small \ptmu bin migrations after applying 
the initial corrections in the $A^{\prime}$ and $M^{\prime}$ parameter extraction.
The differences are small compared with the statistical uncertainty in the calibration procedure.

The uncertainties propagated to the analysis include 
the statistical uncertainties in the calibration parameters extracted from the \JPsi sample, 
with statistical correlations taken into account, 
as well as the statistical uncertainties in the residual nonclosure between the \JPsi and the \PZ samples
and the systematic uncertainty associated with the reference measurement of the \PZ boson mass~\cite{PDG2024}.
Although these uncertainties account for the limited size of the measured and simulated event samples
in the \JPsi calibration procedure and closure tests,
as well as for the uncertainty in the world-average \PZ boson mass, 
other systematic effects might be present, 
related to weak modes with different sensitivity in \jmm and \umm events, trigger biases, or other sources.  
Remaining systematic effects that are not explicitly accounted for 
are assessed from the closure test between the \JPsi calibration and the momentum scale from the \PZ sample.  
The statistical compatibility of this test is assessed for different \etamu binning choices 
and considering several possible correlated patterns of biases. 
To cover all possible biases with a reduced $\chi^2$ smaller than unity,
the statistical uncertainty in the \JPsi calibration parameters is scaled by a factor of~2.1, as shown in Fig.~\ref{fig:zclosure}.

For the momentum resolution, the relative agreement between the measured and simulated samples, 
especially in the tails of the momentum response distribution, 
is affected by a different pixel hits efficiency after the tighter quality requirements imposed in the CVH fit.
To account for this, a systematic uncertainty is evaluated 
by reweighting the simulated pixel hit multiplicity distribution
to match data differentially in \etamu and taking the full difference as an uncertainty.
Since the nominal resolution corrections are also affected by this issue, 
we assign a systematic uncertainty to cover the residual disagreement. 
This uncertainty is expressed in terms of the statistical uncertainty of the resolution correction 
parameters, which are scaled by a factor of 10 to cover the observed differences. 
Because the statistical uncertainty in the resolution correction is small, 
and because the \mw measurement is not sensitive to small changes in the resolution, 
these scaled resolution uncertainties only contribute 1.4\MeV to the uncertainty in \mw.

The uncertainties in \ptmu from the momentum scale and resolution calibrations are 
several orders of magnitude smaller than the 1\GeV \ptmu bin width of our likelihood function. 
If the impact of the \ptmu variation is evaluated using event-level shifts in \ptmu, 
the \etaptqmu template shapes are determined only by events where the 
varied \ptmu is assigned to a different histogram bin than its nominal value. 
The probability of such bin migration depends on the relative size of the \ptmu shift compared 
to the bin width, which is $\mathcal{O}(10^{-4})$ for a typical selected muon. 
As a result, the effective number of events contributing to the template shapes 
is very small, which leads to large statistical fluctuations in the uncertainty templates.
To avoid this issue, the variations are evaluated by
reweighting events in terms of the muon momentum response distribution, resulting in smooth variation templates.
The breakdown of muon momentum calibration uncertainties is shown in Table~\ref{tab:calunc}.
The total contribution of the muon momentum calibration to the \mw uncertainty is 4.8\MeV. 
This uncertainty has been validated 
by applying the difference in scale between the \jmm and \zmm events 
to the \PW boson simulation to build a biased prediction that is tested as pseudo-data in the fit. 
The resulting shift in \mw from this procedure is covered by the corresponding calibration uncertainties.
\insupp{Figure~1 in the \suppmat}{Figure~\ref{fig:mll}} shows the \zmm dimuon mass distributions 
after correcting the muon momentum scale by the calibration parameters extracted from fits to the \JPsi events.

\begin{table*}[ht]
  \centering
  \topcaption{\natcap{Breakdown of muon momentum calibration uncertainties.}}
  \label{tab:calunc}
    \begin{tabular}{lcc}
      \multirow{2}{*}{Source of uncertainty} & Nuisance  & Uncertainty \\
      &  parameters & in \mw (\MeVns) \\ \hline
      \JPsi calibration stat.\ (scaled $\times$2.1) & 144 & 3.7 \\
      \PZ closure stat. & 48 & 1.0 \\
      \PZ closure (LEP measurement) & 1 & 1.7 \\
      Resolution stat.\ (scaled $\times$10) & 72 &  1.4\\
      Pixel multiplicity & 49 & 0.7 \\ \hline
      Total & 314 & 4.8
    \end{tabular}
  \end{table*}

\subsection{Modeling of the \texorpdfstring{\PW}{W} and \texorpdfstring{\PZ}{Z} boson transverse momentum distributions}
\label{sec:ptmodeling}

To achieve the best accuracy in modeling the \ptv spectra, 
we correct the generator-level \ptv and \yv distributions 
in \MiNNLO to state-of-the-art calculations in QCD, 
including the resummation of logarithmically-enhanced contributions at small \ptv 
and a model for nonperturbative effects also at small \ptv.
We use the \scetlib code~\cite{Billis_2021, Ebert:2020dfc, Billis:2024dqq},
which performs \ptv resummation as formulated using soft-collinear
effective theory (SCET)~\cite{Bauer:2000yr, Bauer:2001yt, Chiu:2012ir},
using deterministic numerical integration routines 
to provide predictions with high numerical accuracy.
The resummed predictions from \scetlib are matched to the fixed-order 
calculation from \dyturbo~\cite{Camarda:2019zyx}, at $O(\alphaS^2)$ in the QCD coupling constant \alphaS, 
to achieve \NtLLpNNLO accuracy.
The correction is derived from the ratio of the \scetlibdyturbo and \MiNNLO predictions
for a fixed value of \mv
(after the parton shower but before final-state photon
radiation) in the full phase space of the decay lepton kinematics.
The corrections are binned in 1\GeV bins of \ptv, up to 100\GeV,
and 0.25 wide bins in the $\abs{\yv} < 5.0$ range. 
They are applied to the \MiNNLO simulation by sampling the binned corrections per event 
with the generator-level $\abs{\yv}$ and \ptv
to obtain event-level weights that are propagated through the full experimental analysis.
This procedure allows us to maintain the statistical power of the \MiNNLO MC sample while improving its accuracy at small \ptv.
After the correction, the dependence of the \ptv distribution on the parton shower and tune is negligible.
We have compared the predictions using \scetlibdyturbo 
with those using \dyturbo v1.4.0~\cite{Camarda:2019zyx,Camarda:2021ict}, 
\matrixradish v1.0.0~\cite{Grazzini:2017mhc,Bizon:2019zgf}, 
and CuTe-\mcfm v10.2~\cite{Campbell:2019dru,Becher:2020ugp}, 
at equivalent or higher perturbative order.
After propagating those predictions through the analysis as binned corrections in \ptv,
we find that the expected shifts in \mw are within the \scetlibdyturbo uncertainties.

\begin{figure}[th!]
\centering
\includegraphics[width=0.48\textwidth]{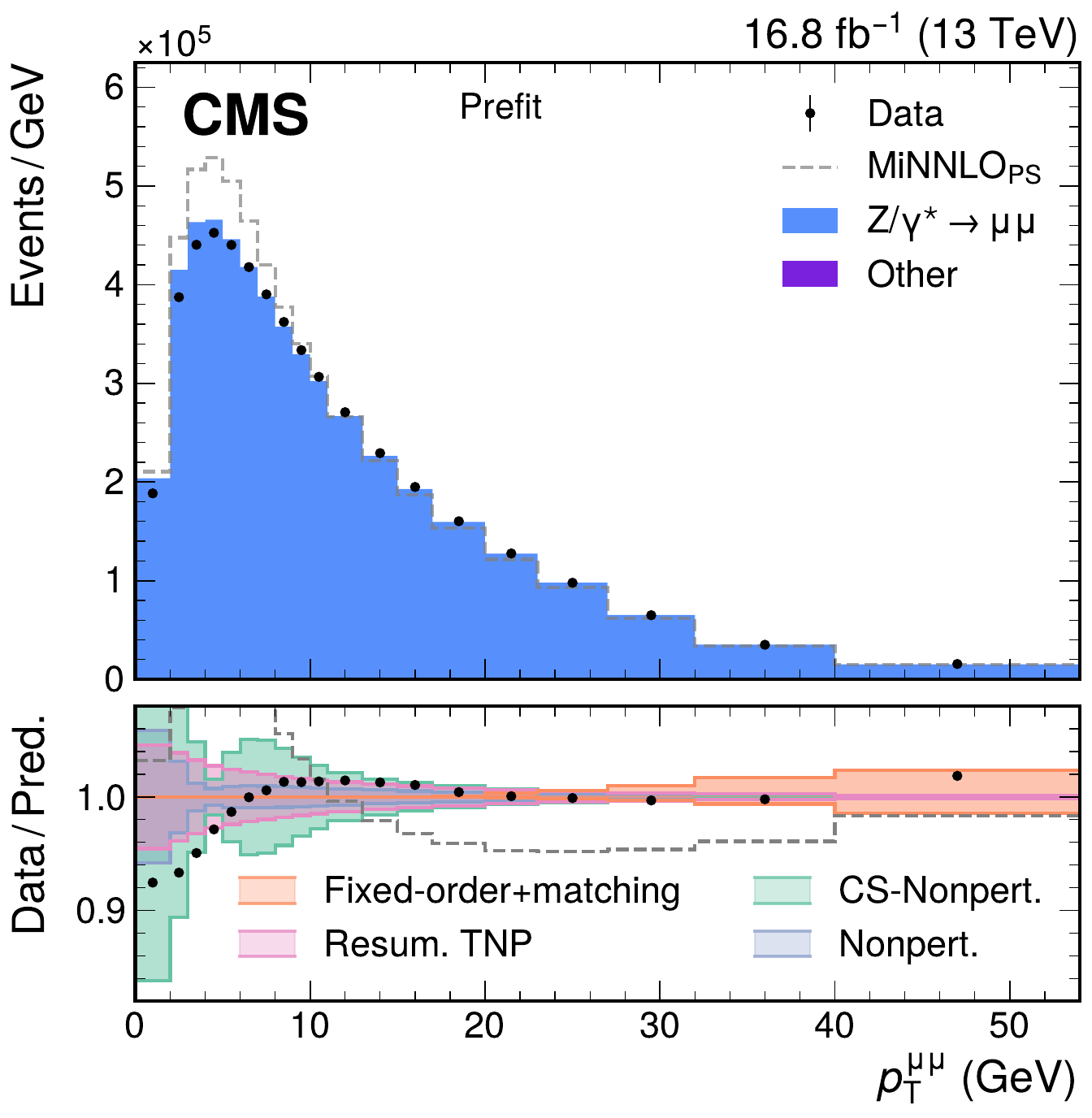}
\includegraphics[width=0.49\textwidth]{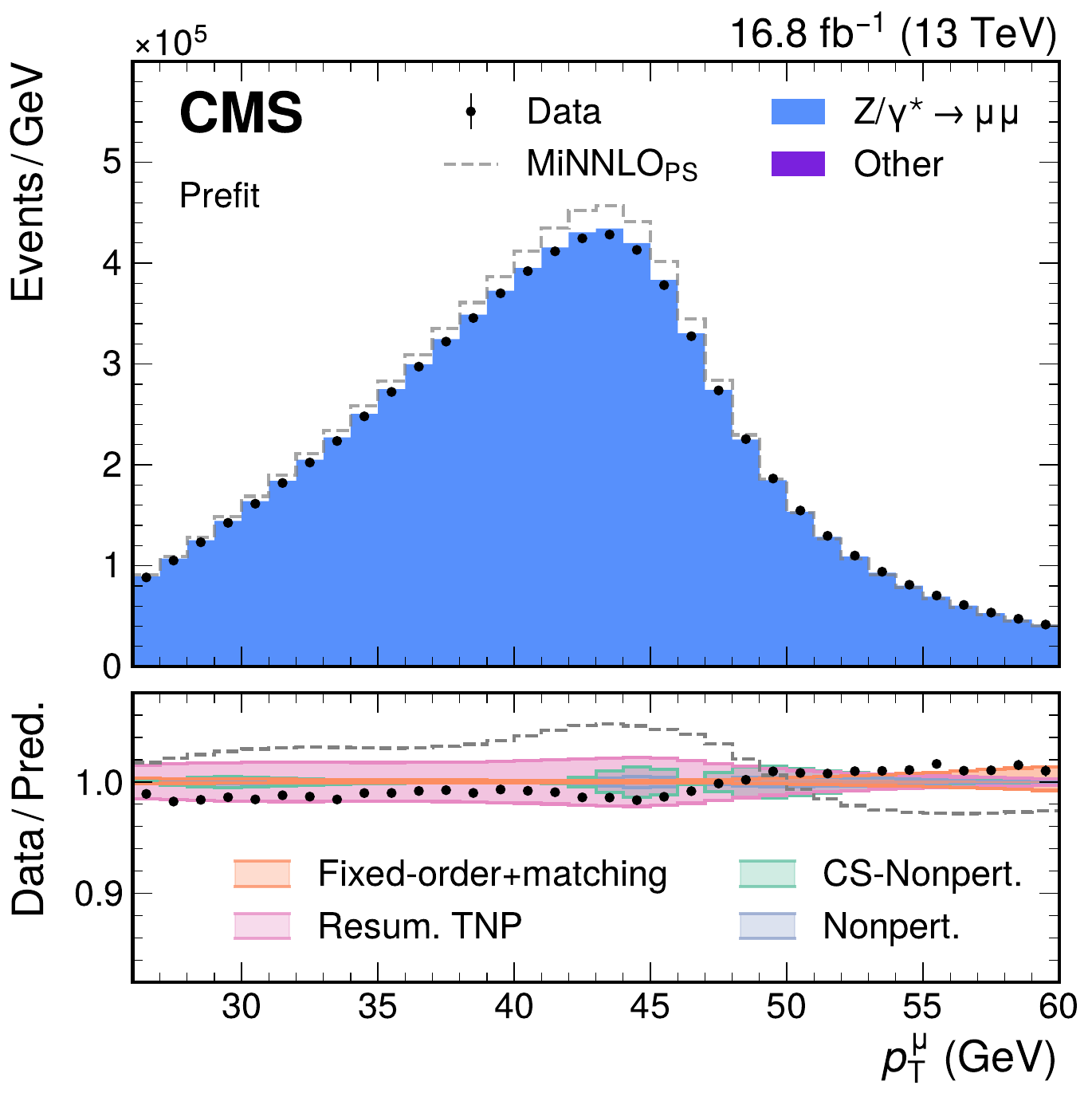}
\caption{\natcap{Measured and simulated \ptmumu (\cmsLeft) and \ptmu (\cmsRight) distributions in selected \zmm events.} 
The standalone uncorrected \MiNNLO predictions are shown by the dashed gray line.
The nominal predictions (blue) correct the \pMiNNLO \ptv with \scetlibdyturbo at \NtLLpNNLO,
as described in the text.
The vertical bars represent the statistical uncertainties in the data.
The bottom panel shows the ratio of the number of events observed in data
to that of the total nominal prediction, as well as the relative impact of variations of the predictions.
Different sources of uncertainty are shown as solid bands in the lower panel: 
the fixed-order uncertainty and the uncertainty in the resummation and fixed-order matching (orange), 
resummed prediction using TNPs (pink), the Collins--Soper (CS) kernel nonperturbative uncertainty (green), 
and other nonperturbative uncertainties (light blue). 
Additional sources of experimental and theoretical uncertainty that impact the agreement
with the data are not shown.}
\label{fig:ptmodeling}
\end{figure}

As illustrated in Fig.~\ref{fig:ptmodeling}, 
the \scetlibdyturbo correction substantially improves the description of \ptmumu and \ptmu data 
in selected \zmm events when compared with the standalone \MiNNLO predictions.
Uncertainties in the \ptw prediction, particularly those impacting the low-\ptw region,
can shift the peak of the \ptmu distribution in a way similar to a variation of \mw. 
Therefore, the sensitivity of the analysis to \mw critically relies on 
differentiating the uncertainty in \ptw and its impact on the \ptmu distribution from 
\mw variations.
As can be appreciated from Fig.~\ref{fig:ptmodeling}, 
different sources of uncertainty contribute predominantly to different \ptv regions.
The nonperturbative uncertainty is most pronounced at low \ptv.
Uncertainties in the resummation calculation and in the matching of the resummed and fixed-order
calculations are relevant up to $\ptv \approx 40\GeV$.
The nonperturbative and resummation uncertainties are most pronounced 
near the peak of the \ptmu distribution, the region most 
sensitive to the \mw value.
Consequently, their contributions have an important impact on the measurement of \mw.
The perturbative uncertainties in fixed-order QCD, which are dominant at high \ptv,
have a small impact on \ptmu in the region sensitive to \mw.
The uncertainties are estimated by varying the relevant parameters of the \scetlibdyturbo calculation
to obtain alternative predictions that are propagated through the full experimental analysis via event-level weights.

\begin{figure}[ht]
\centering
\includegraphics[width=0.45\textwidth]{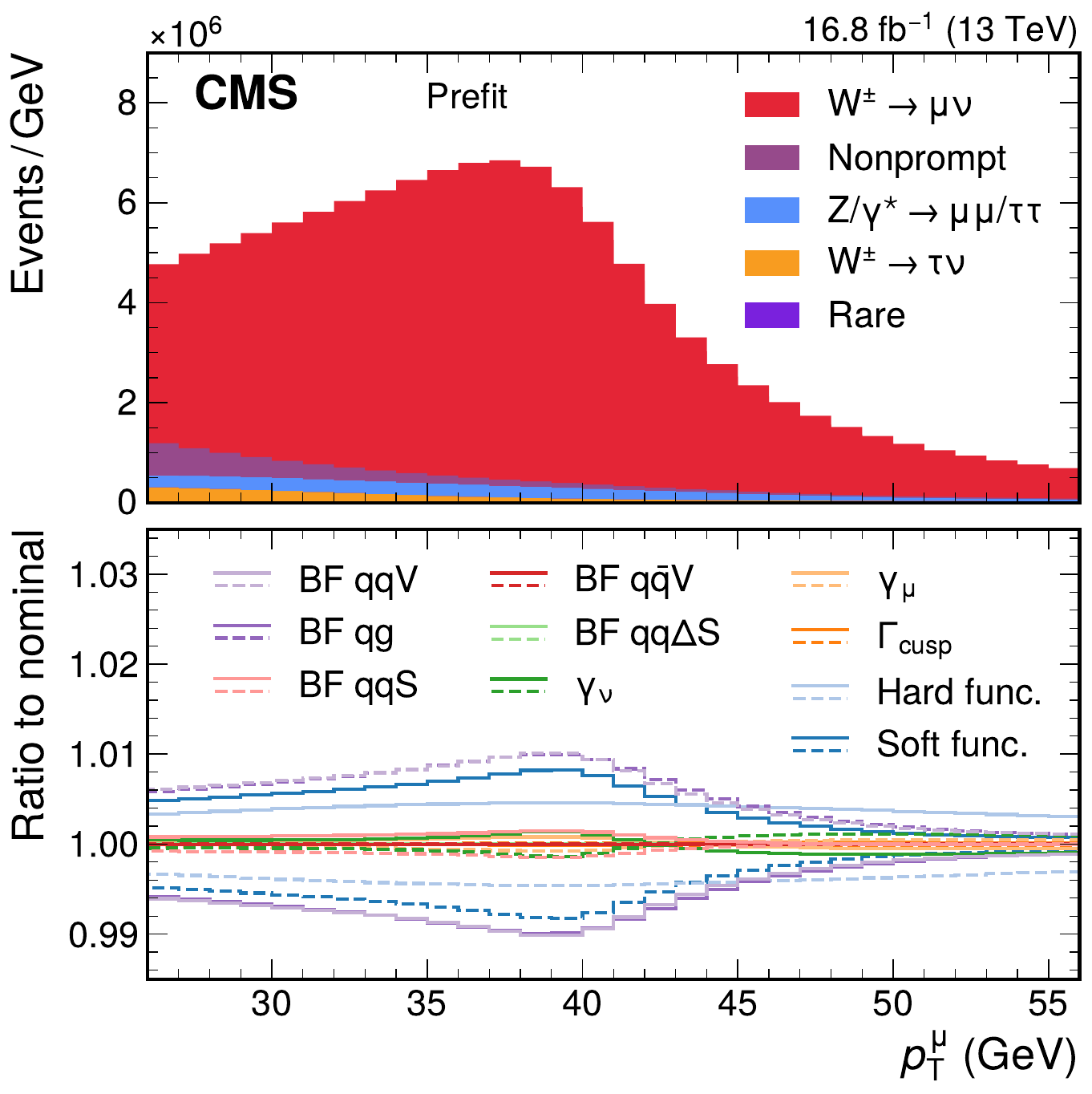}
\caption{\natcap{The predicted \ptmu distribution 
for selected \wmn events, before the maximum likelihood fit.}
The lower panel shows the ratio of the ten TNP variations described in the text
to the nominal prediction, illustrating their impact on the spectrum.}
\label{fig:tnp}
\end{figure}

Perturbative uncertainties in the resummed predictions are evaluated using the TNP approach
recently proposed in Ref.~\cite{TNPs},
which exploits the known all-order perturbative structure of the resummed calculation
and is implemented in \scetlib.
In the SCET formalism used here, there are three perturbative ingredients in the \ptv resummation:
the ``hard function'' that describes the hard virtual corrections for \PW and \PZ production,
the ``proton beam functions'' that extend the PDFs to include collinear radiation, 
and the ``soft function'' describing soft radiation. 
All these functions share a system of renormalization group equations
whose solution yields the all-order resummation of logarithms of $\ptv/\mv$.
In the TNP approach, the minimal independent set of ingredients 
that would be needed at the next perturbative order 
are identified and parameterized in terms of common nuisance parameters.
Specifically, there are six sources of TNPs: 
the three fixed-order boundary conditions of each of the hard (H), soft (S), and beam (B) functions, 
and three anomalous dimensions governing their renormalization group evolution, 
namely the cusp anomalous dimension ($\Gamma_\text{cusp}$) 
and the virtuality and rapidity noncusp anomalous dimensions ($\gamma_\mu$ and $\gamma_\nu$).
The TNPs of the hard and soft functions and the three anomalous dimensions are numerical constants. 
As such,
they are propagated as scalar variations around their known values.
The beam functions (BF) comprise five one-dimensional functions of the Bjorken-$x$
for the different partonic splitting channels. The $\Pq\Pq\PV$ and $\Pq\Pg$ BF
contain the dominant quark to quark ($\Pq\to\Pq$) and gluon to quark ($\Pg\to\Pq$) channels, while the others 
($\Pq\PAQq\PV$, $\Pq\Pq\mathrm{S}$, and $\Pq\Pq\Delta\mathrm{S}$) 
correspond to specific nondiagonal $\Pq\to\Pq^{\prime}$ contributions that are present at higher orders~\cite{Billis_2021}. 
We use their known functional shape 
and treat their normalization as a scalar TNP for each partonic channel. 
The TNPs have a true, but unknown, value that can be varied according to their expected typical magnitude. 
As a result,
the TNP approach provides a robust prediction for the correlation of the uncertainties 
due to the missing higher orders across \ptv, \yv, and \mv, 
which can be consistently used in the profile maximum likelihood fit used to extract \mw.
Figure~\ref{fig:tnp} shows the impact of the ten TNPs,
propagated through the analysis via the event-weighting procedure, on the \ptmu spectrum in \wmn events.

The \scetlib program implements different configurations for the TNPs, 
in terms of the logarithmic accuracy of the prediction and 
the perturbative order at which the TNPs are included. 
The order of the calculation is expressed following the notation of Ref.~\cite{TNPs}.
An \NmpkLL prediction is built from a full calculation of the first $m$ logarithmic terms in the resummation series,
with a further $k$ terms with unknown coefficients used to estimate the theory uncertainty.
We use the \NtpzLL scheme, where the prediction has \NtLL accuracy and the TNPs representing 
the unknown HO corrections are estimated from multiplicative variations of the same \NtLL terms. 
We also consider two alternate schemes, \NtpoLL and \NfpzLL.
The \NtpoLL scheme implements the full \NfLL perturbative structure, 
combining the known values for the parameters up to \NtLL with best estimates of the HO terms 
and their variations to define the TNP variations.
The \NfpzLL scheme follows the same approach as the \NtpzLL, but is based on the \NfLL calculation. 
We have verified that using the \NtpoLL or the \NfpzLL schemes 
has a negligible impact on the results with respect to the nominal \NtpzLL scheme.
As discussed in Sections~\ref{sec:addValidation} and~\ref{sec:wmass-details},
we validate the robustness of this approach against the measured \ptmumu distribution
and the \PW-like measurement of \mz,
where the impact of the \ptz modeling uncertainty is treated in the same way.

The perturbative uncertainty in the fixed-order matching correction of the unpolarized calculation 
is assessed from 7-point variations of the factorization and renormalization scales, \muR and \muF, 
in the \dyturbo calculation. 
The uncertainty is correlated for the different \PW (and \PZ) boson decay channels, 
and between \PWp and \PWm boson production, 
but uncorrelated between \PW and \PZ boson production.
This uncertainty is profiled in the maximum likelihood fit. If the uncertainty
is excluded from the profiling procedure 
and estimated by repeating the maximum likelihood fit after varying \muR and \muF,
or if the variations are correlated between \PW and \PZ
boson production, the shift in the measured \mw value is $<$0.4\MeV.
The impact of both choices on the total uncertainty in \mw is negligible.
An uncertainty due to the matching procedure is evaluated 
by varying the transition scale (the midpoint of the transition function defined in Ref.~\cite{Billis:2024dqq}) between the resummation and the fixed-order regime 
from the nominal value of 0.5\,\mv to 0.35\,\mv and 0.75\,\mv.

Nonperturbative effects, such as the residual transverse motion of the partons inside the
proton, impact the \ptv distribution. 
These effects are expected to scale as
$(\Lambda_{\text{QCD}}/\ptv)^2$~\cite{Collins:1981uk}, where $\Lambda_{\text{QCD}} \approx 200\MeV$ 
is the QCD vacuum expectation value.
As such, their impact is dominant at low \ptv and less relevant for $\ptv \gtrsim 10$\GeV.
The predictions considered here implement phenomenological models that require tuning to data
to describe these nonperturbative effects.
Two sources of nonperturbative effects impact \ptv. First,
there can be nonperturbative corrections to the CS rapidity anomalous 
dimension~\cite{Collins:1981uk},
which are universal for \PW and \PZ boson production.
Second, there are nonperturbative contributions to the beam (and soft) functions,
which account for the ``intrinsic \kt" of the partons inside the protons, 
that are not universal as they can 
depend on the flavor and Bjorken-$x$ of the interacting parton. 
As shown in Ref.~\cite{Ebert:2022cku}, 
the leading nonuniversal dependence can be captured by a
single effective model function that only depends on the vector boson rapidity
for each given vector boson type, 
besides the helicity cross section and the collision center-of-mass energy.
The \scetlib program implements a corresponding nonperturbative model for both these
sources~\cite{Billis:2024dqq}, where the model parameters effectively determine
the first two powers in an expansion in $(\Lambda_{\text{QCD}}/\ptv)^2$
together with a parameter that determines the overall asymptotic behavior for $\ptv\to 0$.
For the intrinsic \kt, the effective model amounts to a (rapidity-dependent)
Gaussian smearing in the Fourier conjugate of \ptv.

In our analysis, the five parameters of the \scetlib model are loosely constrained around nominal values
that correspond to minimal nonperturbative smearing. 
The CS anomalous rapidity parameters are correlated between \PZ and \PW boson production, 
whereas the Gaussian smearing terms are uncorrelated between the two.
Their best fit values are extracted from the maximum likelihood fits to the measured distributions.
The values obtained from a direct maximum likelihood fit to the \ptmumu distribution
are consistent with those resulting from the \PW-like fit to the \ptmu distribution in \zmm events.

The \scetlibdyturbo and \MiNNLO calculations are performed 
in a fixed-flavor scheme with massless quarks.
Calculations with \PQb and \PQc quark masses have not been performed
at a comparable perturbative accuracy. The impact of quark masses is expected to be mostly
at the scale of the \PQb and \PQc masses.
Their impact is partially estimated by varying the heavy-quark thresholds 
using charm and bottom quark mass variations of the MSHT20 PDF set~\cite{Cridge:2021qfd}.
In addition, the loose initial constraints of our nonperturbative uncertainty model provide
flexibility to cover such sources of uncertainty at low \ptv. Quark
mass effects are expected to differ between \PW and \PZ production because of the different flavor
contributions to their production. This difference is captured by the variations of the PDF threshold
and by our independent treatment of nonperturbative uncertainties between \PZ and \PW production.
The sufficiency of our model to capture these effects is confirmed by the likelihood fits to data
discussed in Section~\ref{sec:addValidation}.

The total impact on \mw from the perturbative, nonperturbative, 
and quark mass threshold uncertainties is 2.0\MeV,
the three components yielding comparable contributions. 
A summary of nuisance parameters in the maximum likelihood fit that represent these
uncertainties is given in \insupp{Table~2 in the \suppmat}{Table~\ref{tab:nuisParamFit}}.
Section~\ref{sec:addValidation} further discusses the validation of the model and its uncertainty. 

\subsection{Modeling of the angular distributions in \texorpdfstring{\PW}{W} and \texorpdfstring{\PZ}{Z} boson leptonic decays}
\label{se:angular}

The differential cross section for the production and decay of the spin-1 \PW and \PZ bosons 
can be decomposed in terms of spherical harmonics into nine helicity-dependent states~\cite{mirkes:Wangular},
\ifthenelse{\boolean{cms@external}}
{
\begin{multline}
\frac{\ud\sigma}{\ud \pt^2 \, \ud m \, \ud y \, \ud \! \cos \theta^* \, \ud \phi^*}
= \\
=\frac{3}{16\pi}
\frac{\ud \sigmaUL}{\ud \pt^2 \, \ud m \, \ud y}
\Bigl[ (1+\cos^2\theta^*) \\
+ \sum_{i=0}^7 A_i(\pt, m, y) P_i(\cos\theta^*,\phi^*) \Bigr].
\label{eq:ACdec}
\end{multline}
}{
  \begin{equation}
    \frac{\ud\sigma}{\ud \pt^2 \, \ud m \, \ud y \, \ud \! \cos \theta^* \, \ud \phi^*}
    = \frac{3}{16\pi}
    \frac{\ud \sigmaUL}{\ud \pt^2 \, \ud m \, \ud y}
    \Bigl[ (1+\cos^2\theta^*) + \sum_{i=0}^7 A_i(\pt, m, y) P_i(\cos\theta^*,\phi^*) \Bigr].
    \label{eq:ACdec}
    \end{equation}  
}
We choose the CS reference frame~\cite{CS-frame}, where $\cos\theta^*$ and $\phi^*$ correspond to the polar and azimuthal angles of the muon 
emitted in the \PW boson decay.
The angular coefficients $A_i$ depend on the boson charge, rapidity \yv, \ptv, and \mv.
Combined with the unpolarized cross section \sigmaUL, 
they describe the relationship between the boson production 
and the kinematic distributions of the decay muons.
The $P_i$ spherical harmonics 
describe the kinematic distributions of the daughter muon, 
which depend on the properties of the \PW or \PZ boson. 

The nominal predictions for the angular distributions, from \MiNNLO, are NNLO accurate in QCD.  
Uncertainties in the predicted angular coefficients impact the \ptmu and \etamu distributions
by modifying the polarization of the \PW boson.
Uncertainties in the angular coefficients are assessed by varying \muR and \muF in the \MiNNLO predictions. 
The correlations of HO corrections across phase space and processes are not well known.
Therefore, we consider these variations uncorrelated among the $A_i$ coefficients 
and in ten \ptv bins, but correlated across \yv, and between $\PW^+$, $\PW^-$, and \PZ.

We have verified that the \MiNNLO predictions and uncertainties for the angular coefficients 
are consistent with NNLO fixed order calculations, 
and that the $A_i$ coefficients predicted at \NtLL, 
assessed with both \scetlib and \dyturbo, 
are consistent with the \MiNNLO predictions within the assigned uncertainties.
We have validated that the $A_i$ coefficients predicted by \MiNNLO
are consistent with those measured in data for \PZ boson production.
Measurements of the $A_i$ coefficients for \PW boson production recently
reported by the ATLAS Collaboration~\cite{ATLAS:2025mlt} confirm the
accuracy of the NNLO predictions.
The isotropic smearing of the colliding partons due to the intrinsic \kt model of \PYTHIA
induces a modest change to the angular coefficients, 
in particular $A_1$ and $A_3$ at low \PW or \PZ boson transverse momentum.
The full difference between the angular coefficients before and after the \PYTHIA8 shower and intrinsic \kt 
is taken as an additional systematic uncertainty, 
fully correlated across angular coefficients, phase space, and \PW and \PZ boson production.
The total uncertainty in \mw due to the uncertainty in the predicted angular coefficients is 3.2\MeV,
with the largest contributions coming from $A_{0}$, $A_{2}$, and $A_{4}$.
If the uncertainty in the angular coefficients defined by variations of \muR and \muF is excluded from the profiling 
procedure and estimated by repeating the maximum likelihood fit for each variation,
the measured value of \mw shifts by $+2.5\MeV$.
This increases the uncertainty in the angular coefficients to 3.9\MeV.

\subsection{Parton distribution functions}
\label{sec:pdf}

\insupp{Figure~2 in the \suppmat}{Figure~\ref{fig:pdfs_w}} shows the \etamu distribution for \PWp and \PWm boson events,
compared with the predictions obtained with the CT18Z PDF set and its uncertainties,
as well as with the central predictions for several other PDF sets.
The consistency among the five PDF sets and the observed data
is determined by performing likelihood fits to these distributions for each PDF set under consideration.
Fits are performed including all the uncertainties of the nominal \mw fit, 
as well as removing the PDF$+$\alphaS or the theory uncertainties.
The impact of \alphaS is evaluated from alternative PDF fits 
with \alphaS shifted by $\pm 0.015$ from its nominal value of 0.118. 
The change of \alphaS is propagated through the matrix element calculation in \MiNNLO 
and the \scetlibdyturbo corrections. 
The corresponding saturated likelihood goodness-of-fit values
are reported in \insupp{Table~1 in the \suppmat}{Table~\ref{tab:pdf_fits_yll_etal_q}}.

To test the dependence of the result on the choice of PDF set, 
we performed the \mw measurement with the 
NNPDF3.1~\cite{NNPDF:2017mvq}, NNPDF4.0~\cite{NNPDF40}, CT18~\cite{Hou:2019efy}, 
MSHT20~\cite{Bailey_2021}, and PDF4LHC21~\cite{Ball_2022} sets at NNLO, 
and the approximate \NtLO set MSHT20aN3LO~\cite{Cridge:2023ryv}. 
To assess the consistency of the PDF sets
we perform studies where the MC simulation for the \PW and \PZ boson production, and the corresponding PDF uncertainties,
are obtained from a given PDF set while another PDF set provides pseudo-data. 
We then evaluate if the \mw value extracted from the fit 
lies within the uncertainty predicted by the PDF set under test. 
In the case of the CT18Z, CT18, and PDF4LHC PDF sets, 
their uncertainty covers the \mw value extracted with all other PDF sets. 
This does not happen for the remaining PDF sets and, hence,
we test the impact of increasing their PDF uncertainty 
by scaling all eigenvectors by a constant value
until the difference in the extracted \mw is within the postfit $\sigma_\text{PDF}$. 
The scale factors determined with this procedure are reported in Table~\ref{tab:mw_pdf},
which also shows that the total \mw (unscaled) uncertainty due to the alternative PDF sets 
ranges from 2.4 to 4.6\MeV.
The scaling of the PDF uncertainty has only a moderate impact on the total \mw uncertainty.
Results for each PDF set, obtained with and without the scaling factors, 
are reported in Section~\ref{sec:otherPDFs}. 
When the PDF uncertainty is increased, the degree to which the fit is constrained to the 
predictions of the global PDF fit is relaxed, 
and the relative importance of this dataset is increased,
leading to a better consistency between the measured \mw values for different PDF sets.
Given its agreement with data, relatively large uncertainty, 
and consistency with the other PDF sets, we select the CT18Z PDF set for the nominal prediction. 
The PDF uncertainty in \mw from the CT18Z set is 4.4\MeV.

\subsection{Impact of missing higher-order electroweak  corrections}
\label{sec:EW}

By interfacing \MiNNLO with \photospp, 
QED final-state radiation (FSR) is considered at LL accuracy, 
including matrix-element corrections and the effect of lepton pair production.
The uncertainty in the QED FSR modeling is evaluated by 
comparing the predictions of {\MiNNLO}{+}{\photospp}
to the prediction in which the matrix-element corrections of \photospp are switched off.
In addition, we evaluate the difference between the {\MiNNLO}\allowbreak{+}{\photospp} prediction and 
the prediction from \horace v3.2~\cite{CarloniCalame:2003ux,CarloniCalame:2005vc}
with the QED FSR modeled at LL in the collinear approximation.
For each comparison, the difference between the two predictions is
evaluated from the two-dimensional distribution of the \pt and mass of the dilepton system,
with the charged lepton momentum defined after FSR.
This difference is applied to the nominal {\MiNNLO}{+}{\photospp} prediction and propagated through the analysis as a systematic uncertainty.
The impact on \mw is $<$0.3\MeV.

The QED initial-state radiation (ISR) is modeled by the \PYTHIA8 parton shower at LL accuracy.
The uncertainty is evaluated by comparing to a sample with ISR photon radiation switched off. 
The modifications on the \ptv and \yv distributions are propagated through the analysis 
and found to have a negligible impact on \mw.
Besides the photonic corrections, we consider the impact of EW virtual corrections.
For the neutral-current Drell--Yan process, 
the separation between weak and photonic corrections can be performed in a gauge-invariant way.
The virtual EW corrections are calculated at NLO 
with the  Z\_ew process in the \POWHEG-\textsc{box-v2} (rev.\ 3900) program~\cite{Barze:2013fru, Chiesa:2024qzd} 
including universal HO corrections. 
The ratios between the \NLOpHO EW and LO EW predictions of the \PZ boson mass, 
and of the rapidity and $\cos\theta^*$ distributions,
are used to define a systematic variation to the nominal \MiNNLO prediction.

For \PW boson production, the splitting into virtual and photonic
corrections is not gauge invariant and is, hence, ambiguous.
Nonetheless, it is possible to separate the two contributions, 
to reproduce the QED FSR given by \photospp.
This separation is implemented in ReneSANCe~1.3.11~\cite{Bondarenko:2022mbi}, 
and the uncertainty in weak
virtual corrections is defined as the 
ratios between the \NLOpHO EW and LO EW predictions
of the \PW boson mass, and of the 
rapidity and $\cos \theta^*$ distributions. 
We cross-checked that the full NLO EW corrections (QED plus weak) agree at the 0.3\% level
between the \POWHEG-\textsc{box-v2}~\cite{Barze:2012tt} and the ReneSANCe programs,
also confirming previous benchmarks~\cite{Alioli:2016fum}. 
The uncertainty from the virtual EW corrections has an impact on \mw of 1.9\MeV.

\subsection{Additional validation of theoretical modeling}
\label{sec:addValidation}

In order to validate the uncertainties in our theoretical predictions 
and to quantify the sensitivity of our result to alternative \ptv modeling approaches, 
we performed several additional checks to demonstrate the stability of the results
when modifying the treatment of theoretical predictions and their uncertainties in the analysis.
To facilitate this, we correct our dimuon data sample 
for the effect of the detector response by ``unfolding" the two-dimensional \ptymumu distribution, extending the study reported in Ref.~\cite{CMS:2019raw}. 
The \zgmm production cross section is extracted inclusively 
in the kinematics of the decay muons, defined before final-state photon radiation,
and for $60 < \mz <120\GeV$, $\ptz < 54\GeV$, and $\ayz < 2.5$.
The unfolding is performed via a maximum likelihood fit to the two-dimensional distribution of \ptymumu without regularization. 
The unfolded \ptz and \ayz distributions, shown in \insupp{Fig.~2 and \suppmat Fig. 3}{Figs.~\ref{fig:ptll_gen_postfit} and \ref{fig:yll_gen_postfit}},
are obtained by integrating over the other dimension of the measured two-dimensional distribution.

We have repeated the \wlike \mz and \mw measurements 
using predictions from \scetlib at different perturbative orders, 
matched to \dyturbo, as well as different approaches to incorporate the TNPs. 
When using \NtpoLL and \NfpzLL
predictions and uncertainties, 
the measured value of \mw is shifted by less than the 2.0\MeV of total \ptw-modeling uncertainty of the nominal result.
Although the approximate \NfLL predictions~\cite{Billis:2024dqq} 
give slightly reduced \ptw uncertainties, 
the difference in the total uncertainty in \mw is negligible 
when compared with the nominal result.
In addition, we have tested the impact of extracting \mw 
with an uncertainty model 
that corresponds to a simplified and more constrained version of the helicity fit
rather than relying on the \scetlib TNPs.
In this approach, the \ptv modeling uncertainty is defined by varying all scales in the \scetlib
calculation independently and taking the envelope of the variations (the maximum per bin of the
distribution) as uncertainty. This uncertainty is uncorrelated
across 10 quantiles of \ptv, 
such that the model is sufficiently
flexible to describe the true \ptv distribution in data.
The \mw value measured with
this configuration is shifted by $<$0.2\MeV with respect to the nominal result.

To further assess the impact of missing HO corrections, we have
performed the analysis with the \scetlib calculation matched 
to the $O(\alphaS^3)$ \NtLO predictions from \nnlojet~\cite{Huss:2025iov,Gehrmann-DeRidder:2017mvr}. 
The \scetlibnnlojet predictions, using the CT18Z NNLO PDF set,
are introduced into the analysis via two-dimensional corrections
with the same procedure as for \scetlibdyturbo.
Given the very high complexity of the calculation, the numerical precision of the results
is significantly worse than that of the \MiNNLO MC sample, and applying this correction
introduces non-negligible fluctuations in the predicted templates.
These fluctuations are partially 
accounted for by propagating the statistical uncertainty of the \scetlibnnlojet predictions
into the \etaptqmu observable,
but the full impact of these fluctuations has not been 
rigorously assessed. 
The \mz results obtained from the \wlike fit with the \MiNNLO predictions corrected to \NtpoLLpNtLO and \NfpzLLpNtLO are shifted 
down by about 3\MeV 
with respect to the nominal \NtpzLLpNNLO result, respectively. Due to the complexity of the calculation, we have only assessed the impact of the \NtLO prediction for \PWm production.
The value of \mwm is extracted from a fit to the \etaptqmu distribution with only negatively charged muons selected.
For the nominal configuration using \scetlibdyturbo at \NtpzLLpNNLO, the total uncertainty in \mwm is 19.4\MeV, with a \ptw modeling uncertainty of 3.1\MeV.
The \mwm results at \NtpoLLpNtLO and \NfpzLLpNtLO differ from the \NtpzLLpNNLO result
by less than the \ptw modeling uncertainty.
These checks show that our results are not significantly impacted
by HO predictions for \PZ and \PW boson production. 

As an additional test of the measurement dependence on the accuracy of the \ptv modeling,
we have performed the \wlike \mz measurement 
with the \MiNNLO MC sample reweighted such that the predicted \ptz matches the measured \ptmumu
distribution.
The corrections are derived from the unfolding measurement shown in \insupp{Fig.~\ref{fig:ptll_gen_postfit} and \suppmat Fig.~3}{Figs.~\ref{fig:ptll_gen_postfit} and \ref{fig:yll_gen_postfit}} and applied directly to the \MiNNLO prediction. 
The value of \mz extracted from this configuration differs by $-1.8$\MeV with respect to the nominal result, to be compared with the 1.7\MeV uncertainty due to the \ptz modeling. 
The stability of these results, and their consistency with the independently measured value of \mz, 
supports the use of the \scetlibdyturbo prediction and its uncertainties. 
We also tested the impact of applying the same corrections,
derived from the ratio of the unfolded data and the \MiNNLO predictions for \ptz, 
to the \PW boson simulation. 
This procedure corresponds approximately to tuning the predictions to reproduce the \ptz spectrum, 
under the assumption that the differences between the data and the \ptz and \ptw predictions arise from the same sources. 
Given the weakness of this assumption, we do not consider this procedure to be an acceptable 
approach for the nominal result.
The resulting shift in \mw with respect to the nominal result 
(based on the \scetlibdyturbo prediction) is smaller than 0.5\MeV 
and the change in the total uncertainty is negligible.

Finally, we test a simultaneous fit to the \etaptqmu distribution in \wmn events 
and the \ptymumu distribution in \zmm events.
The TNPs are correlated across the \PW and \PZ boson processes, 
whereas uncertainties in the matching contributions and angular coefficients are left uncorrelated between the different processes.
For the nonperturbative model, 
the Gaussian smearing parameters are considered independent for \PW and \PZ, 
whereas the CS anomalous rapidity is correlated.
The \mw value extracted in this fit is shifted by $+0.6\MeV$  compared with the nominal result.
The total uncertainty is moderately reduced because of 
additional constraints on theory and experimental uncertainties 
that are correlated across the \PW and \PZ processes.
\insupp{Figure 18 in the \suppmat}{Figure~\ref{fig:theory_model_comparisons}} presents a summary of these results, 
shown as a comparison to the nominal result and its uncertainty.

\begin{figure}[ht]
\centering
\includegraphics[width=0.45\textwidth]{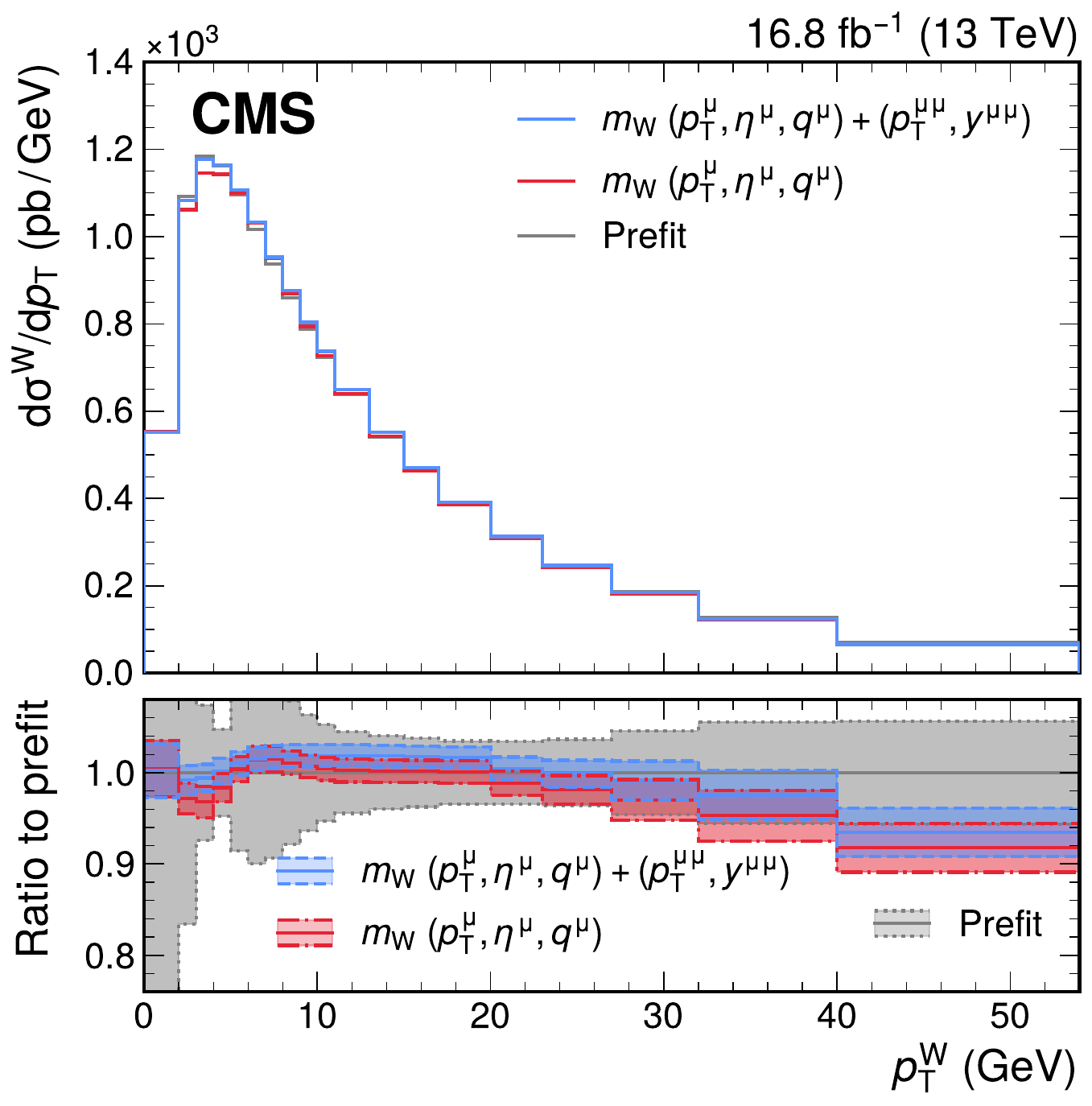}
\caption{\natcap{The generator-level \ptw distribution, 
with three instances of the prediction and their uncertainty: 
before the maximum likelihood fit (``prefit"),
and reflecting the results of the two fits described in the text.}
The distribution and uncertainties obtained from the combined \etaptqmu and \ptymumu fit is shown in red, 
whereas the blue band shows the distribution obtained from the nominal \etaptqmu fit. 
The generator-level distribution predicted by \scetlibdyturbo 
before incorporating in situ constraints 
is shown in gray.
The ratio of the postfit predictions to the prefit prediction (in gray), 
as well as their uncertainties, are shown by the shaded bands in the lower panel.}
\label{fig:ptw_gen_postfit}
\end{figure}

The impact
of including the \ptmumu data in the fit is illustrated in Fig.~\ref{fig:ptw_gen_postfit},
which compares the generator-level \ptw spectrum modified by the best fit values of nuisance parameters 
for the two fits. 
The consistency of the two results supports the conclusion 
that the \ptmumu measurement is not required as an input to describe the \ptw distribution, 
with the added benefit of a reduced model dependence of the result.
In fact, the loose assumptions about the correlation of the nonperturbative parameters 
between \PW and \PZ boson production limit the impact of including the \ptmumu data.

\subsection{Helicity fit}
\label{sec:agnostic}

While the theoretical model and uncertainties
described in Section~\ref{sec:theory}
reflect our best knowledge of QCD and of the proton structure, 
approximations of this model or the presence of new physics
motivates the extraction of \mw using a parallel approach with a reduced model dependence,
which we call ``helicity fit''.
With this technique we extract, from a likelihood fit to the \etaptqmu distribution,
not only the mass of the \PW boson but also, simultaneously, its polarization and the \ptw and \yw spectra.
At the core of this alternative analysis is the observation that \mw
variations induce a uniform scaling of the \ptmu spectrum,
whereas changes in the \PW boson polarization or 
in the $(\ptw, \yw)$ double-differential cross sections
lead to a nonuniform sculpting of the \ptmu and \etamu spectra.
We implement variations in the \PW boson polarization 
and in the $(\ptw, \yw)$ distribution 
as a set of independent nuisance parameters
in the signal likelihood function that is used to fit the 
measured \etaptqmu distribution.
The \PW boson polarization enters into our analysis procedure
through the helicity decomposition of Eq.~(\ref{eq:ACdec}).
We use helicity cross sections, $\sigma_i$,
which correspond to the product of the angular coefficients $A_i$ and the unpolarized cross section,
and we neglect the terms with $i>4$,
predicted to be zero in first approximation and having no effect on our measurement 
(given that we integrate over the $\phi^*$ angle in Eq.~(\ref{eq:ACdec})).

For each muon charge, the analysis covers the \ptv \vs \yv plane with 
nuisance parameters that represent variations of the production cross section, separately for
7 bins in \yv (within $\abs{\yv} < 3$) times 8 bins in \ptv (for $\ptv < 60$\GeV), 
plus 16 overflow bins.
The unpolarized cross section (\sigmaUL) and
five helicity cross sections
are defined for each of those 144 bins, 
leading to a total of 864 nuisance parameters.
The helicity cross sections, $\sigma_{i} \propto \sigma^{U+L}A_{i}$,
are defined for $i = 0$--4 in terms of $A_i$ and $\sigma^{U+L}$ from Eq.~(\ref{eq:ACdec}).
We propagate variations in the helicity amplitudes,
which depend on the unobserved \ptw and \yw, 
into a multitude of \etaptqmu distributions,
obtained by reweighting the simulated events.  
For each individual variation, 
the sum of all contributions is recomputed 
to get a new prediction for the \etaptqmu distribution.
The nuisance parameters are constrained around the theoretical predictions
with uncertainties that are relaxed with respect to their theoretical
values, used for the nominal result.
The \sigmaUL and $\sigma_4$ parameters have very loose initial constraints, 
of $\pm 50\%$ and $\pm 100\%$ of the predicted cross sections, respectively.
The initial uncertainties in the four other helicity terms, 
for which the fit has limited constraining power, 
are defined by the spread of theory predictions (reflecting missing higher orders)
and by uncertainties covering several different PDF sets.
To ensure coverage of all possible correlated variations 
allowed by the theory model used in the baseline analysis,
in addition to the explicit helicity cross section variations
we also retain the PDF and missing HO uncertainties, 
as well as the primordial-\kt smearing and nonperturbative uncertainties in the angular coefficients.
The latter two are also retained in the unpolarized term,
given that their impact on the cross section at low \ptv is significant within the finite-width bins of the helicity cross section variations.  
In contrast, we do not consider uncertainties in the unpolarized cross section 
from resummation, matching, and missing higher orders because they are largely redundant 
with respect to the explicit \sigmaUL variations.
This approach results in a significant reduction in
model-dependent assumptions with respect to the nominal analysis.

We validate the helicity fit approach by measuring a negligible \mw bias 
in pseudo-data samples generated with different PDF sets and \ptw or \yw spectra, 
and by measuring the \PZ boson mass in the \wlike configuration. 
The expected \mw uncertainty is evaluated for different prefit constraints on 
the helicity nuisance parameters. 
We observe only a mild dependence of the \mw uncertainty on all helicity terms, 
except for $\sigma_3$, 
whose variations have a similar impact on \ptmu as those resulting from \mw variations. 
Therefore, 
in the \mw extractions made to verify the stability of the measurement, 
we scale the prefit uncertainties of $\sigma_3$ and of all the other $\sigma_i$ terms
by two independent factors.

\begin{figure}[!th]
\centering
\includegraphics[width=0.50\textwidth]{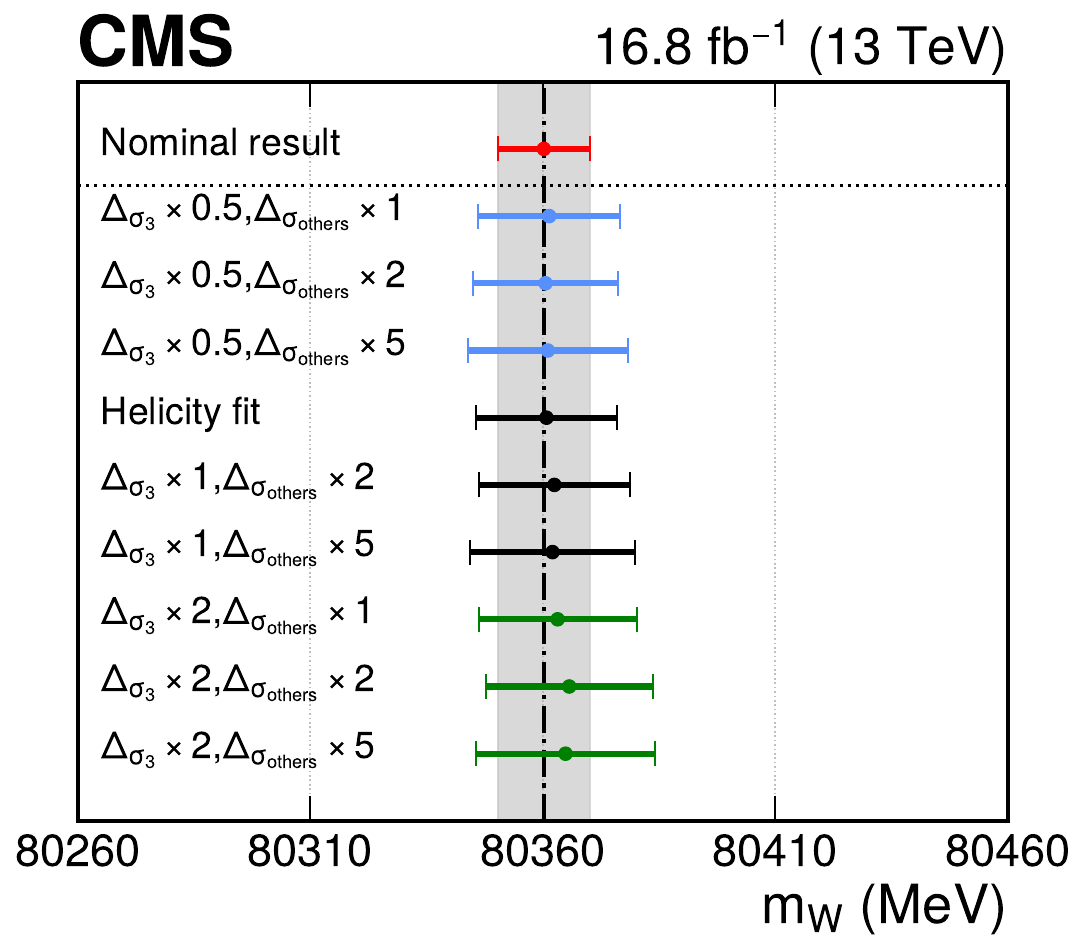}
\caption{\natcap{The \PW boson mass measured with the helicity fit 
for different scaling scenarios of the prefit helicity cross section uncertainties,
denoted by $\Delta_{\sigma_3}$ and $\Delta_{\sigma_{\text{others}}}$
for the $\sigma_3$ and the other components, respectively.}
The points are grouped and colored according to the scaling of $\sigma_3$.
The black line indicates the nominal result, with its uncertainties shown by the gray band.}
\label{fig:hf_mw_inflation}
\end{figure}

Figure~\ref{fig:hf_mw_inflation} shows the \mw values measured
with the helicity fit for different scenarios of the prefit helicity cross section uncertainties. 
We halved or doubled the default $\sigma_3$ prefit uncertainty,
to study possible shifts of the central value 
under more aggressive or conservative theoretical assumptions.
For each of those scenarios, 
we inflated the other helicity cross section uncertainties by factors of 2 or~5
(in addition to the nominal uncertainty). 
All eight extra cases give central \mw values that are stable and consistent with 
both the baseline and helicity fit nominal results.
\insupp{Figure 6 in the \suppmat}{Figure~\ref{fig:helicity_qty}} shows the \PW boson differential cross sections in
 \ptw and \ayw, extracted from the \etaptqmu distributions through the
decomposition of the helicity amplitudes in \ptw and \yw bins.

\subsection{The \texorpdfstring{\wlikens}{W-like} \texorpdfstring{\PZ}{Z} and \texorpdfstring{\PW}{W} boson mass measurements}
\label{sec:wmass-details}

\begin{figure}[!htp]
\centering
\includegraphics[width=0.9\columnwidth]{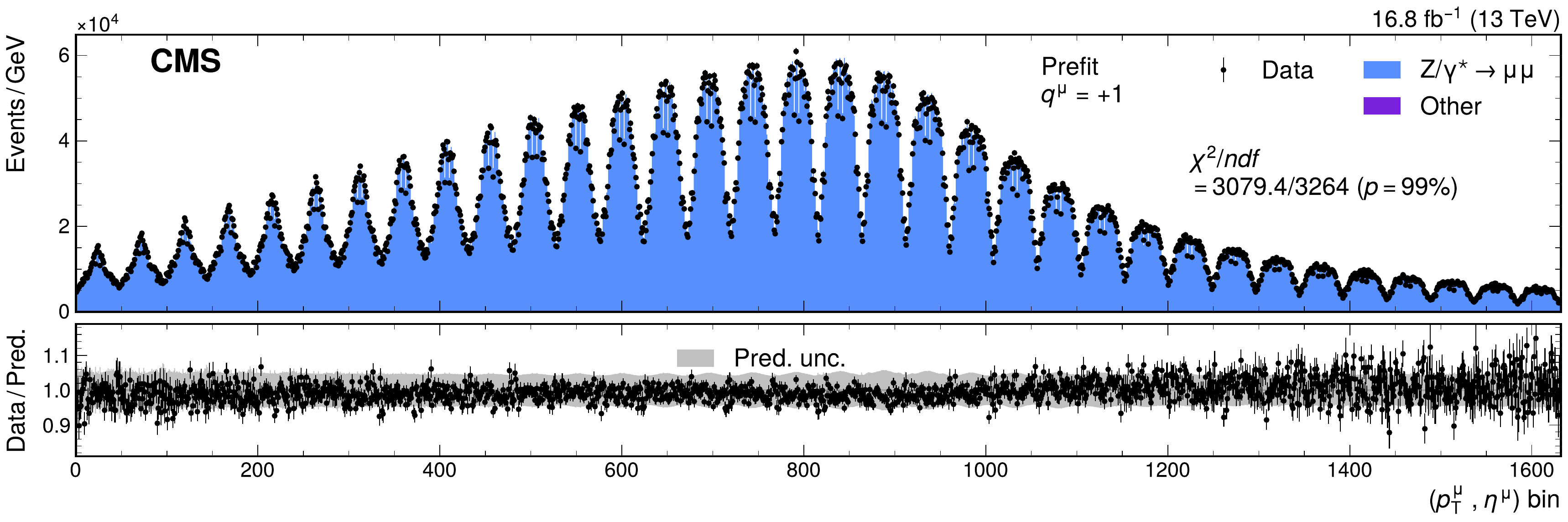}
\includegraphics[width=0.9\columnwidth]{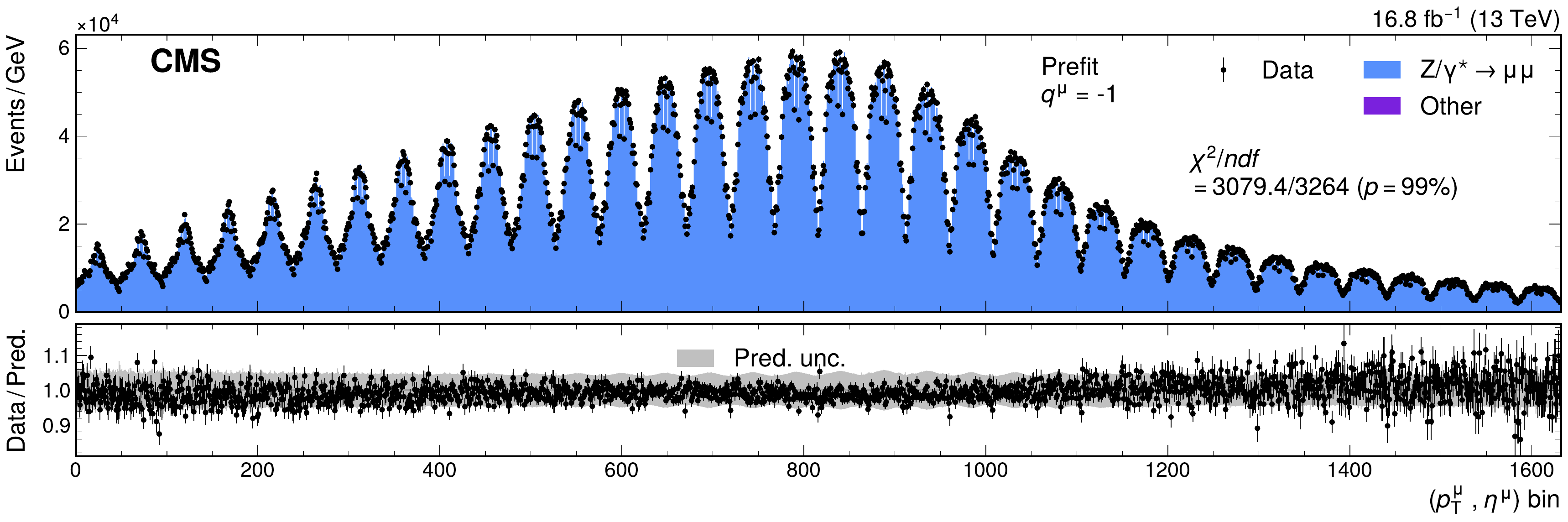}
\includegraphics[width=0.9\columnwidth]{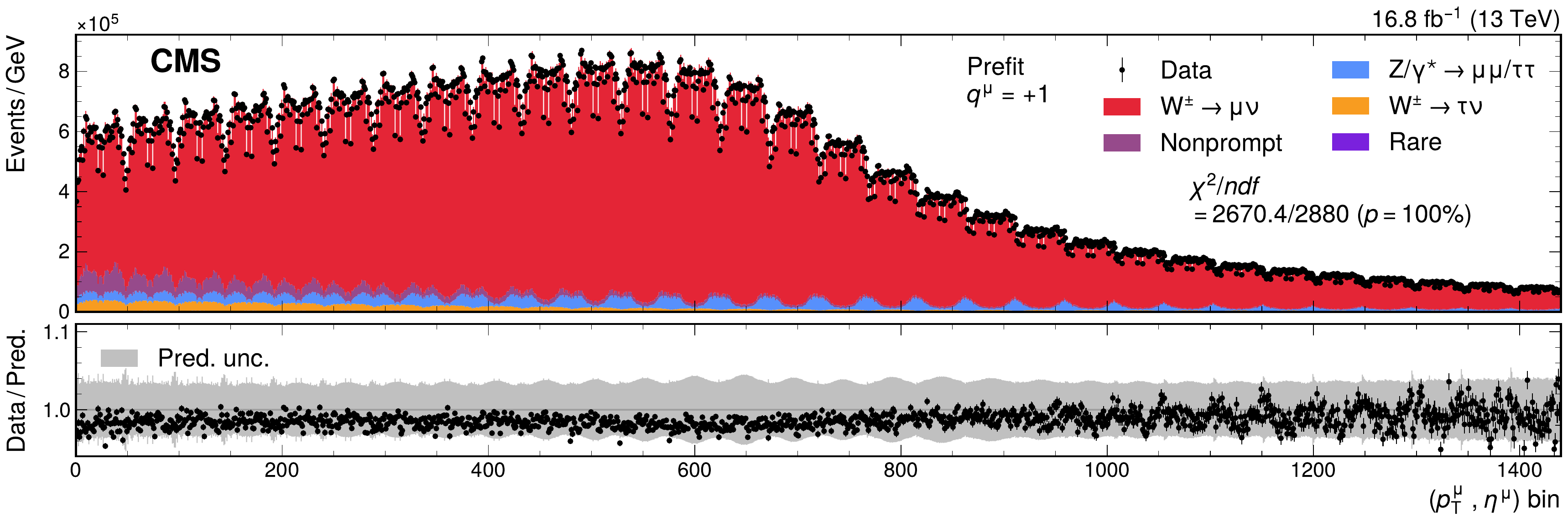}
\includegraphics[width=0.9\columnwidth]{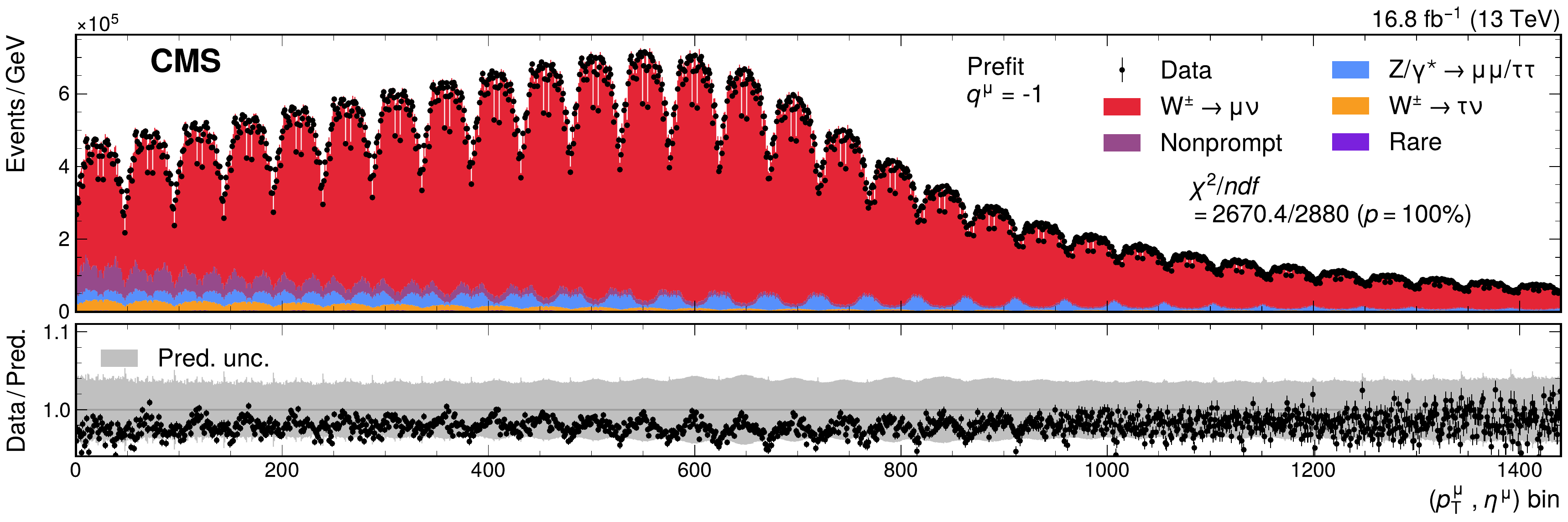}
\caption{\natcap{Measured and predicted \etaptmu distributions used in the \wlike \mz (upper two) and \mw (lower two) measurements
for positively (upper and second from bottom) and negatively (second from top and lower) charged muons.}
The two-dimensional distribution is ``unrolled" such that each bin on the $x$-axis represents one \etaptmu cell. 
The gray band represents the uncertainty in the prediction, before the fit to the data.
The bottom panel shows the ratio of the number of events observed in data to the nominal prediction.
The vertical bars represent the statistical uncertainties in the data.}
\label{fig:pteta}
\end{figure}

\begin{figure}[t]
\centering
\includegraphics[width=0.45\textwidth]{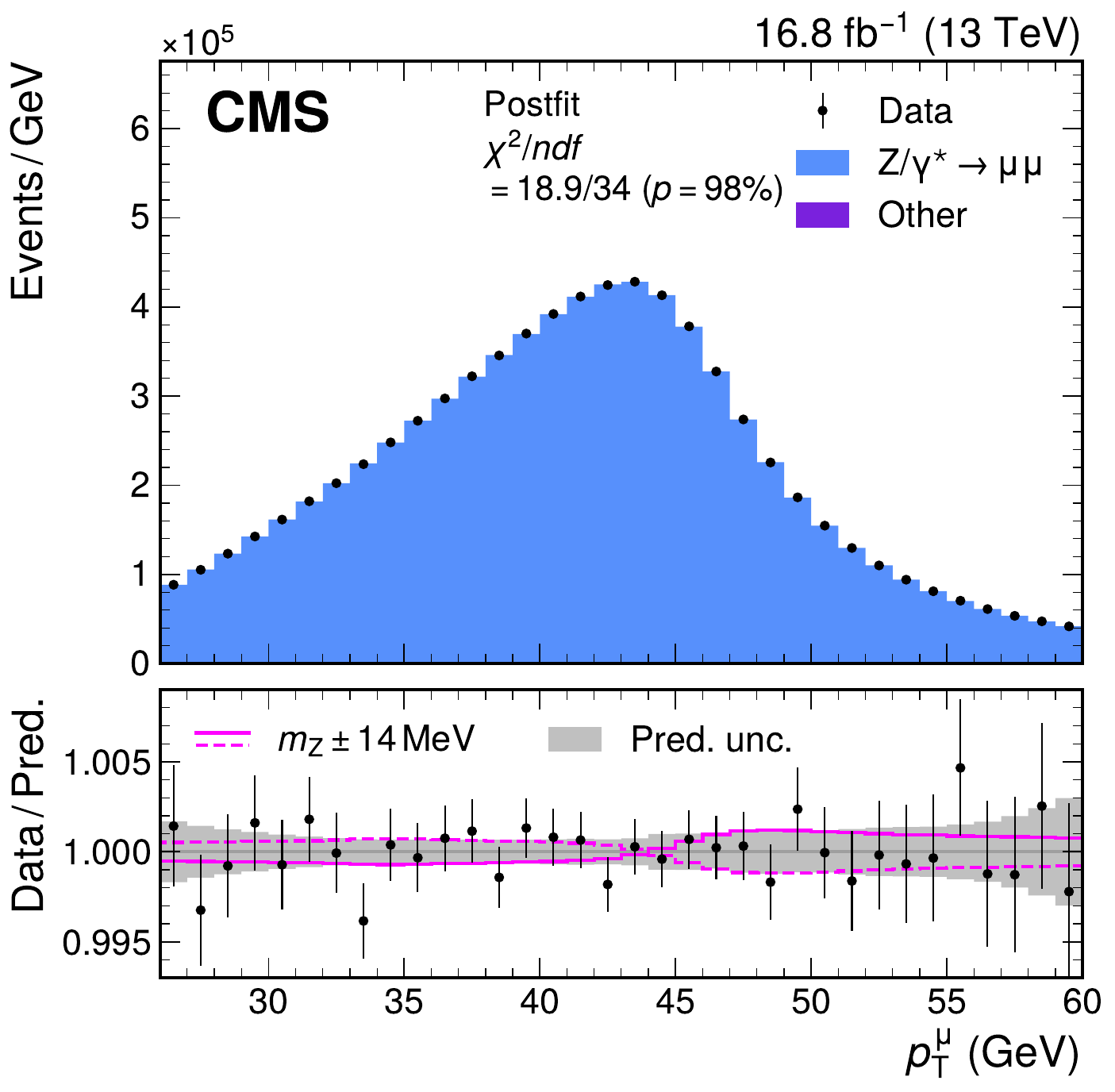}
\caption{\natcap{Measured and simulated \ptmu distributions, 
with the prediction adjusted according to the best fit values of nuisance parameters and of \mz
obtained from the maximum likelihood fit of the \wlike \mz analysis.}
The vertical bars represent the statistical uncertainties in the data.
The bottom panel shows the ratio of the number of events observed in data 
to the nominal prediction.
The solid and dashed purple lines represent, respectively, 
the relative impact of an increase and decrease of \mz by 14\MeV. 
The uncertainties in the predictions, 
after the systematic uncertainty profiling in the maximum likelihood fit, 
are shown by the shaded band.}
\label{fig:ptl_mz}
\end{figure}

\insupp{Figure~\ref{fig:pteta} and \suppmat Fig. 8}{Figures~\ref{fig:pteta} and \ref{fig:ptetapf}} show the \etaptmu distributions
used in the binned maximum likelihood template fits to perform the \wlike \mz and \mw measurements,
before and after the maximum likelihood fit, respectively.
The \etamu binning allows sensitivity to discontinuities in the geometry of the detectors 
and maximally exploits in situ constraints of systematic uncertainties. 
The \ptmu binning roughly corresponds to the \ptmu resolution, 
useful to enhance the sensitivity to the measured mass, 
while avoiding fluctuations in the simulated templates that could potentially lead to 
undercoverage of the estimated uncertainties~\cite{alexe2024undercoverage}.
The predicted and observed \ptmu distribution of the \wlike analysis, 
with the prediction corrected by the best fit values of the nuisance parameters 
after the maximum likelihood fit to the \etaptqmu distribution, is shown in Fig.~\ref{fig:ptl_mz}. 
The impact of a variation of \mz corresponding to the total uncertainty of the measurement is also shown, 
as well as the uncertainties in the prediction, illustrating the precision of the measurement.

The uncertainty due to the size of the simulated samples is estimated via 
the Barlow--Beeston approach~\cite{Barlow:1993dm}, 
as simplified by Conway~\cite{Conway:2011in}.
The estimate is increased by 25\%
to account for the effect of fluctuations in the alternate templates 
used to construct the systematic variations~\cite{alexe2024undercoverage}. 
The 1.25 scaling factor is estimated by evaluating the coverage of the uncertainty
with bootstrap resampling of the simulated samples~\cite{Efron1992}.
The width of the \PW boson, \gw, is varied with the mass 
according to the SM relationship $\gw \propto \mw^{3}$.
The theoretical uncertainty of 0.6\MeV from the EW fit is 
taken as an uncertainty in \gw, but we verified
that using the experimental uncertainty of Ref.~\cite{PDG2024} has a
negligible impact on the results. The uncertainty in \mw from \gw is $<$0.2\MeV.
Systematic uncertainties whose
$+1\sigma$ and $-1\sigma$ variation templates affect the event yields asymmetrically
are decomposed into two symmetric variations, 
defined such that the symmetric and anti-symmetric components are represented separately. 
This procedure preserves the total variance and covariance of the event yields and reduces 
non-linearities in the likelihood, simplifying the evaluation of uncertainties and impacts.

Several additional tests are performed to verify the robustness of the statistical procedure. The likelihood minimization is performed with an independent implementation of the likelihood function and with a different minimization algorithm, 
which yields a value of \mw that is identical to the nominal result within the numerical precision. The minimization was performed ten times with the starting values of \mw and the nuisance parameters
set to random values. The resulting minima of the likelihood function and the associated \mw values are consistent with the nominal result in all cases.
Finally, an estimation of \mw is performed linearizing the likelihood function and absorbing all systematic uncertainties 
into the data covariance matrix, such that the systematic uncertainties are not explicit fit parameters. Instead, they
are treated equivalently to the statistical uncertainty of the data.  In this configuration, the minimum can be evaluated analytically,
without the need for iterative minimization. As discussed in Ref.~\cite{Pinto:2023yob}, this treatment is mathematically equivalent
when the likelihood is purely quadratic. The value of \mw extracted with this procedure is shifted by 1.5\MeV with respect to the nominal result.
The difference is primarily due to the linearization of uncertainties associated with the nonprompt background estimate, which in this approximated treatment are not scaled with the yield of the nonprompt background in case it differs from its prefit value.  While the nominal treatment is a more accurate representation of the uncertainties, 
the stability of the result within the associated uncertainties addresses potential concerns about the robustness of the likelihood minimization~\cite{Kotwal:2024ume}.

The individual systematic uncertainties in the \wlike \mz and \mw measurements 
are presented in Table~\ref{tab:impacts_all}.  
The uncertainty breakdown labeled as ``Nominal impact'' 
is computed according to the procedure described in Ref.~\cite{CMS:SMP-18-012}, 
where the data statistical uncertainty corresponds to 
a hypothetical analysis with no systematic uncertainties. 
The impact for all other sources of uncertainty corresponds to 
the amount by which the total uncertainty would decrease in quadrature 
if that source were removed from the analysis. 
The total uncertainty cannot be calculated as the sum in quadrature of the impacts
because of correlations between the partial uncertainties.

\begin{table*}[th!]
\centering
\topcaption{\natcap{Uncertainties in the \wlike \mz and \mw measurements, with
contributions to the total uncertainty from individual sources separated according to
the ``nominal''~\cite{CMS:SMP-18-012} and ``global''~\cite{Pinto:2023yob} definitions of the impacts.}}
\label{tab:impacts_all}
\begin{tabular}{lcccc}
\multirow{3}{*}{Source of uncertainty} & \multicolumn{4}{c}{Impact (\MeVns)} \\
                                       & \multicolumn{2}{c}{Nominal}  & \multicolumn{2}{c}{Global} \\
                                       &  in \mz & in \mw & in \mz & in \mw\\ \hline
 Muon momentum calibration              & 5.6     &   4.8  &  5.3     &   4.4 \\
 Muon reco.\ efficiency                 & 3.8     &   3.0  &  3.0     &   2.3 \\
 \PW and \PZ angular coeffs.            & 4.9     &   3.3  &  4.5     &   3.0 \\
 Higher-order EW                        & 2.2     &   2.0  &  2.2     &   1.9 \\
 \ptv modeling                          & 1.7     &   2.0  &  1.0     &   0.8 \\
 PDF                                    & 2.4     &   4.4  &  1.9     &   2.8 \\
 Nonprompt-muon background                   & \NA  &   3.2  &  \NA  &   1.7 \\
 Integrated luminosity                  & 0.3     &   0.1  &  0.2     &   0.1 \\
 MC sample size                         & 2.5     &   1.5  &  3.6     &   3.8 \\ \hline
 Data sample size                       & 6.9     &   2.4  &  10.1    &   6.0 \\
 Total uncertainty                                 & 13.5    &   9.9  &  13.5    &   9.9 \\   
\end{tabular}
\end{table*}

This uncertainty breakdown is not directly comparable to that of ATLAS~\cite{ATLAS:2024erm}, 
which uses an alternative method to define the uncertainty contributions, 
referred to as ``global'' impacts~\cite{Pinto:2023yob}.
In that approach, the data statistical uncertainty is, instead, 
computed with the nuisance parameters present in the fit.
If the data constrain the nuisance parameters in situ, beyond the externally imposed constraints, 
then fluctuations in the data and the simulated event samples become correlated with the fitted values of the nuisance parameters, 
which in turn increases the statistical components of the uncertainty.
The impacts of systematic sources
are computed considering fluctuations of the corresponding external measurements
(i.e., of the so-called ``global observables'') within their uncertainties.
In the presence of stronger in situ constraints, 
this method typically leads to smaller impacts than our approach.
These two procedures only differ in 
the split between the statistical and systematic components of the uncertainty;
they do not impact the total uncertainty of the result.
To facilitate the comparison with the uncertainty breakdown of the ATLAS measurement, 
Table~\ref{tab:impacts_all} also reports the leading uncertainties using global impacts.  

\insupp{Table~2 in the \suppmat}{Table~\ref{tab:nuisParamFit}} shows a summary of the number of nuisance parameters 
included in the likelihood for the \wlike \mz and \mw fits. 
The parameters are categorized into groups, 
corresponding to the main sources of uncertainty reported in Table~\ref{tab:impacts_all},
and gathering conceptually related systematic uncertainties. 
Uncertainties specific to \PW bosons, for instance the mass or width variations, 
are not implemented in the \wlike \mz analysis because the \wj background is negligible.

\subsection{Measurement of \texorpdfstring{$\mwp-\mwm$}{mW+ - mW-}}
\label{sec:Wmassdifference}

Our measurement assumes that the \PWp and \PWm bosons have identical masses, $\mwp = \mwm$, 
as required by CPT symmetry. 
This requirement reduces the impact of uncertainties that affect the two charges differently, 
including the PDFs, angular coefficients, and the alignment terms of the muon momentum calibration. 
By relaxing this requirement, we perform a measurement of the mass difference,
\begin{equation*}
\mwp-\mwm = 57.0 \pm 30.3\MeV .
\end{equation*}
The significant increase in the uncertainty compared with the \mw measurement is due to uncertainties 
that have a strong negative correlation between
the two charges.
In particular, 
the muon momentum calibration contributes an uncertainty of 22.0\MeV, 
the angular coefficients contribute 18.7\MeV, 
and the PDF uncertainty is 11.8\MeV. 
The statistical uncertainty of the data is 4.7\MeV. 
The $p$-value indicating the compatibility of this result and
the expectation of $\mwp-\mwm=0$ is 6.0\%, or about $1.9\sigma$.
The charge-independent \mw value measured in this configuration is shifted by 0.3\MeV
with respect to the nominal result, having a negligible effect on the total uncertainty.
The correlation coefficient between \mwp and \mwm is $\rho_{+-} = -0.40$, 
whereas the correlation between the mass difference and mass average is only 0.02. 
The small correlation between \mw and $\mwp-\mwm$ is a consequence of a strong degree 
of anticorrelation for the alignment components of the \ptmu calibration uncertainties, 
and the uncertainties in the $A_3$ angular coefficient.

If the \mwp and \mwm values are measured independently, the total uncertainty
in each is $\approx$20\MeV, with statistical uncertainties of 2 and 3\MeV, respectively.
Because the statistical uncertainties are negligible,
the difference between the \PWp and \PWm production rates does not have a visible effect on the charge-inclusive \mw
measurement, which is the average of the two results. 
The total variance in the sum ($\mwp+\mwm = 2\mw$) or difference ($\mwp-\mwm$) is
$\sigma_{+}^2+\sigma^{2}_{-} + 2c\,\rho_{+-}\sigma_{+}\sigma_{-}$, 
where $\sigma^{2}_{+}$ and $\sigma^{2}_{-}$ are the variances of \mwp and \mwm, respectively, and 
$c=+1$ for the sum and $-1$ for the difference.
The opposite effect of $\rho_{+-}$ on the propagation of uncertainty
results in a factor of $\sim\!1/3$ between the \mw uncertainty and that of
$\mwp-\mwm$.

As a validation of this result, 
we also perform the corresponding measurement in the case of the \wlike \mz measurement
using the positively and negatively charged muons. 
In this case, the two leptons are from the same object and, therefore, 
the comparison is purely a validation of the theoretical and experimental inputs. 
The result when selecting positively charged muons in odd event-number events is
\begin{equation*}
m_{\PZ^{+}} - m_{\PZ^{-}} = 30.9 \pm 32.5\MeV,
\end{equation*}
and for the reversed muon charge \vs event number selection we get
\begin{equation*}
m_{\PZ^{+}} - m_{\PZ^{-}} = 6.4 \pm 32.3\MeV.
\end{equation*}
Apart from the PDFs, which are not relevant for this measurement, 
the breakdown of uncertainties is similar to the $\mwp-\mwm$ case. 
The muon momentum scale and the angular coefficients contribute
uncertainties of 23.1 and 14.5\MeV, respectively. 
The statistical uncertainty of the data is 13.9\MeV.

\insupp{Table~3 in the \suppmat}{Table~\ref{tab:impacts_WlikeZandW_chargeDiff}}
shows the impacts on the difference between the measured mass 
with positive or negative muons in the \wlike \mz and \mw analyses, 
comparing with the nominal result from the simultaneous fit to both charges, 
and using nominal impacts. 
The breakdown of uncertainties from the global definition of the impacts 
is also reported, for completeness. 

We have performed several additional checks that confirm that the small tension 
with the expectation of $\mwp=\mwm$ does not reflect a bias 
or an underestimation of our uncertainties that would impact our result. 
The alignment components of the muon momentum scale calibration 
and the $A_3$ angular coefficient uncertainties 
are the dominant sources affecting the $\mwp-\mwm$ measurement.
Therefore, we repeat both measurements after varying the central value 
of these parameters by $1\sigma$
such that the charge difference is reduced, keeping their relative uncertainty fixed. 
Up variations of the \ptmu scale alignment terms, 
and down variations of the $A_3$ coefficient uncertainties, 
each reduce $\mwp-\mwm$. 
The maximum shift in $\mwp-\mwm$ when varying these two terms, either
independently or coherently, moves the result towards zero by $1.2\sigma$, 
compared with the $\mwp-\mwm$ measurement with the nominal uncertainty model.
If the alignment parameters of the muon momentum scale calibration are extracted from \zmm events, 
the measured $\mwp-\mwm$ shifts towards zero by 16\MeV, with a total uncertainty of 25\MeV.
In the extreme configuration where the alignment term is varied by $3\sigma$, 
resulting in $\mwp-\mwm \approx 0$, the extracted \mw differs from our nominal result by 0.6\MeV.

\subsection{Results with alternative parton distribution functions}
\label{sec:otherPDFs}

We performed the \mw measurement using alternative PDF sets, 
with and without scaling factors, following the procedure described in Section~\ref{sec:pdf}. 
The results are shown in Table~\ref{tab:mw_pdf} and \insupp{\suppmat Fig. 7}{Fig.~\ref{fig:mw_pdf}}. The scaling procedure, combined with the uncertainty profiling and in situ data constraints, 
improves the consistency between the \mw results obtained with the different PDF sets. 
If no uncertainty scaling is used, the results vary by 8.1\MeV, 
from $80\,355.1 \pm 9.3\MeV$ (NNPDF4.0) to $80\,363.2 \pm 9.9\MeV$ (PDF4LHC21).
If we use the uncertainty scaling, the spread of results is reduced to 6.2\MeV,
ranging from $80\,357.0 \pm 10.8\MeV$ (NNPDF4.0 with uncertainties scaled by 5.0) 
to $80\,363.2 \pm 9.9\MeV$ (PDF4LHC21 with no scaling factor).
The spread of these values is within the total PDF uncertainty of our nominal measurement (performed with CT18Z).

\begin{table*}[tb]
\centering
\topcaption{\natcap{The \mw values measured for different PDF sets.}
The second column indicates the prefit uncertainty scaling factors required to cover
the central predictions of the considered PDF sets, as described in the text. The
observed values,
with uncertainties scaled following the procedure described in Section~\ref{sec:pdf}
and with the default unscaled uncertainties, are shown in the last two columns.
The uncertainty in \mw from the PDF is indicated in parenthesis following the total uncertainty.}
\label{tab:mw_pdf}
\begin{tabular}{lccc}
\multirow{2}{*}{PDF set} & \multirow{2}{*}{$\sigma_\text{PDF}$ scale fact.} & \multicolumn{2}{c}{Extracted \mw ($\sigma_\text{PDF}$) (\MeVns)} \\
 & & Original $\sigma_\text{PDF}$ &  Scaled $\sigma_\text{PDF}$	\\ \hline
        CT18Z       & \NA & \multicolumn{2}{c}{$80\,360.2 \pm 9.9~~\,(4.4)$}  \\ 
        CT18        & \NA & \multicolumn{2}{c}{$80\,361.8 \pm 10.0 \,(4.6)$}  \\ 
        PDF4LHC21   & \NA & \multicolumn{2}{c}{$80\,363.2 \pm 9.9~~\,(4.1)$} \\
        MSHT20      & 1.5 & $80\,361.4 \pm 10.0$   (4.3) & $80\,361.7 \pm 10.4 $ (5.1)    \\
        MSHT20aN3LO & 1.5 & $80\,359.9 \pm 9.9~~ $ (4.2) & $80\,359.8 \pm 10.3 $ (4.9)    \\
        NNPDF3.1    & 3.0 & $80\,359.3 \pm 9.5~~ $ (3.2) & $80\,361.3 \pm 10.4 $ (5.3)    \\
        NNPDF4.0    & 5.0 & $80\,355.1 \pm 9.3~~ $ (2.4) & $80\,357.0 \pm 10.8 $ (6.0)    \\
\end{tabular}
\end{table*}

We also tested the impact of excluding the PDF uncertainties from the profiling procedure in the maximum likelihood fit,
and estimated their impact by repeating the fit for each PDF eigenvector variation. 
The total uncertainty is defined as the sum in quadrature of the PDF eigenvector variation uncertainties and the total profiled uncertainty.
In this configuration, when using the CT18Z PDF set, the measured \mw value decreases by less than 2\MeV,
within the PDF uncertainty of the nominal result using profiling.
The PDF uncertainty estimated from the CT18Z set increases by a factor of $\sim$2.5
when the PDF uncertainty is not profiled, and the total uncertainty 
in \mw increases by $\approx\!3\MeV$. The shift in the measured
\mw value is within $1\sigma$ of the profiled PDF uncertainty for NNPDF4.0, MSHT20, MSHT20aN3LO, and PDF4LHC, and
within $2\sigma$ for NNPDF3.1 and CT18. The PDF uncertainty increase varies between
a factor of 1.5 (NNPDF4.0) and a factor of 3 (CT18).
We have explicitly verified the coverage of the uncertainty, for both the profiled and the unprofiled cases, 
using pseudo-experiments whose pseudo-data are obtained from a random draw of counts as predicted 
from a random realization of the statistical model parameters.
This method corresponds to the ``frequentist toys'' approach described in Ref.~\cite{CMS:2024onh}.

\subsection{Additional validation checks of experimental inputs}

A number of additional tests were performed to ensure that the analysis is robust with respect to variations in the selections used.

The \mz extraction from \zmm events is performed in subsets of events defined by the relative location of the two muons in the CMS detector.
\insupp{Figure~9 in the \suppmat}{Figure~\ref{fig:decorrMassShiftRegions}} shows that the nominal \mz result is compatible with the results obtained when both muons are central ($\aetamu<0.9$), one is central and one is forward, or both are forward.
The same exercise is performed depending on \etamu, by requiring both muons on the $\etamu<0$ half of the detector, both on the positive half, and one in each half of the detector, with the same conclusions.
Concerning the \wlike \mz and \mw analyses, the \mv extraction is performed in 24 bins of \etamu.
The results are shown in \insupp{Fig.~11 in the \suppmat}{Fig.~\ref{fig:decorrMassShiftWZ}},
where the compatibility with the nominal result can also be appreciated.
The \etamu dependence of the result is also assessed by evaluating the difference between the \mv values measured when selecting muons in the central ($\aeta < 0.9$) and forward ($0.9 < \aeta < 2.4$) 
regions of the detector. 
The difference between the \mw values measured with central muons and with forward muons is $15.3\pm14.7\MeV$. 
For the \wlike \mz analysis, the corresponding result is $22.8\pm21.1\MeV$.

We test the stability of the measurement by performing the \mw extraction in separate data collection periods
and by dividing the dataset into five regions according to the number of reconstructed vertices.
The number of reconstructed vertices is strongly correlated with the number of pileup interactions, 
and the average pileup increased with time during data taking. 
Therefore, these checks confirm that the measurement is stable with respect to pileup,
as well as other time dependent changes in operational conditions.
Because of the limited sample size, we do not derive independent efficiency SFs per period. Instead, we use the nominal SFs measured for the inclusive dataset. We estimate the impact of time-dependent effects that are not modeled by the simulation and apply them as uncertainties in the SFs per data collection period.
An independent \mw parameter is assigned to each bin, and they are fitted simultaneously.
Most of the experimental uncertainties are uncorrelated across the independent bins.
The results are presented in \insupp{Fig.~13 and Fig.~14 in the \suppmat}{Fig.~\ref{fig:decorr_massShiftW_runBins} and Fig.~\ref{fig:decorr_massShiftW_nRecoVtxBins}}, which show good compatibility with the nominal measurement. 

Additionally, we evaluate the effect of splitting the data by the muon azimuthal angle \phimu. 
For this study, we use the nominal SFs and apply uncertainties to account for the estimated variation of the efficiency SFs across the \phimu regions.
As shown in \insupp{Fig.~12 in the \suppmat}{Fig.~\ref{fig:decorr_massShiftW_phiBins}}, the results are consistent with the nominal measurement.
We asses the impact of the \mtw selection in the \mw measurement
by repeating the measurement with different \mtw thresholds, from 30 to 50\GeV. 
The summary is shown in \insupp{Fig.~15 in the \suppmat}{Fig.~\ref{fig:massShiftW_varyMTcut}}. 
The largest deviations from the nominal value correspond to $-2.8\MeV$ and $3.3\MeV$, 
comparable to the total nonprompt-muon background uncertainty. 
The total uncertainty varies inversely with the \mtw threshold, from 10.1 to 9.7\MeV, because of the reduced impact of the nonprompt-muon background uncertainty when the \mtw threshold is increased. However, the nonprompt-muon background estimation and 
the recoil calibration and uncertainties have not been independently optimized for the varied thresholds.

Other tests that were performed and also showed no incompatibility with the corresponding nominal result include:
\begin{enumerate} 
\item Performing the \wlike \mz and \mw analyses splitting events by the sign of \etamu, 
for each half of the CMS detector separately.
\item Performing the \mw analysis reducing the \ptmu range considered 
by removing 4\GeV on the high end, on the low end, and on both ends.
\item Performing the \mw analysis treating the normalization of the \PW signal process unconstrained. 
A scaling factor of $0.979\pm 0.026$ is obtained, in agreement with the SM expectation of unity.
\end{enumerate}

Tabulated results are provided in the HEPData record for this work~\cite{HEPData}. 
\clearpage

\begin{dataavailability}
Release and preservation of data used by the CMS Collaboration as the basis for publications is guided by the  \href{https://cms-docdb.cern.ch/cgi-bin/PublicDocDB/RetrieveFile?docid=6032&filename=CMSDataPolicyV1.2.pdf&version=2}{CMS data preservation, re-use and open access policy}.
\end{dataavailability}
  
\begin{codeavailability}
The CMS core software is publicly available in our \href{https://github.com/cms-sw/cmssw}{GitHub repository}.
The data and simulation were processed with v.10\_6\_26 to produce output in the NanoAOD~\cite{Peruzzi_2020} format.
The processing of the data into histograms was performed with the WRemnants framework, which is
publicly available in a \href{https://github.com/WMass/WRemnants}{GitHub repository}. The version tagged ``published'' has been used for the results in the paper.
The statistical analysis was performed with the Rabbit software framework
(\href{https://github.com/WMass/rabbit}{GitHub repository}), which was extensively
validated against the results of the CMS Combine~\cite{CMS:2024onh} statistical analysis tool.
\end{codeavailability}
  
\bibliography{auto_generated} 
  
\begin{acknowledgments}
We congratulate our colleagues in the CERN accelerator departments for the excellent performance of the LHC and thank the technical and administrative staffs at CERN and at other CMS institutes for their contributions to the success of the CMS effort. In addition, we gratefully acknowledge the computing centers and personnel of the Worldwide LHC Computing Grid and other centers for delivering so effectively the computing infrastructure essential to our analyses. Finally, we acknowledge the enduring support for the construction and operation of the LHC, the CMS detector, and the supporting computing infrastructure provided by the following funding agencies: SC (Armenia), BMBWF and FWF (Austria); FNRS and FWO (Belgium); CNPq, CAPES, FAPERJ, FAPERGS, and FAPESP (Brazil); MES and BNSF (Bulgaria); CERN; CAS, MoST, and NSFC (China); MINCIENCIAS (Colombia); MSES and CSF (Croatia); RIF (Cyprus); SENESCYT (Ecuador); ERC PRG, RVTT3 and MoER TK202 (Estonia); Academy of Finland, MEC, and HIP (Finland); CEA and CNRS/IN2P3 (France); SRNSF (Georgia); BMBF, DFG, and HGF (Germany); GSRI (Greece); NKFIH (Hungary); DAE and DST (India); IPM (Iran); SFI (Ireland); INFN (Italy); MSIP and NRF (Republic of Korea); MES (Latvia); LMTLT (Lithuania); MOE and UM (Malaysia); BUAP, CINVESTAV, CONACYT, LNS, SEP, and UASLP-FAI (Mexico); MOS (Montenegro); MBIE (New Zealand); PAEC (Pakistan); MES and NSC (Poland); FCT (Portugal); MESTD (Serbia); MCIN/AEI and PCTI (Spain); MOSTR (Sri Lanka); Swiss Funding Agencies (Switzerland); MST (Taipei); MHESI and NSTDA (Thailand); TUBITAK and TENMAK (Turkey); NASU (Ukraine); STFC (United Kingdom); DOE and NSF (USA).

\hyphenation{Rachada-pisek} Individuals have received support from the Marie-Curie program and the European Research Council and Horizon 2020 Grant, contract Nos.\ 675440, 724704, 752730, 758316, 765710, 824093, 101115353, 101002207, 101001205, and COST Action CA16108 (European Union); the Leventis Foundation; the Alfred P.\ Sloan Foundation; the Alexander von Humboldt Foundation; the Science Committee, project no. 22rl-037 (Armenia); the Belgian Federal Science Policy Office; the Fonds pour la Formation \`a la Recherche dans l'Industrie et dans l'Agriculture (FRIA-Belgium); the F.R.S.-FNRS and FWO (Belgium) under the ``Excellence of Science -- EOS" -- be.h project n.\ 30820817; the Beijing Municipal Science \& Technology Commission, No. Z191100007219010 and Fundamental Research Funds for the Central Universities (China); the Ministry of Education, Youth and Sports (MEYS) of the Czech Republic; the Shota Rustaveli National Science Foundation, grant FR-22-985 (Georgia); the Deutsche Forschungsgemeinschaft (DFG), among others, under Germany's Excellence Strategy -- EXC 2121 ``Quantum Universe" -- 390833306, and under project number 400140256 - GRK2497; the Hellenic Foundation for Research and Innovation (HFRI), Project Number 2288 (Greece); the Hungarian Academy of Sciences, the New National Excellence Program - \'UNKP, the NKFIH research grants K 131991, K 133046, K 138136, K 143460, K 143477, K 146913, K 146914, K 147048, 2020-2.2.1-ED-2021-00181, TKP2021-NKTA-64, and 2021-4.1.2-NEMZ\_KI (Hungary); the Council of Science and Industrial Research, India; ICSC -- National Research Center for High Performance Computing, Big Data and Quantum Computing and FAIR -- Future Artificial Intelligence Research, funded by the NextGenerationEU program (Italy); the Latvian Council of Science; the Ministry of Education and Science, project no. 2022/WK/14, and the National Science Center, contracts Opus 2021/41/B/ST2/01369 and 2021/43/B/ST2/01552 (Poland); the Funda\c{c}\~ao para a Ci\^encia e a Tecnologia, grant CEECIND/01334/2018 (Portugal); the National Priorities Research Program by Qatar National Research Fund;  MCIN/AEI/10.13039/501100011033, ERDF ``a way of making Europe", and the Programa Estatal de Fomento de la Investigaci{\'o}n Cient{\'i}fica y T{\'e}cnica de Excelencia Mar\'{\i}a de Maeztu, grant MDM-2017-0765 and Programa Severo Ochoa del Principado de Asturias (Spain); the Chulalongkorn Academic into Its 2nd Century Project Advancement Project, and the National Science, Research and Innovation Fund via the Program Management Unit for Human Resources \& Institutional Development, Research and Innovation, grant B39G670016 (Thailand); the Kavli Foundation; the Nvidia Corporation; the SuperMicro Corporation; the Welch Foundation, contract C-1845; and the Weston Havens Foundation (USA).  
\end{acknowledgments}

\begin{authorcontributions}
All authors have contributed to the publication, being variously involved in the design  and the construction of the detectors, in writing software, calibrating subsystems,  operating the detectors and acquiring data, and finally analysing the processed data.  The CMS Collaboration members discussed and approved the scientific results. The  manuscript was prepared by a subgroup of authors appointed by the collaboration and  subject to an internal collaboration-wide review process. All authors reviewed and  approved the final version of the manuscript. 
\end{authorcontributions}

\begin{interests}
The authors declare no competing interests.
\end{interests}

\ifthenelse{\boolean{cms@external}}{}{
  \providecommand{\cmsTable}[1]{#1}
  \section{Supplementary information: additional figures and tables}
\label{sec:supplemental}

\begin{figure}[ht]
\centering
\includegraphics[width=0.48\textwidth]{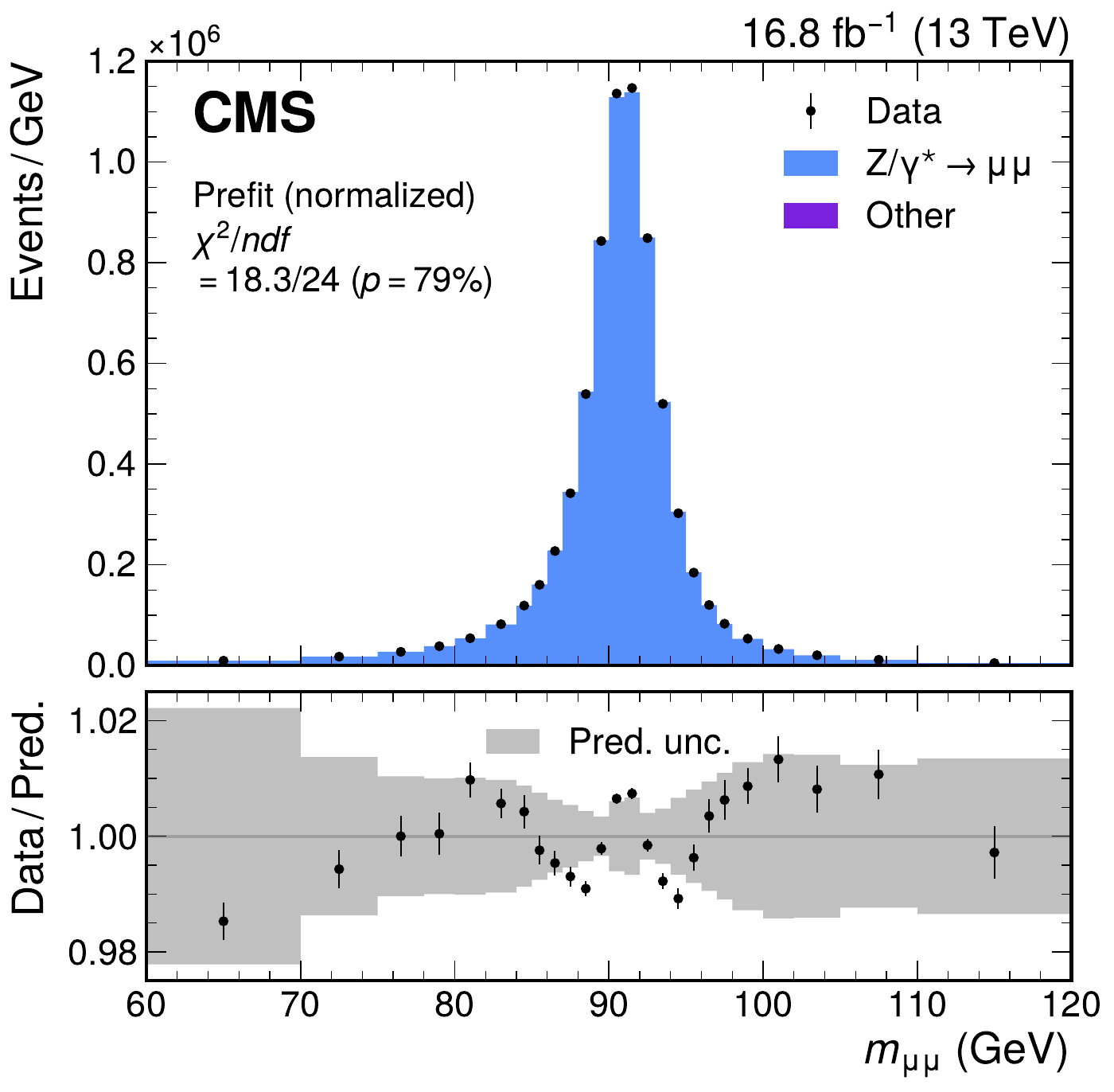}
\caption{Measured and simulated \zmm dimuon mass distributions,
after applying the muon momentum scale and resolution corrections. 
The simulated predictions and uncertainties are scaled to match the number of observed data events.
The vertical bars represent the statistical uncertainties in the data.
The bottom panel shows the ratio of the number of events observed in data
and of variations in the predictions
to that of the total nominal prediction.}
\label{fig:mll}
\end{figure}

\begin{figure}[!ht]
\centering
\includegraphics[width=0.48\textwidth]{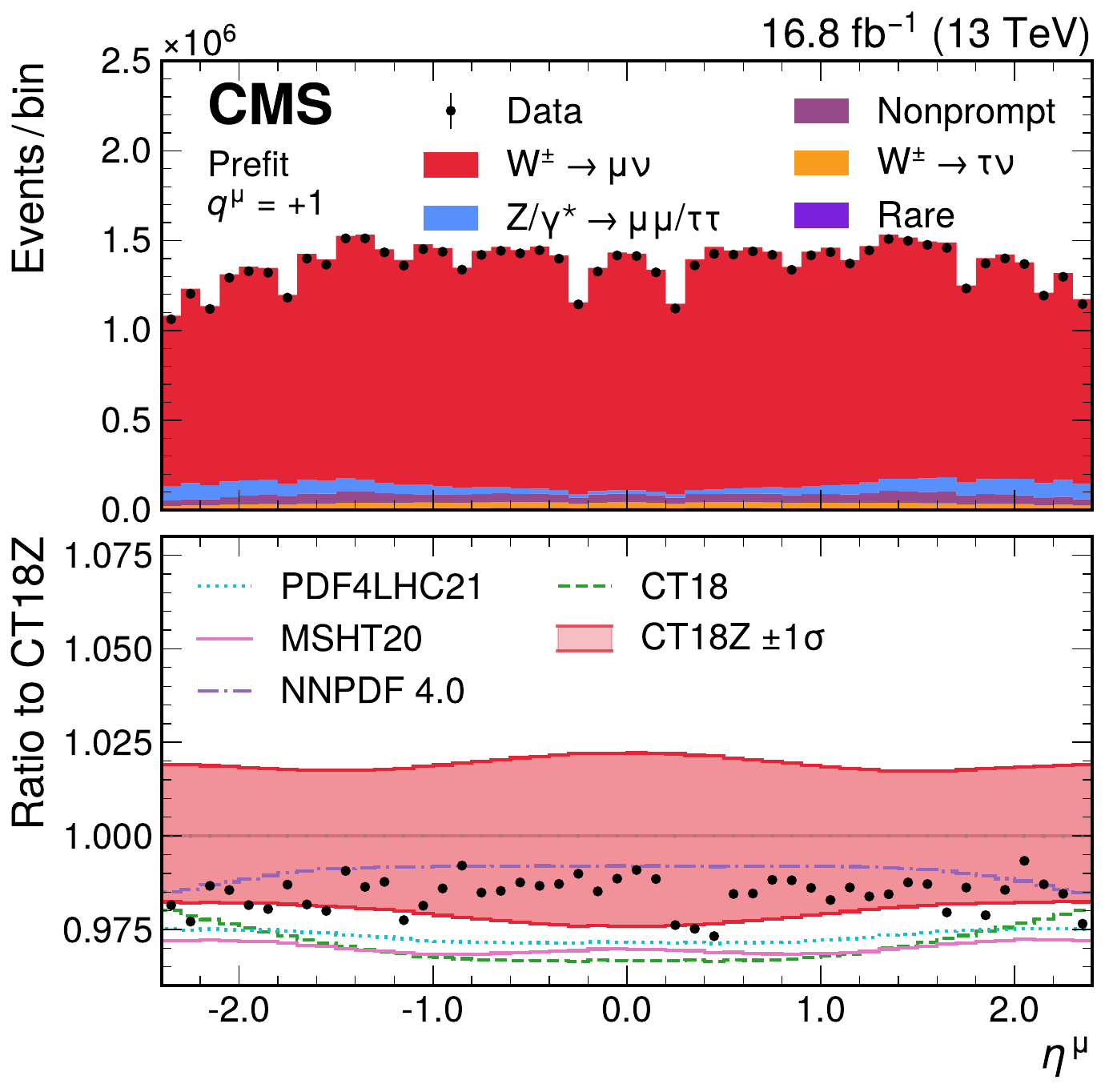}
\includegraphics[width=0.48\textwidth]{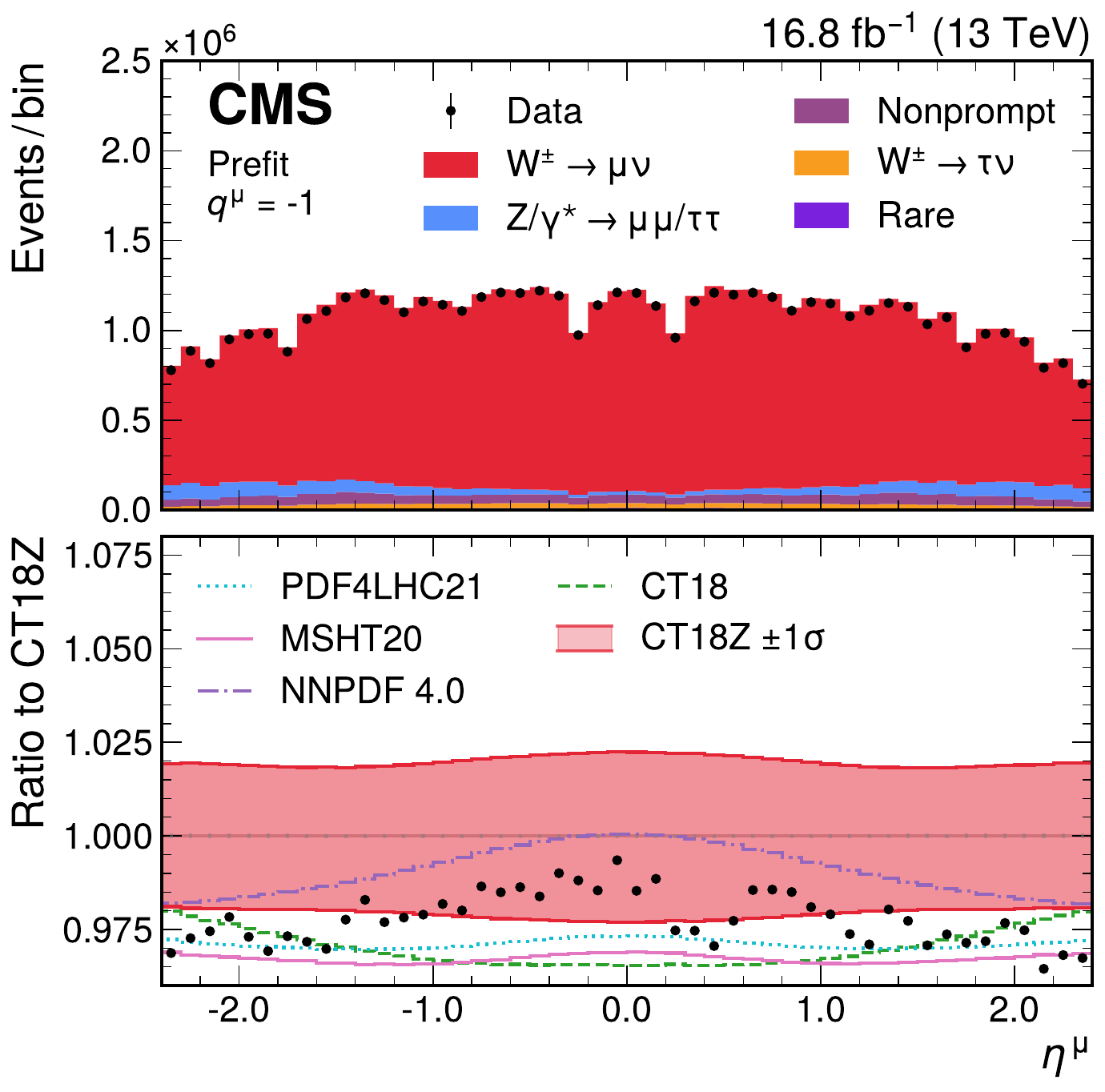}
\caption{Measured and predicted \etamu distributions 
for positively (\cmsLeft) and negatively (\cmsRight) charged muons.
The nominal prediction, obtained with the CT18Z PDF set, is shown in filled red. 
The uncertainty, evaluated as the sum of the eigenvector variation sets,
is represented by the filled band in the lower panel.
The predictions using the PDF4LHC21, MSHT20, NNPDF4.0, and CT18 sets 
are also shown (without uncertainty bands).
The vertical bars represent the statistical uncertainties in the data.
The bottom panel shows the ratio of the number of events observed in data 
and of variations in the predictions
to that of the nominal prediction.}
\label{fig:pdfs_w}
\end{figure}

\begin{figure}[ht]
\centering
\includegraphics[width=0.45\textwidth]{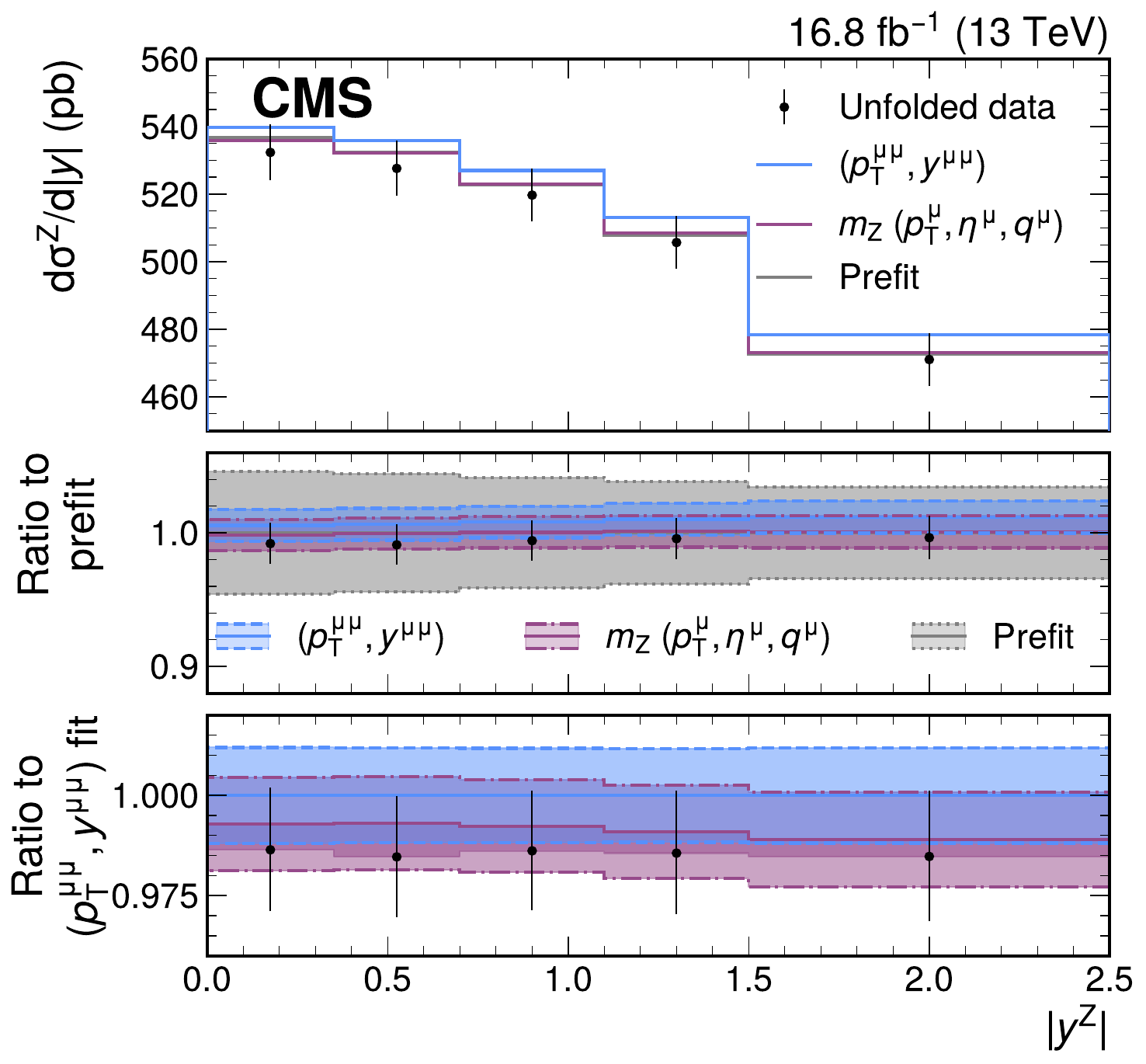}
\caption{
Unfolded measured \ayz distribution (points) compared with the generator-level \scetlibminnlo predictions
before (prefit, gray) 
and after adjusting the nuisance parameters to the best fit values
obtained from the \wlike \mz fit (magenta) or from the direct fit to the \ptmumu distribution (blue).
The results are obtained with the selection $\ayz < 2.5$
and $\ptz < 54\GeV$.
The center panel shows
the ratio of the predictions and unfolded data to the prefit prediction.
The uncertainty in the prefit prediction is shown by the shaded gray area.
The bottom panel shows the ratio of the predictions and unfolded data
to the postfit prediction from the
fit to the \ptymumu distribution. The postfit uncertainties in the predictions
are shown in the shaded
magenta and blue bands.
The vertical bars represent the total uncertainty in the unfolded data.}
\label{fig:yll_gen_postfit}
\end{figure}

\begin{figure}[t!]
\centering
\includegraphics[width=0.5\textwidth]{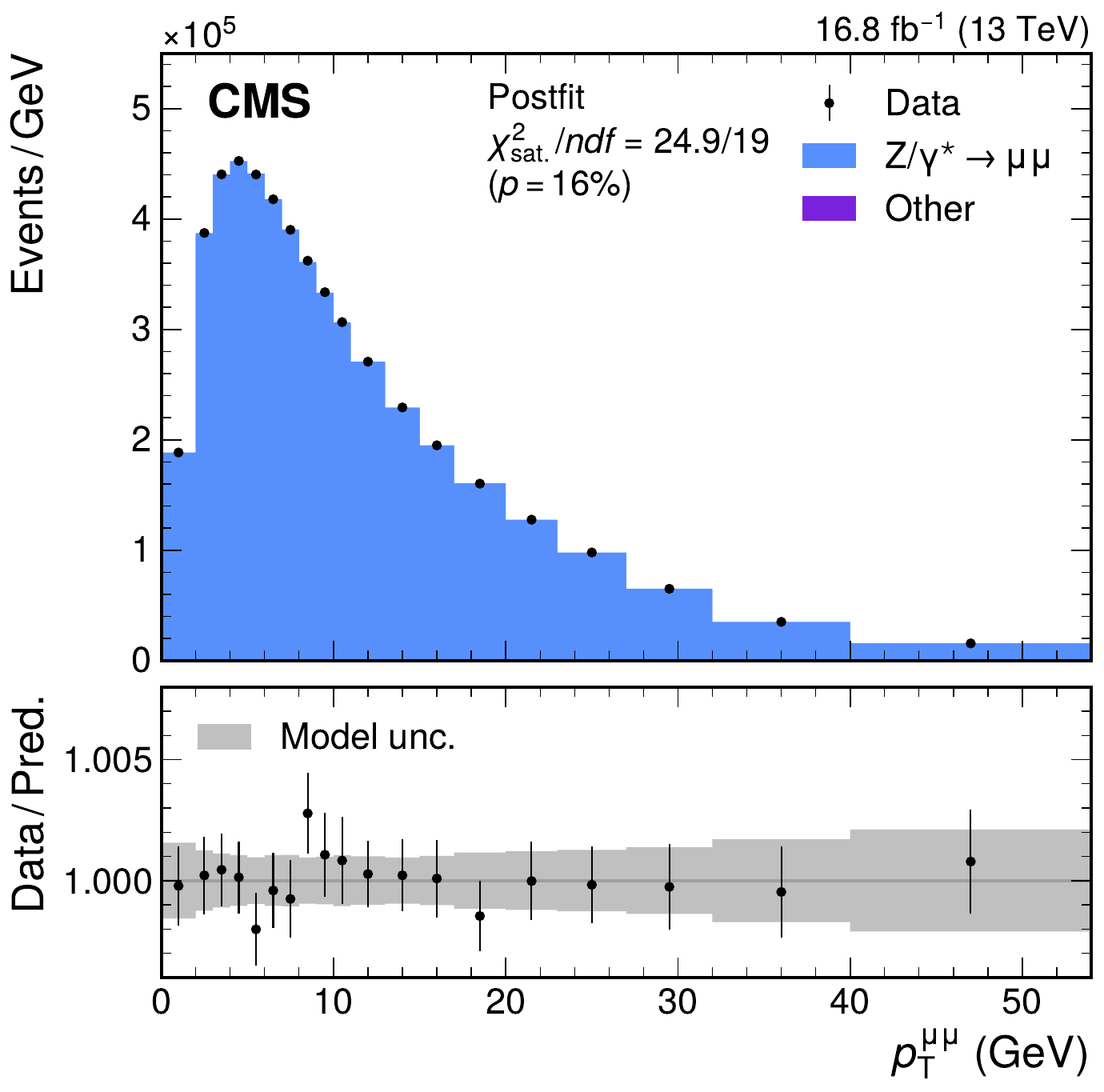}
\caption{Measured and simulated \ptmumu distributions in selected \zmm events,
with the normalization and uncertainties of the prediction set 
to the postfit values.
The gray band represents the total systematic uncertainty.
The vertical bars represent the statistical uncertainties in the data.
The bottom panel shows the ratio between the number of events observed in data, 
including variations in the predictions,
and the nominal prediction.}
\label{fig:ptll_postfit}
\end{figure}

\begin{figure}[ht]
\centering
\includegraphics[width=0.45\textwidth]{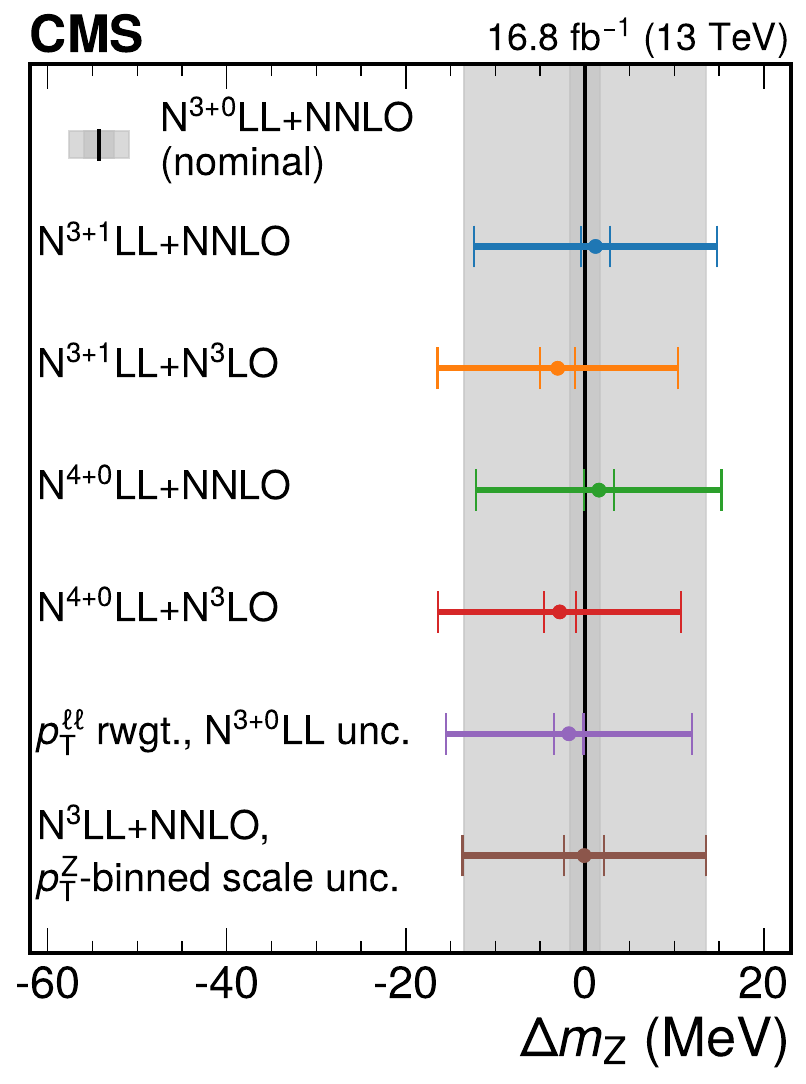}
\includegraphics[width=0.45\textwidth]{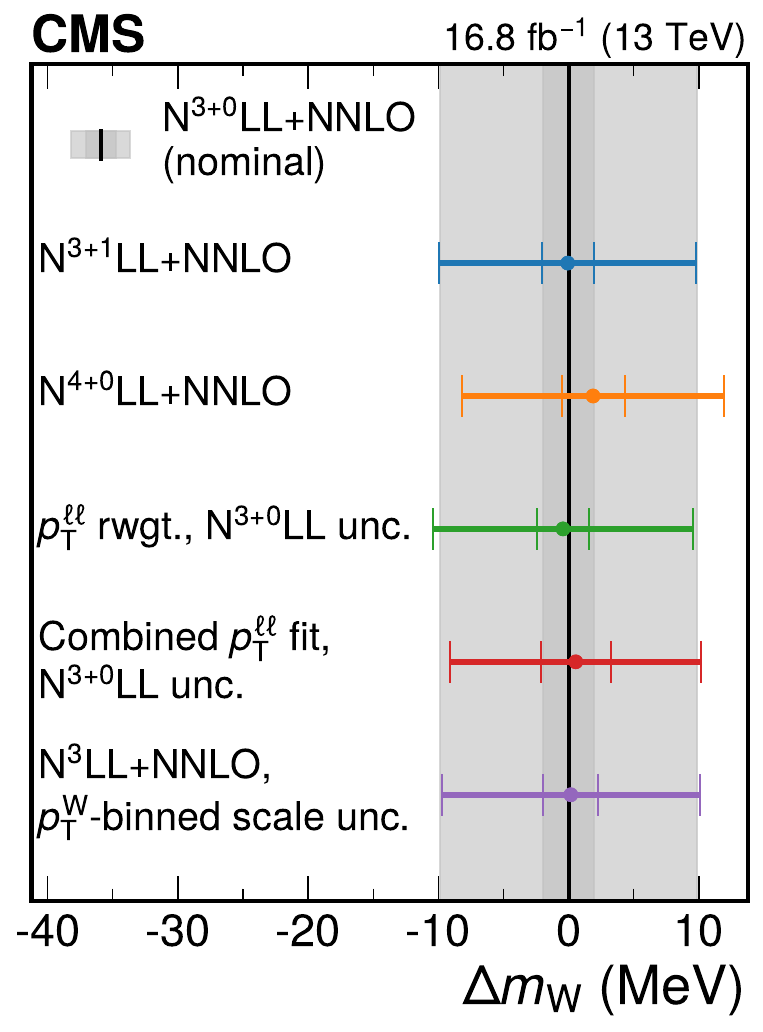}
\caption{Comparison of the nominal result and its theory uncertainty
for the \wlike \mz measurement (left) and the \mw measurement (right), 
using \scetlibdyturbo at \NtpzLLpNNLO,
with the difference in \mv measured when using alternative approaches to the \ptv modeling
and its uncertainty.
The results from alternative approaches to the \ptv modeling and uncertainty are shown as points.
The solid black line represents the nominal result,
the inner shaded gray band shows the \ptv modeling uncertainty,
and the outer shaded gray band shows the total uncertainty in the nominal result.
The \ptv modeling uncertainties are shown as the inner bars
while the outer bars denote the total uncertainties.}
\label{fig:theory_model_comparisons}
\end{figure}

\begin{figure}[!ht]
\centering
\includegraphics[width=0.45\textwidth]{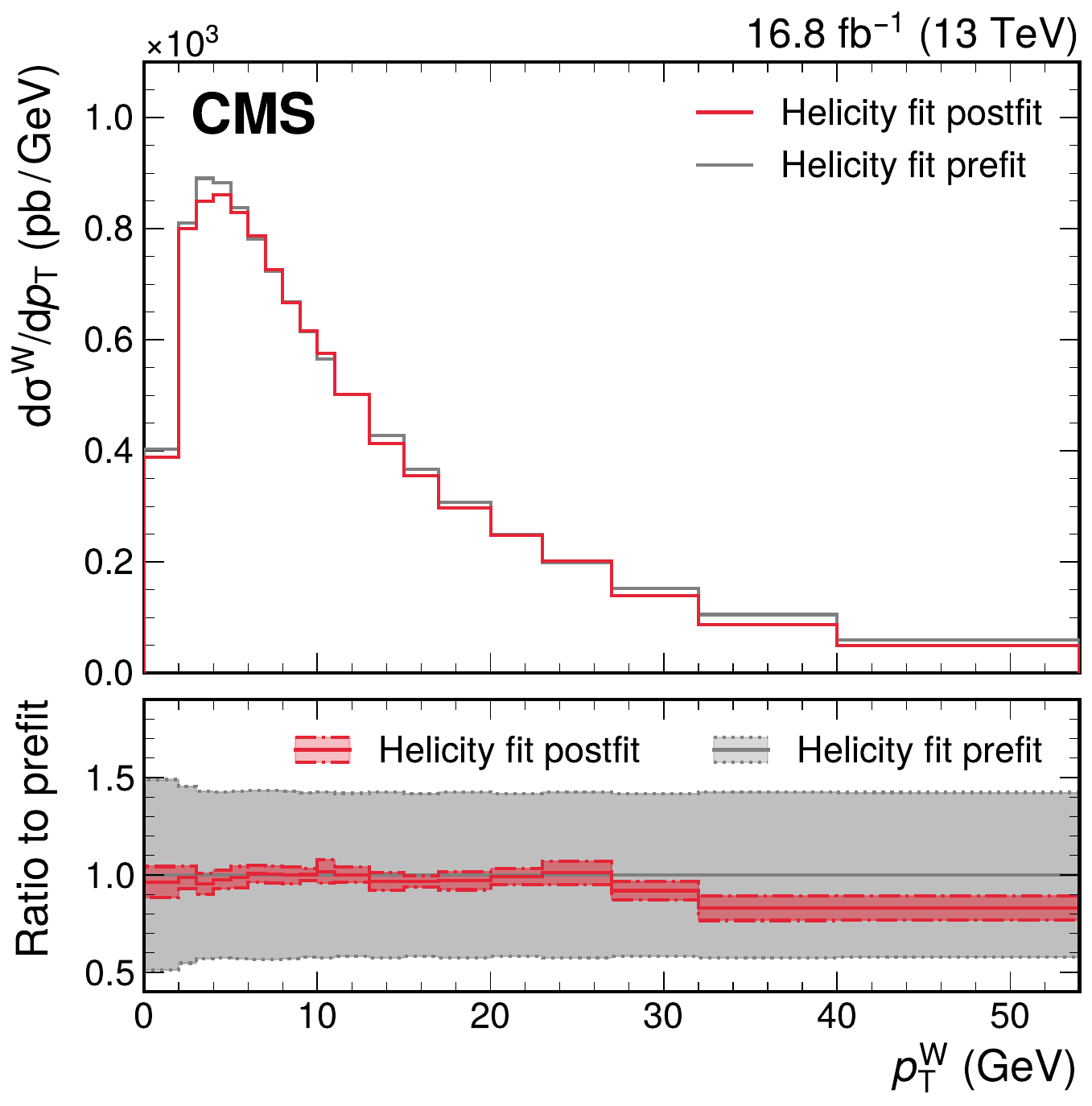}
\includegraphics[width=0.47\textwidth]{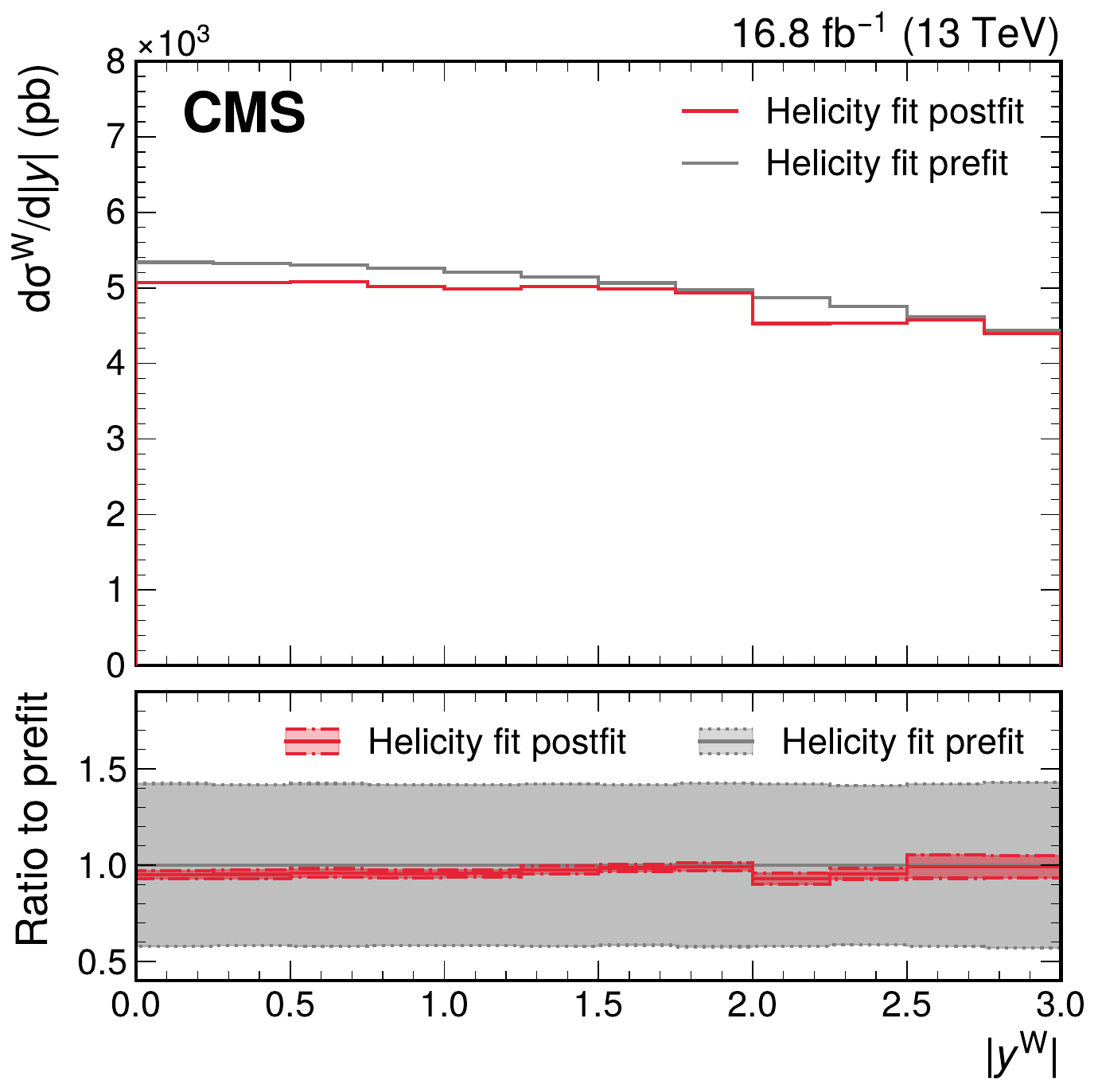}
\caption{Differential \PW boson production cross section, in the \wmn decay channel,
in \ptw (\cmsLeft) and \ayw (\cmsRight), 
measured from the \etaptqmu distributions using the helicity fit approach (in red). 
The \scetlibdyturbo generator-level predictions, 
before incorporating in situ constraints, are also shown (in gray).
The results are shown for the selection $\ayw < 3.0$ and $\ptw < 54\GeV$.
The lower panel shows the ratio between the postfit and prefit spectra.}
\label{fig:helicity_qty}
\end{figure}

\begin{figure}[t]
\centering
\includegraphics[width=0.45\textwidth]{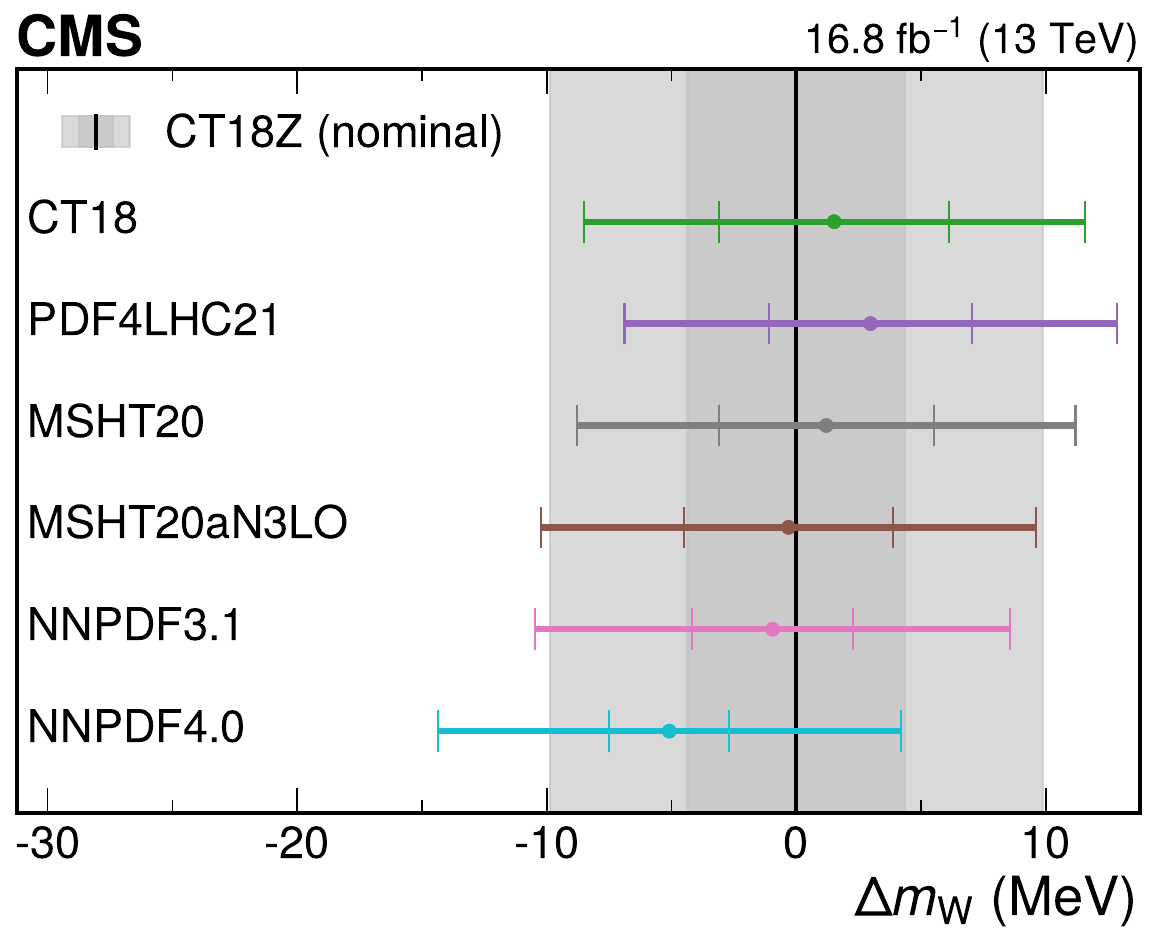}
\includegraphics[width=0.45\textwidth]{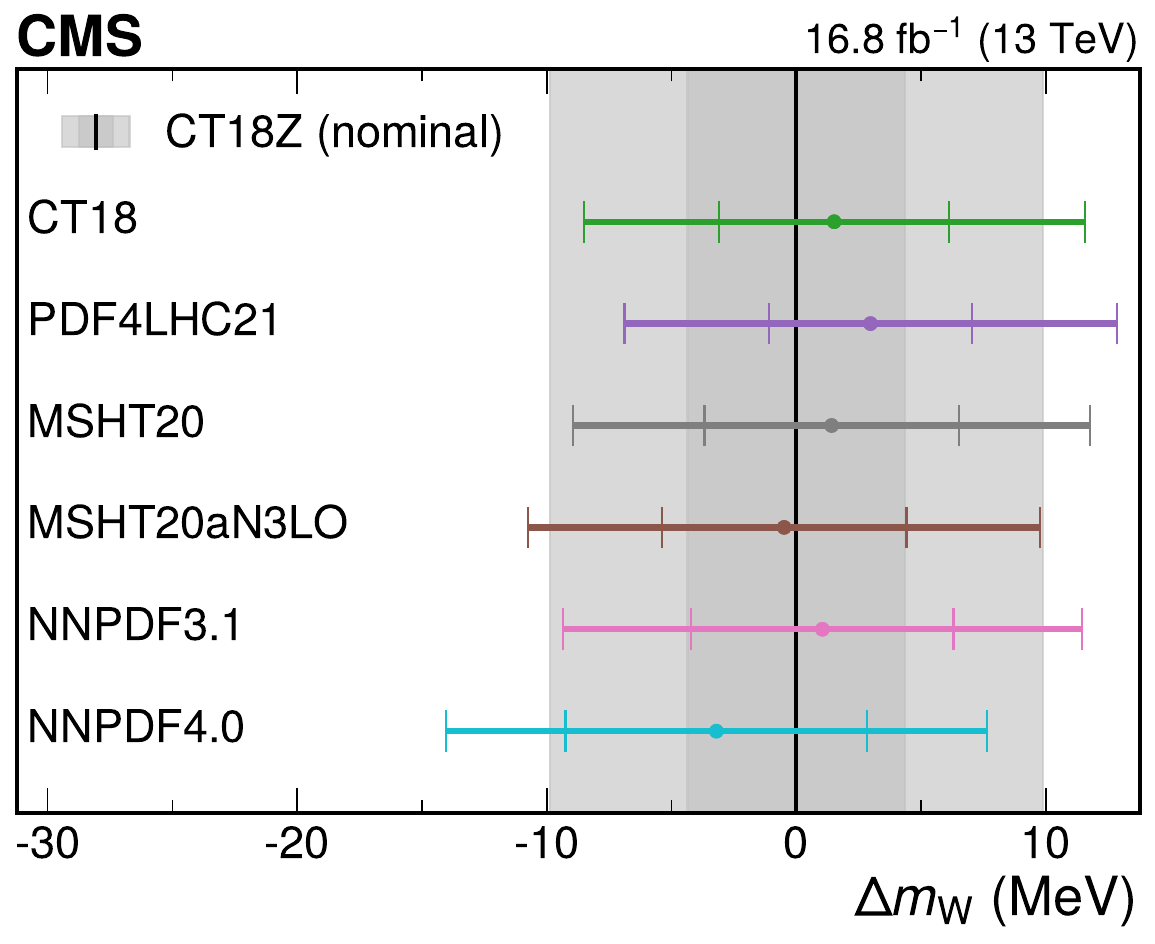}
\caption{Difference in \mw values for six alternative recent PDF sets,
when using the original uncertainty for the given set (left)
and when the uncertainties are scaled to accommodate the central prediction of the other sets (right).
Each point corresponds to the result obtained when using the indicated PDF set 
and its uncertainty for the simulated predictions. 
The inner bar shows the uncertainty from the PDF and the outer bar the total uncertainty.
The nominal result, using CT18Z, is shown by the black line, 
with the CT18Z PDF and total uncertainty shown in dark and light gray, respectively.
The uncertainty scaling procedure described in \insupp{Section of 8 the Methods section in the main work}{Section~\ref{sec:pdf}} improves the consistency of the \mw values 
across the PDF sets and with the nominal result.}
\label{fig:mw_pdf}
\end{figure}

\begin{figure*}[!htp]
\centering
\includegraphics[width=\textwidth]{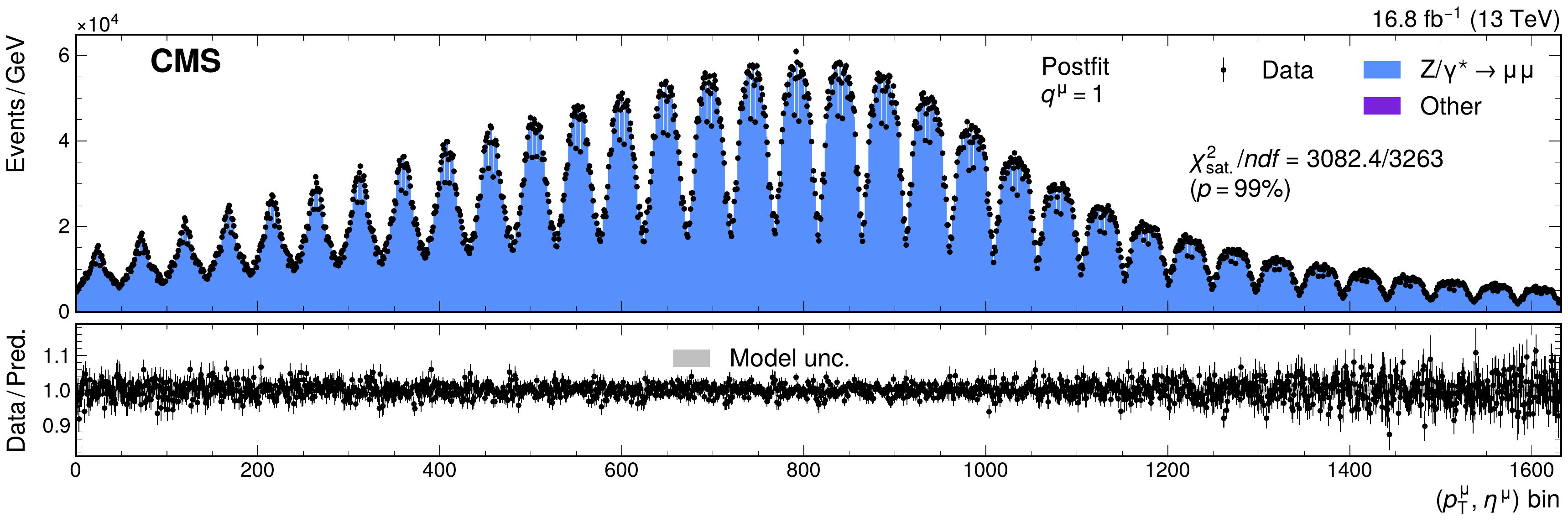}
\includegraphics[width=\textwidth]{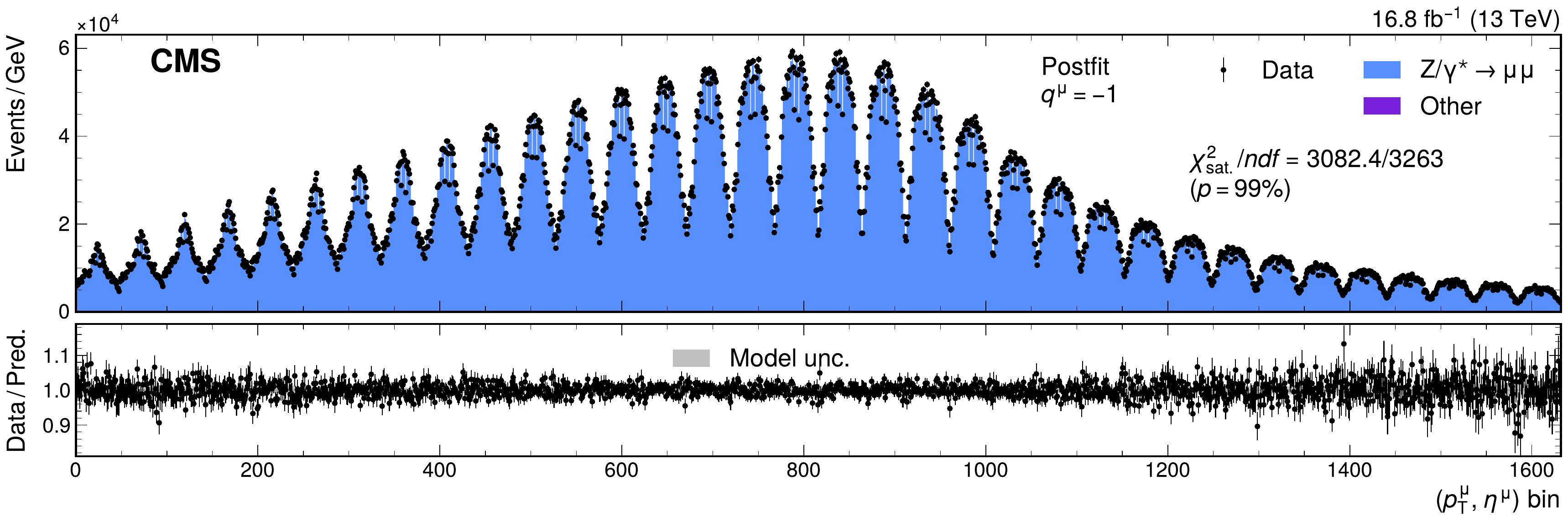}
\includegraphics[width=\textwidth]{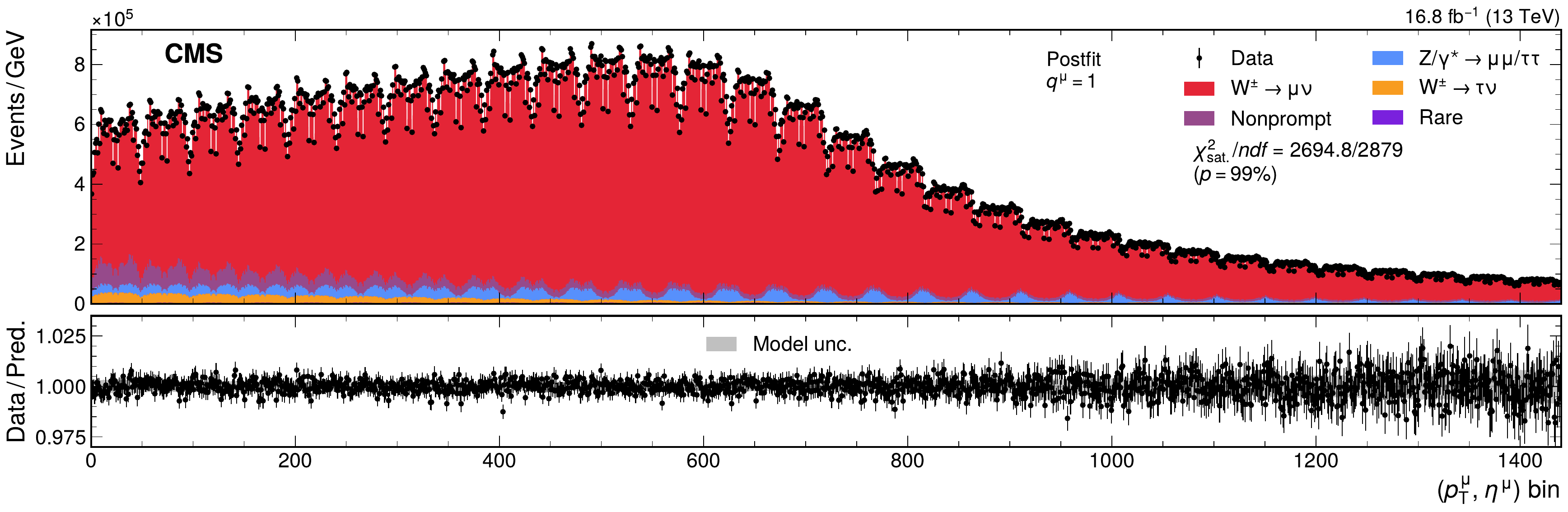}
\includegraphics[width=\textwidth]{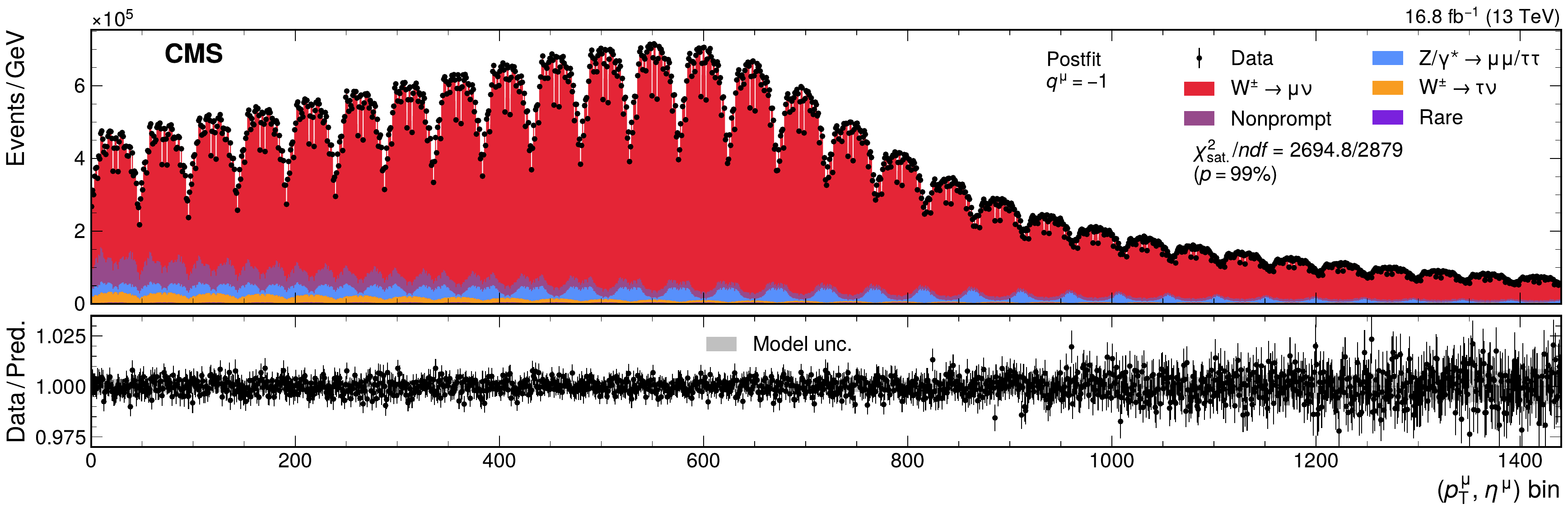}
\caption{The postfit \etaptmu distribution compared to the observed data for the \wlike \mz (upper two) and \mw (lower two) measurements
for positively (upper and second from bottom) and negatively (second from top and lower) charged muons.
The predictions and their uncertainties are adjusted to the best fit values obtained from the maximum likelihood fit.
The two-dimensional distribution is ``unrolled" such that each bin on the $x$-axis represents one \etaptmu cell. 
The gray band represents the uncertainty in the prediction, before the fit to the data.
The bottom panel shows the ratio of the number of events observed in data to the nominal prediction.
The vertical bars represent the statistical uncertainties in the data.}
\label{fig:ptetapf}
\end{figure*}

\begin{figure}[ht]
\centering
\includegraphics[width=0.48\textwidth]{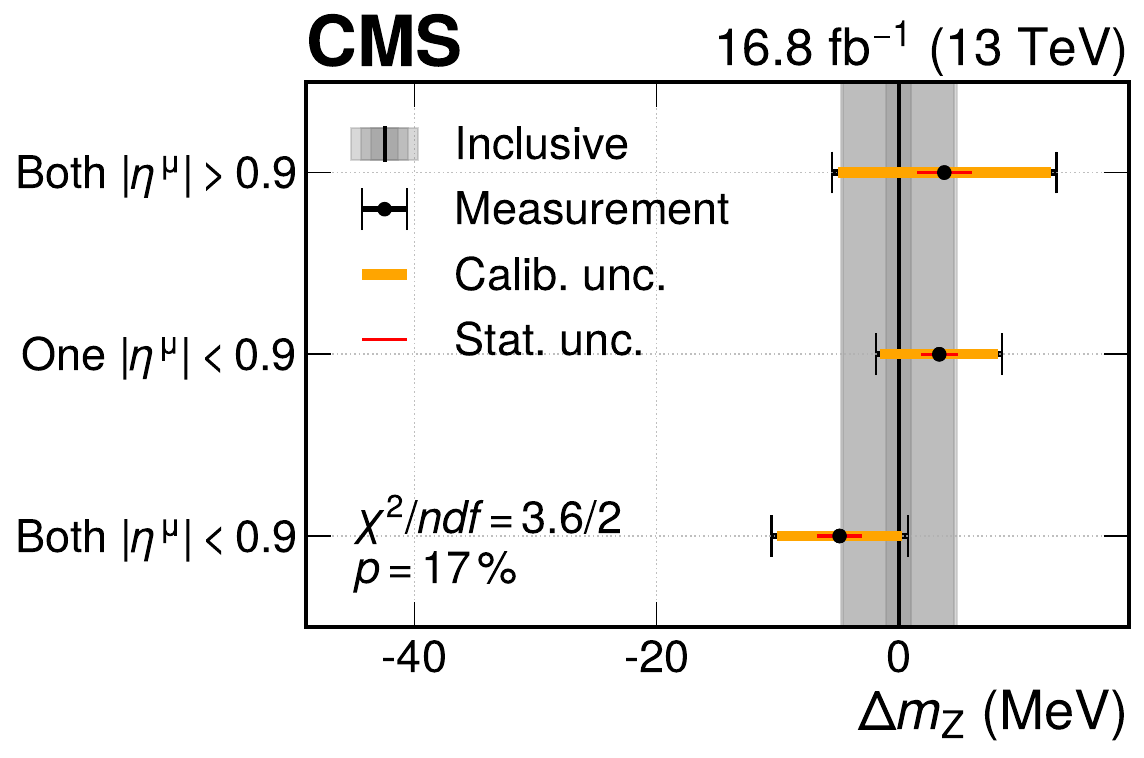}
\includegraphics[width=0.455\textwidth]{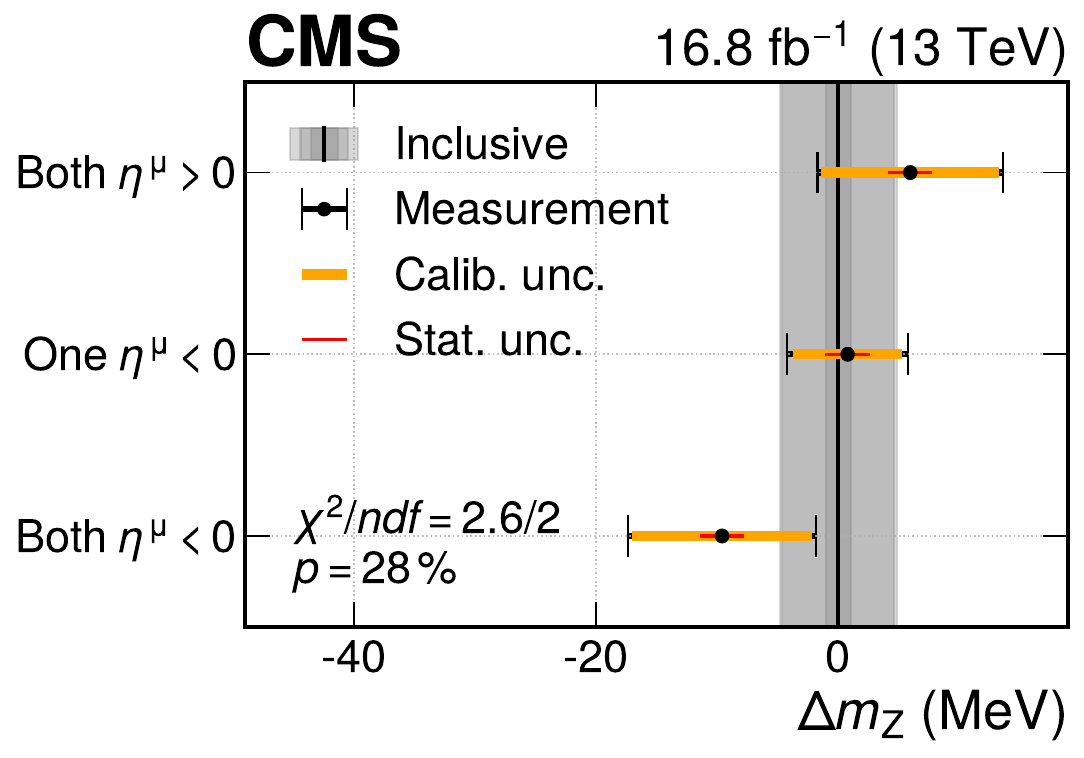}
\caption{The difference between the nominal \mz value measured from the \zmm events
and the result when \mz is allowed to vary, in three regions of 
the \etamu of the two muons. The results binned in \aetamu (both central, one central and one forward, and both forward) are shown on the left 
and results binned in \etamu (both negative, one positive and one negative, and both positive) are shown on the right.
The result of a fit with three \mz parameters is compared with the result with a single \mz parameter
and the compatibility of the results is also shown, 
as assessed via the saturated goodness-of-fit test.
The points show the \mz result for the indicated \etamu region and the horizontal bars represent the calibration (orange line), statistical (red line), and total (black line) uncertainties.
The black vertical line represents the result with a single \mz parameter, with the three shaded gray bands representing the statistical (dark grey), calibration (intermediate grey), and total (light grey) uncertainties.}
\label{fig:decorrMassShiftRegions}
\end{figure}

\begin{figure}[ht]
\centering
\includegraphics[width=0.48\textwidth]{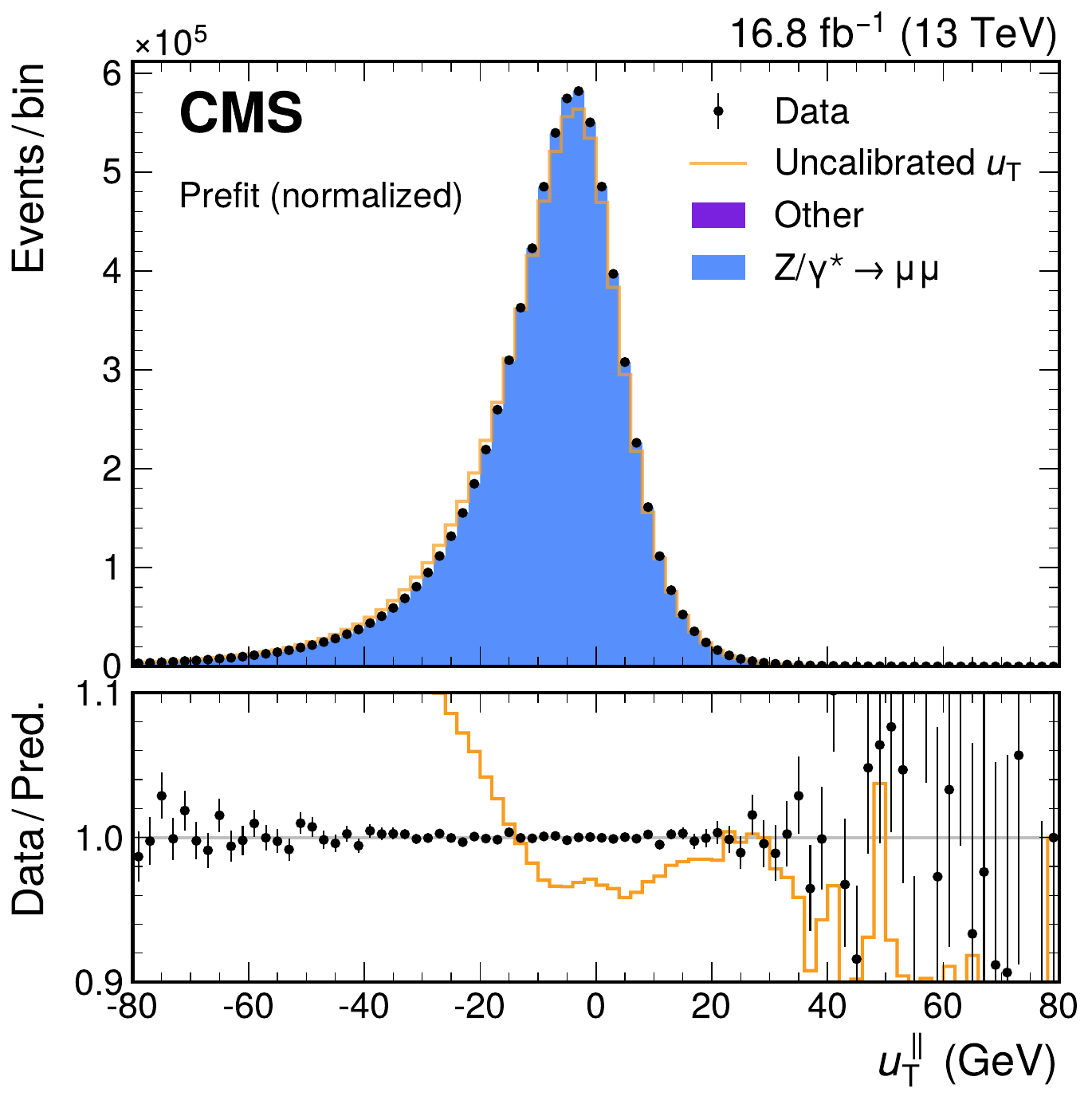}
\includegraphics[width=0.48\textwidth]{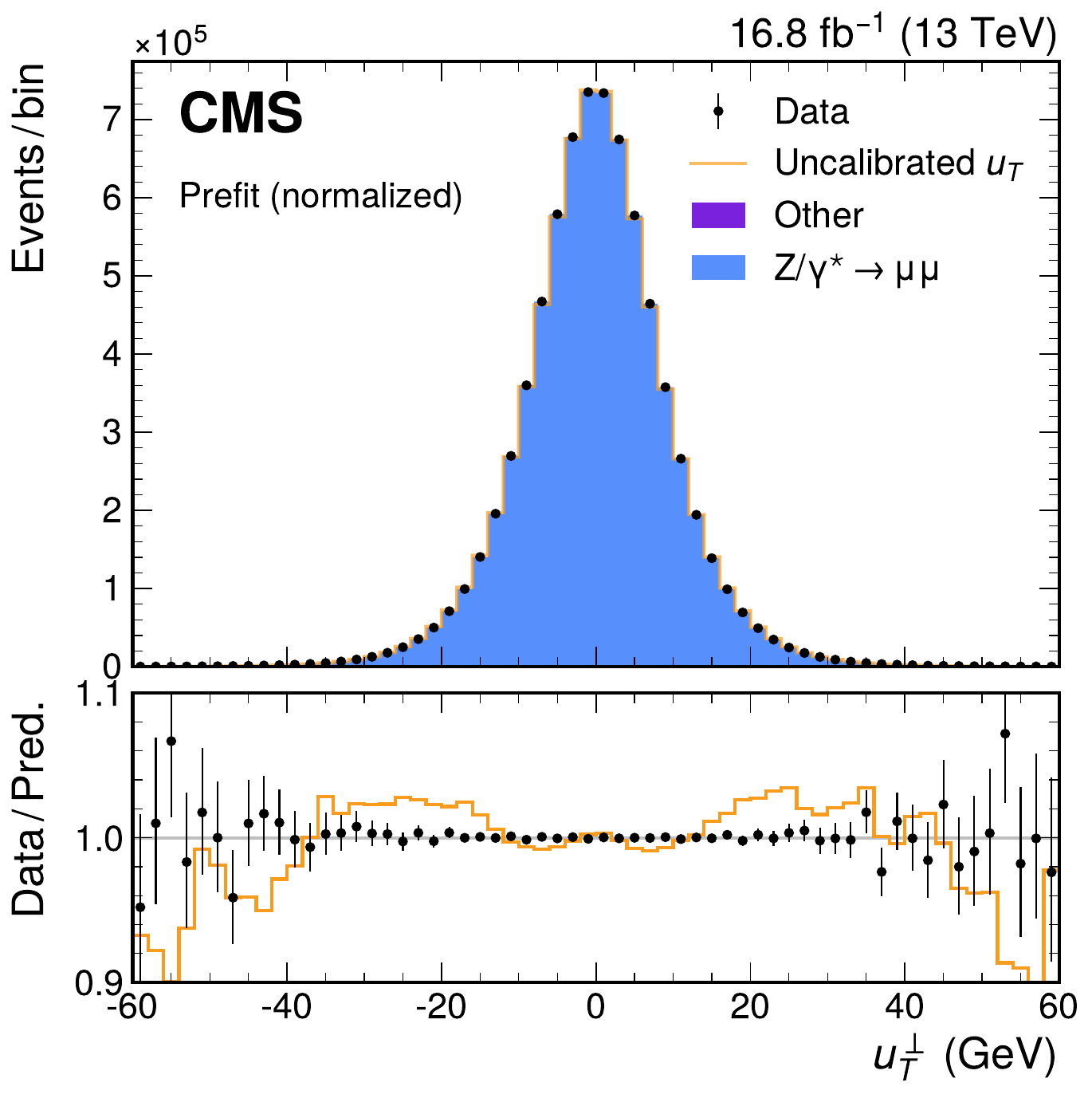}
    \caption{Comparison of the prediction and observed data for the parallel ($\ut^{\parallel}$, left) and perpendicular ($\ut^{\perp}$, right)
    components of the hadronic recoil. The filled histograms show the simulation with the hadronic recoil corrected 
    according to the procedure described in the text. The orange line shows the predicted distribution before the hadronic recoil corrections.
    The uncertainties in the predictions are not shown.
    The bottom panel shows the ratio of the number of events observed in data (black point) and the 
    uncorrected prediction (orange line) to the recoil-corrected prediction.
}
\label{fig:recoil_calib}
\end{figure}

\begin{figure}[hbtp]
\centering
\includegraphics[width=0.49\textwidth]{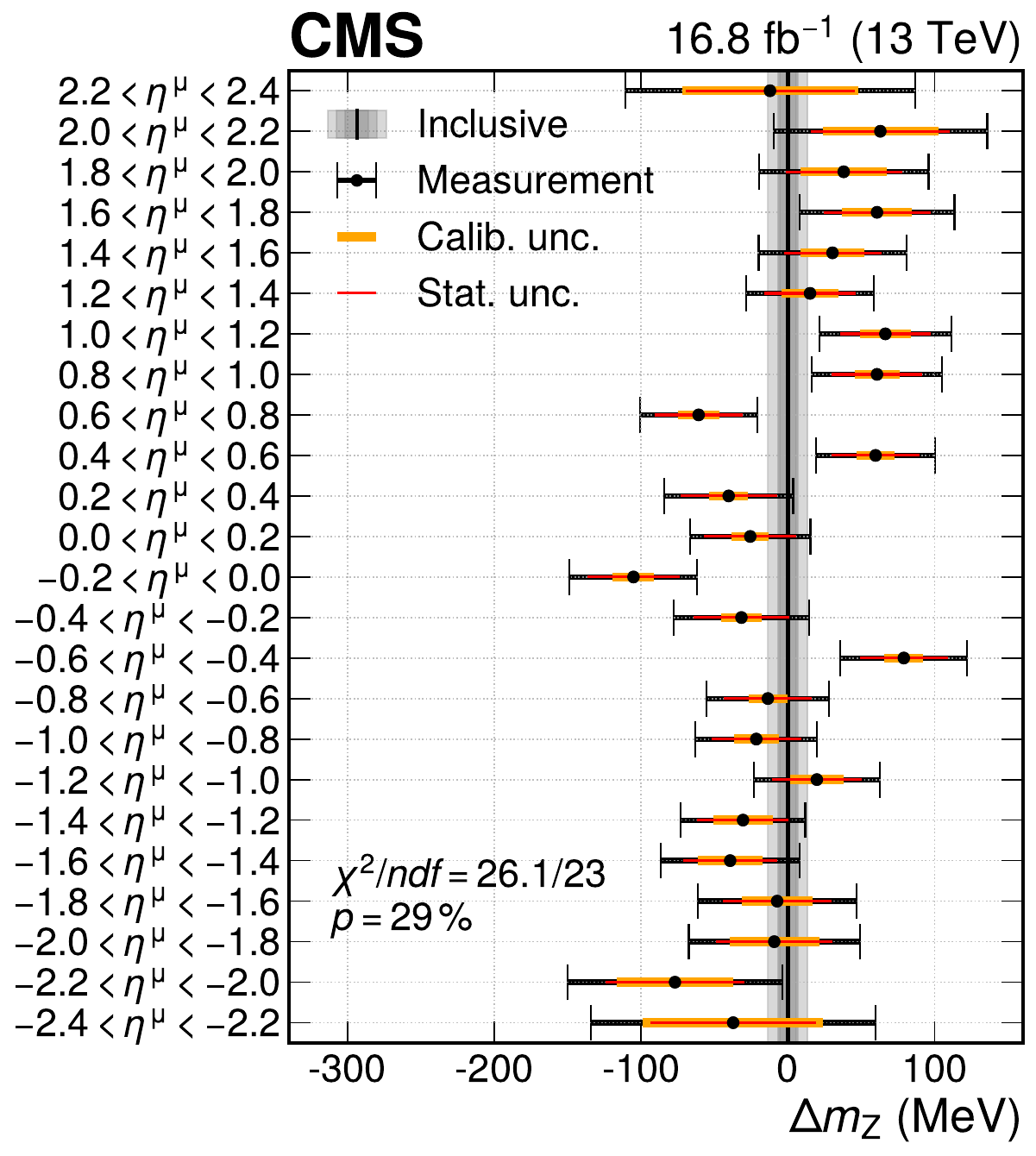}
\includegraphics[width=0.49\textwidth]{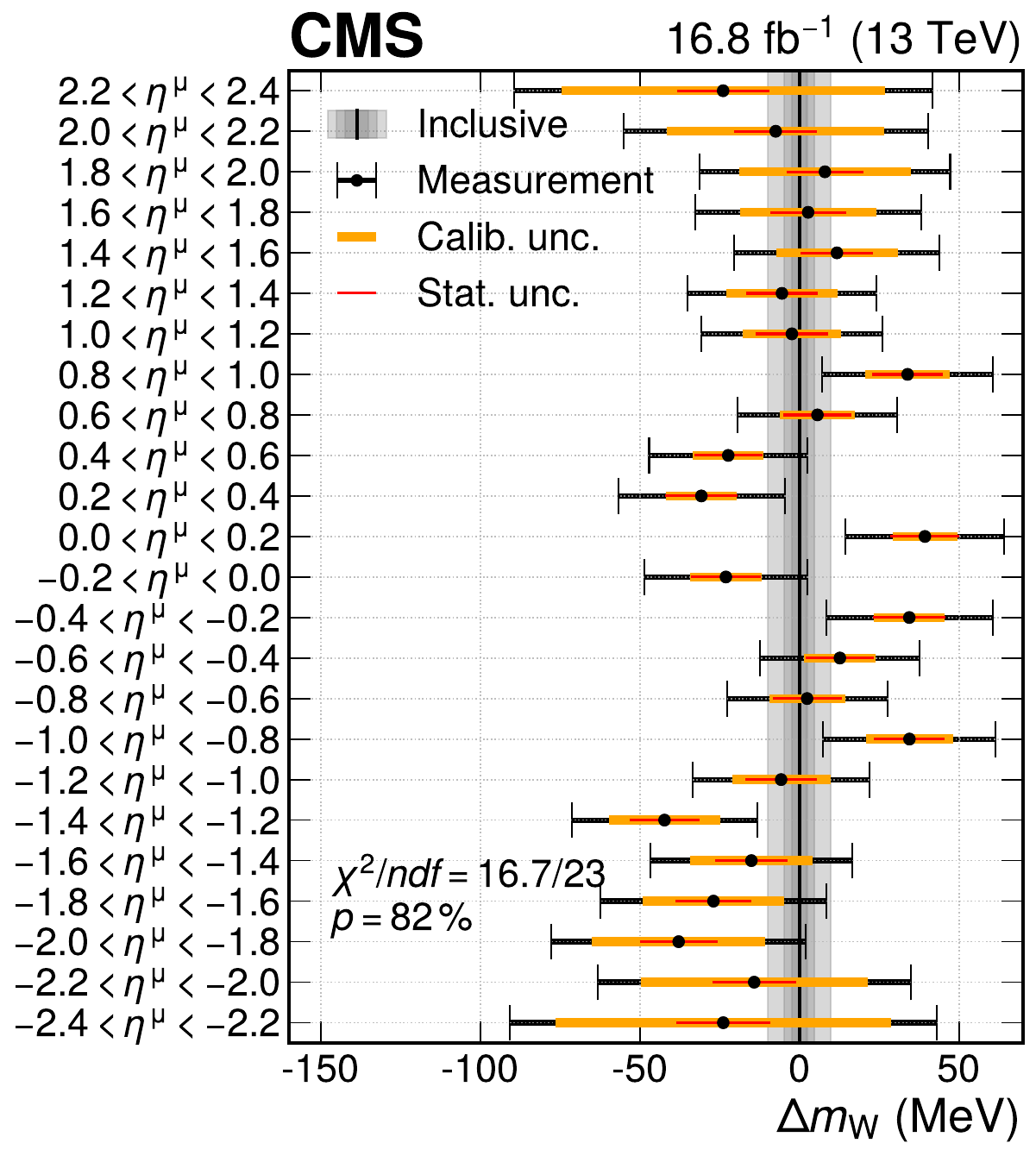}
\caption{For the \wlike \mz analysis (left) and the \mw measurement (right) the result of a fit with 24 \mv parameters corresponding to different \etamu ranges is compared with the nominal \mv fit result.
The $\chi^2$-like compatibility of the two fits is also shown, 
assessed via the saturated goodness-of-fit test.
The points show \mv result for the indicated \etamu region,
and the horizontal bars represent the calibration (orange line), statistical (red line), and total (black line) uncertainties.
The black vertical line shows the result with a single \mv parameter, with the shaded gray bands representing its statistical, calibration, and total uncertainties.}
\label{fig:decorrMassShiftWZ}
\end{figure}

\begin{figure}[hbtp]
\centering
\includegraphics[width=0.6\textwidth]{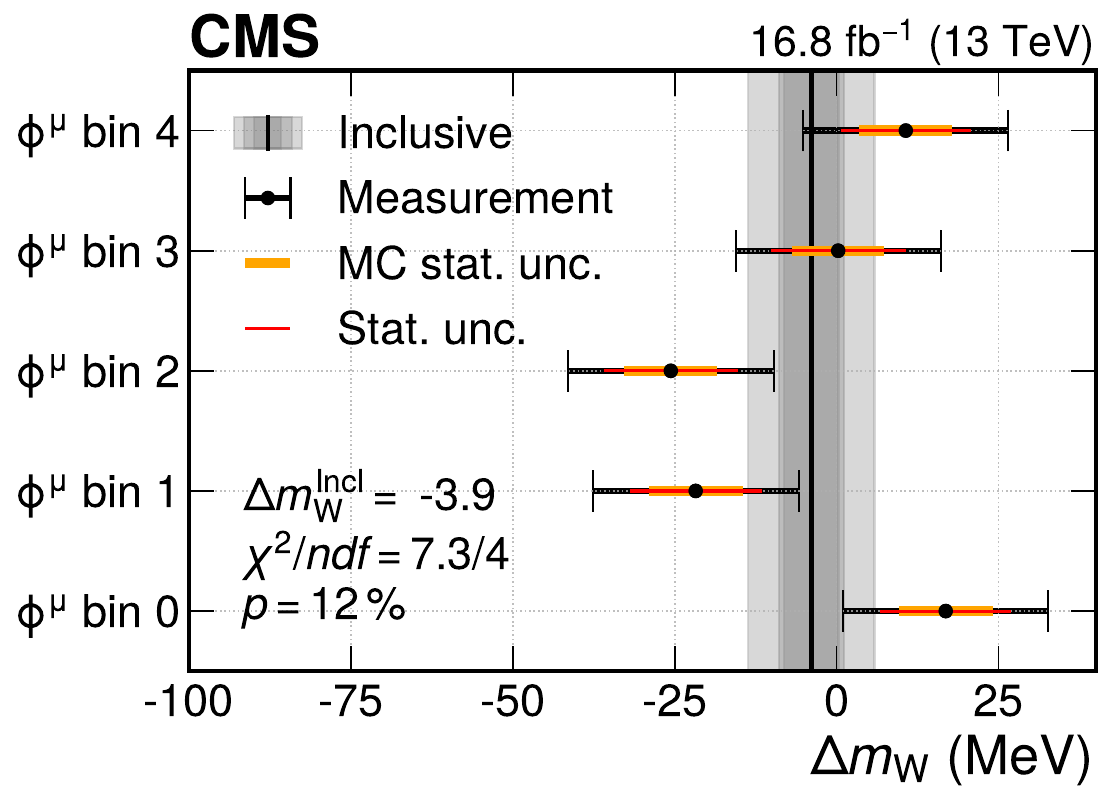}
\caption{Measured value of \mw after splitting the analyzed data and simulated samples in five uniformly spaced bins of muon $\phi^{\PGm}$ from $-\pi$ to $\pi$. 
The points show the \mw measurement for the indicated bin,
and the horizontal bars represent the MC (orange line), data statistical (red line), and total (black line) uncertainties.
The $\chi^2$-like compatibility of the measurements is also shown, 
assessed via the saturated goodness-of-fit test.
The mutual correlation of the five measurements is accounted for in the $\chi^2$, and is about 30\% accounting for the common theoretical uncertainties. 
Most of the experimental uncertainties are uncorrelated across the five bins.
The black vertical line shows the combined result from a simultaneous fit of the five bins with a single \mw parameter, with the shaded gray bands representing its data or MC statistical uncertainty and the total uncertainty. The zero of the horizontal axis corresponds to the nominal measured value. The partial uncertainties are defined using the ``global'' impacts.}
\label{fig:decorr_massShiftW_phiBins}
\end{figure}

\begin{figure}[hbtp]
\centering
\includegraphics[width=0.6\textwidth]{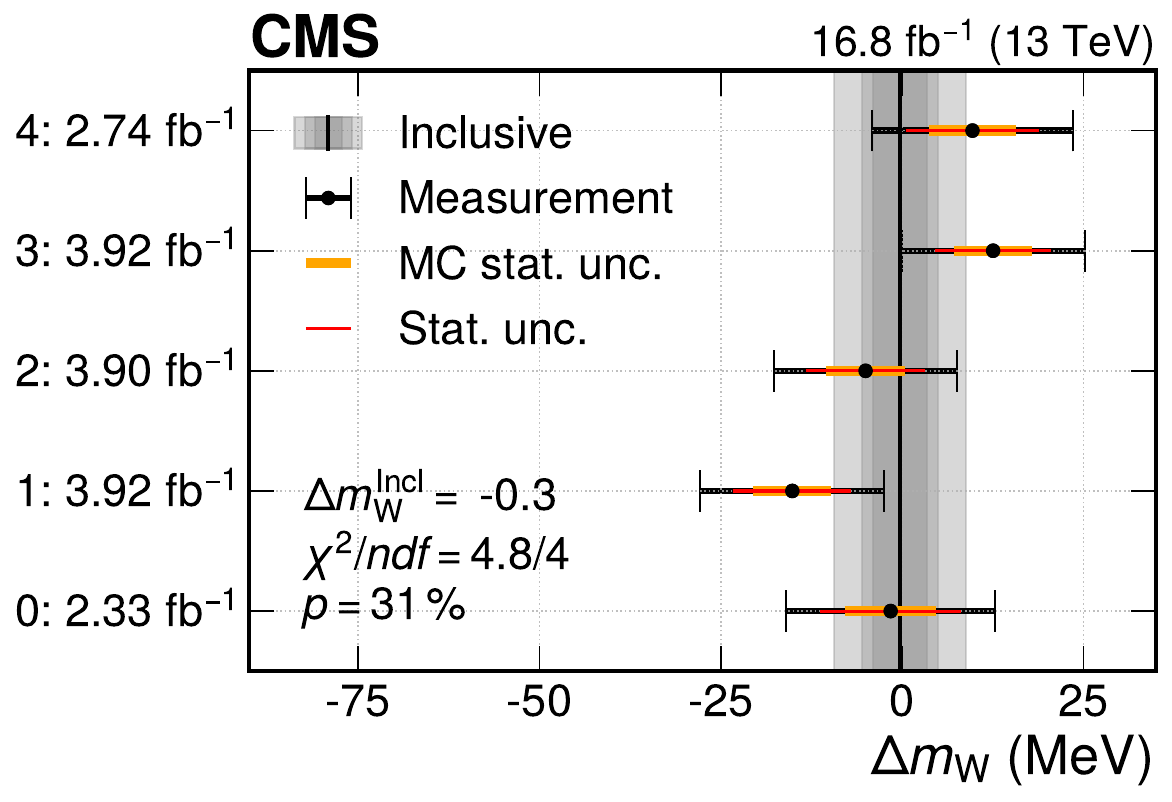}
\caption{Measured value of \mw after splitting the analyzed data and simulated samples in five independent subsets. 
The points show the \mw measurement for the indicated integrated luminosity,
and the horizontal bars represent the MC (orange line), data statistical (red line), and total (black line) uncertainties.
The integrated luminosity of each bin follows the discrete pattern of the data-taking runs. The five bins gather data collected from the beginning to the end of the data taking from bottom to top. 
Since the average pileup increased with time during 2016, this splitting approximately corresponds to a categorization in bins of pileup as well.
The $\chi^2$-like compatibility of the measurements is also shown, 
assessed via the saturated goodness-of-fit test.
The mutual correlation of the five measurements is accounted for in the $\chi^2$, and is about 30\% accounting for the common theoretical uncertainties. 
Most of the experimental uncertainties are treated as uncorrelated across the five bins.
The black vertical line shows the combined result from a simultaneous fit of the five bins with a single \mw parameter, with the shaded gray bands representing its data or MC statistical uncertainty and the total uncertainty. The zero of the horizontal axis corresponds to the nominal measured value. The partial uncertainties are defined using the ``global'' impacts.}
\label{fig:decorr_massShiftW_runBins}
\end{figure}

\begin{figure}[hbtp]
\centering
\includegraphics[width=0.6\textwidth]{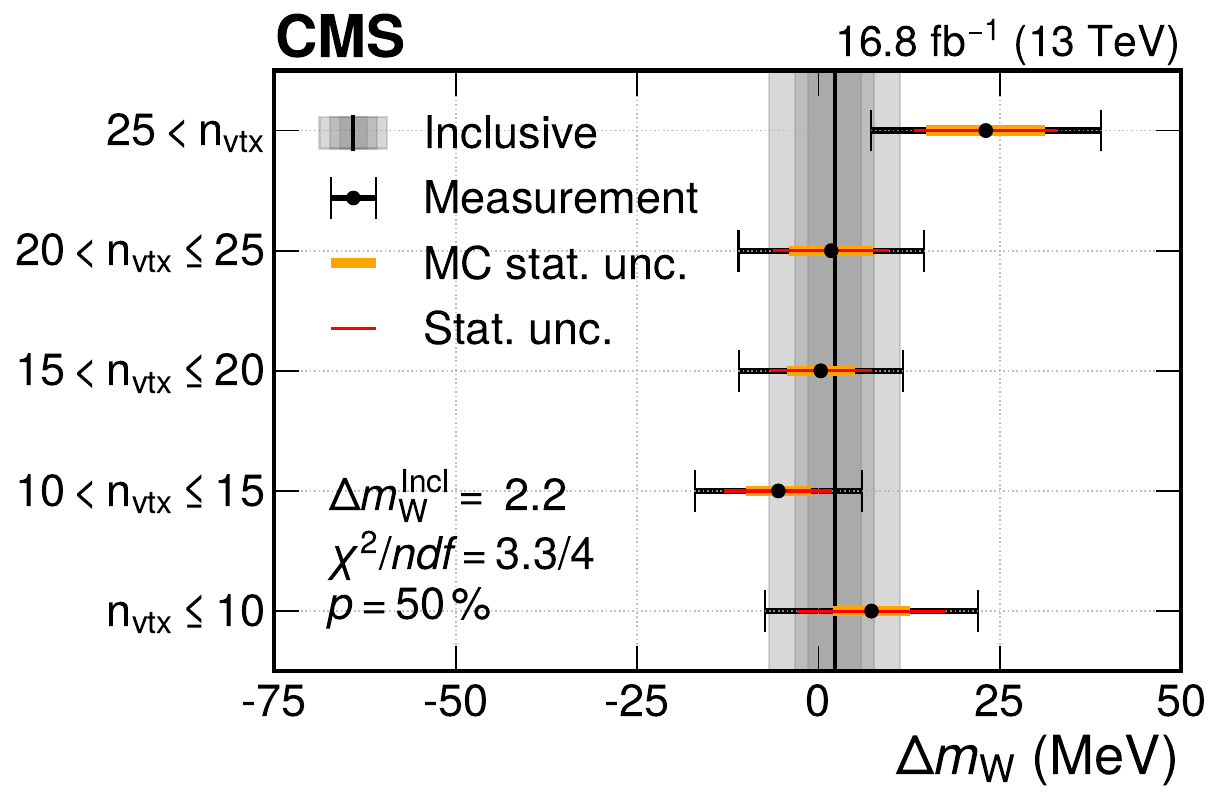}
\caption{Measured value of \mw after splitting the analyzed data and simulated samples in five independent subsets based on the number of reconstructed vertices (\nvtx). 
The points show the \mw measurement for the indicated \nvtx region,
and the horizontal bars represent the MC (orange line), data statistical (red line), and total (black line) uncertainties.
The $\chi^2$-like compatibility of the measurements is also shown, 
assessed via the saturated goodness-of-fit test.
The mutual correlation of the five measurements is accounted for in the $\chi^2$, and is about 30\%, accounting for the common theoretical uncertainties.
The black vertical line shows the combined result from a simultaneous fit of the five bins with a single \mw parameter, with the shaded gray bands representing its data or MC statistical uncertainty and the total uncertainty. The zero of the horizontal axis corresponds to the nominal measured value, $\mw = 80\,360.2$\MeV. The partial uncertainties are defined using the ``global'' impacts.}
\label{fig:decorr_massShiftW_nRecoVtxBins}
\end{figure}

\begin{figure}[hbtp]
\centering
\includegraphics[width=0.6\textwidth]{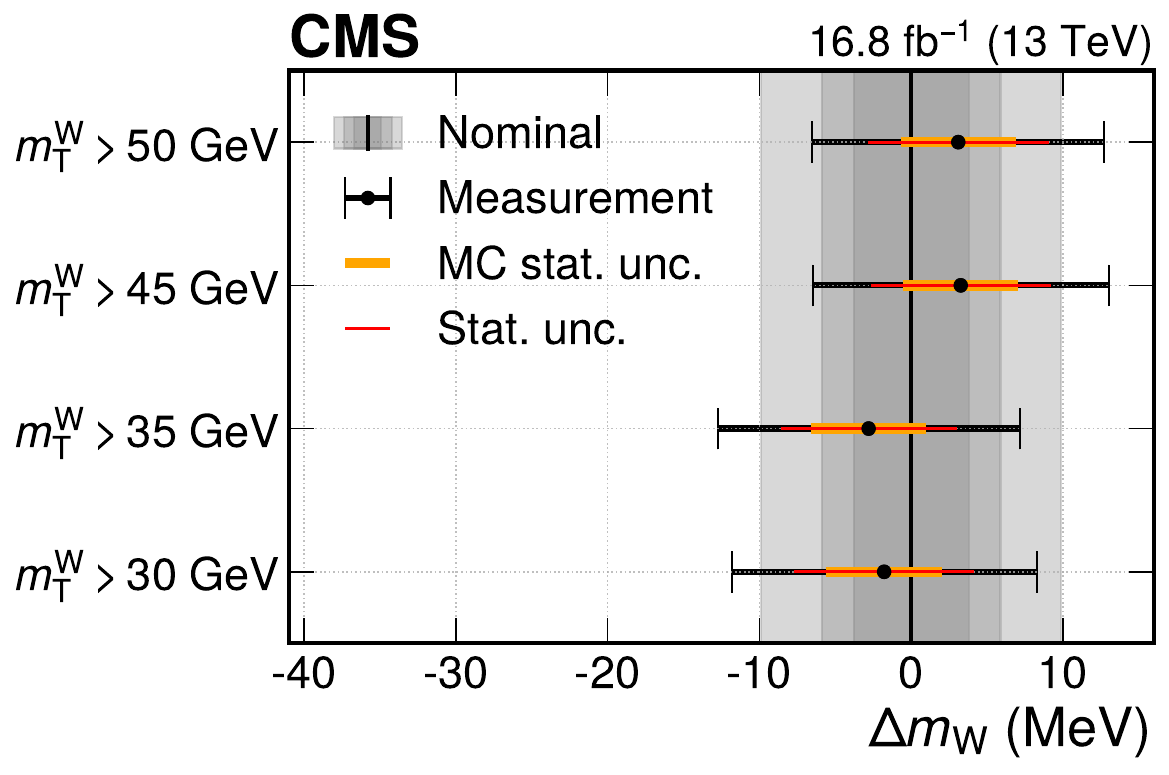}
\caption{Measured value of \mw after modifying the threshold in the transverse mass \mt.
The points show the \mw measurement for the indicated threshold,
and the horizontal bars represent the MC (orange line), data statistical (red line), and total (black line) uncertainties.
The partial uncertainties are defined using the ``global'' impacts.
The black vertical line shows the nominal measured value, for which the \mt threshold is 40\GeV. 
The measurements are not statistically independent.
}
\label{fig:massShiftW_varyMTcut}
\end{figure}

\begin{figure}[hbtp]
\centering
\includegraphics[width=0.45\textwidth]{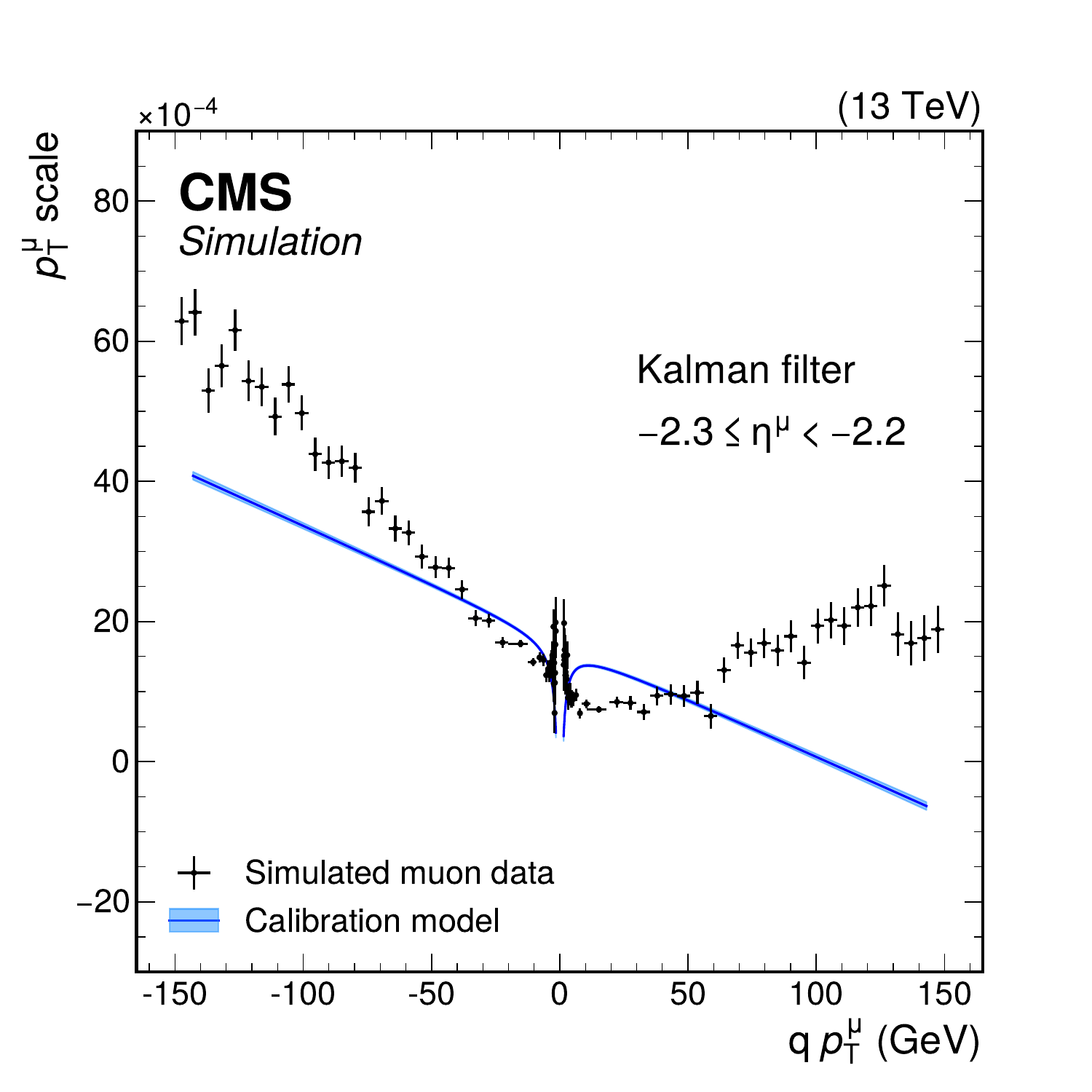}\quad
\includegraphics[width=0.45\textwidth]{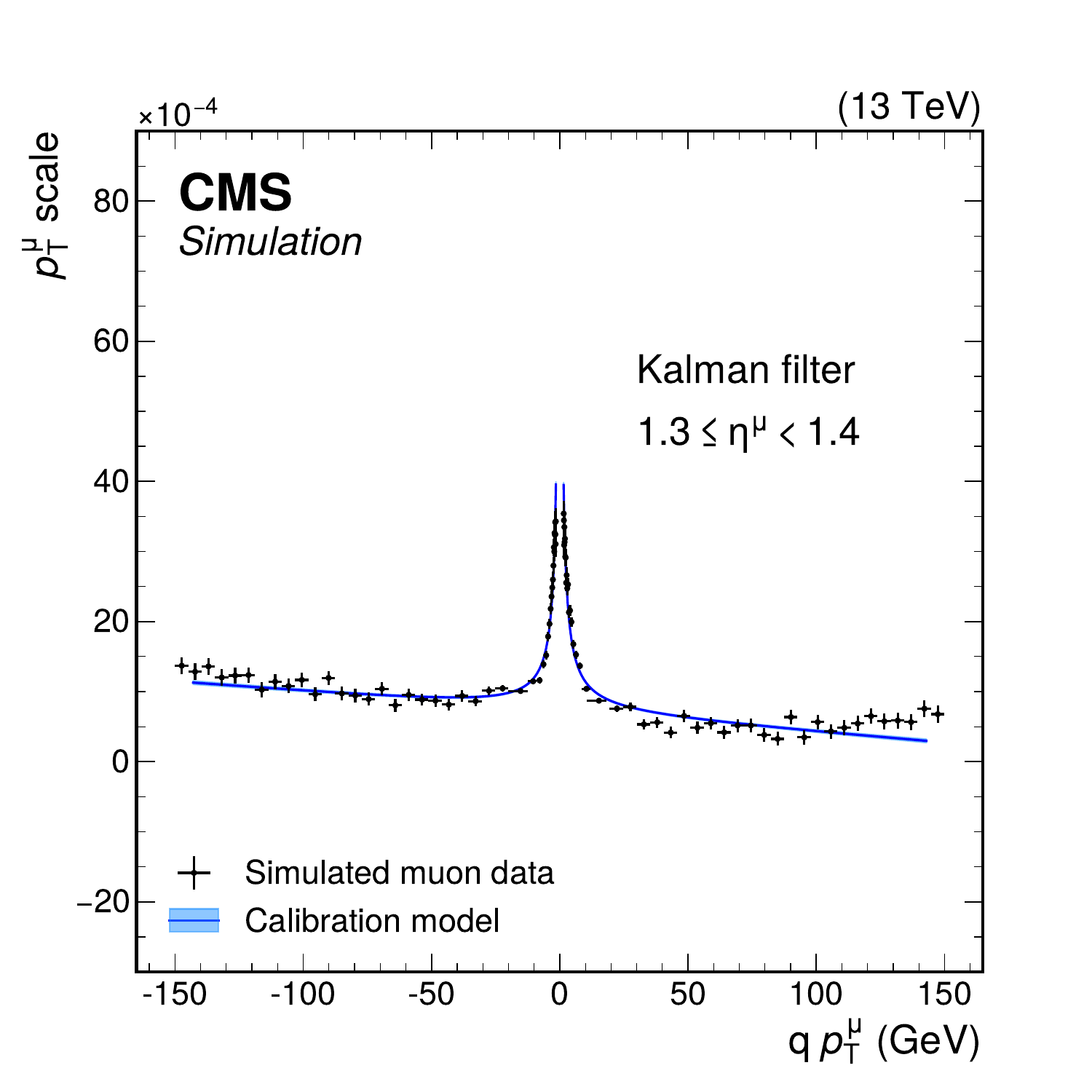}\\
\includegraphics[width=0.45\textwidth]{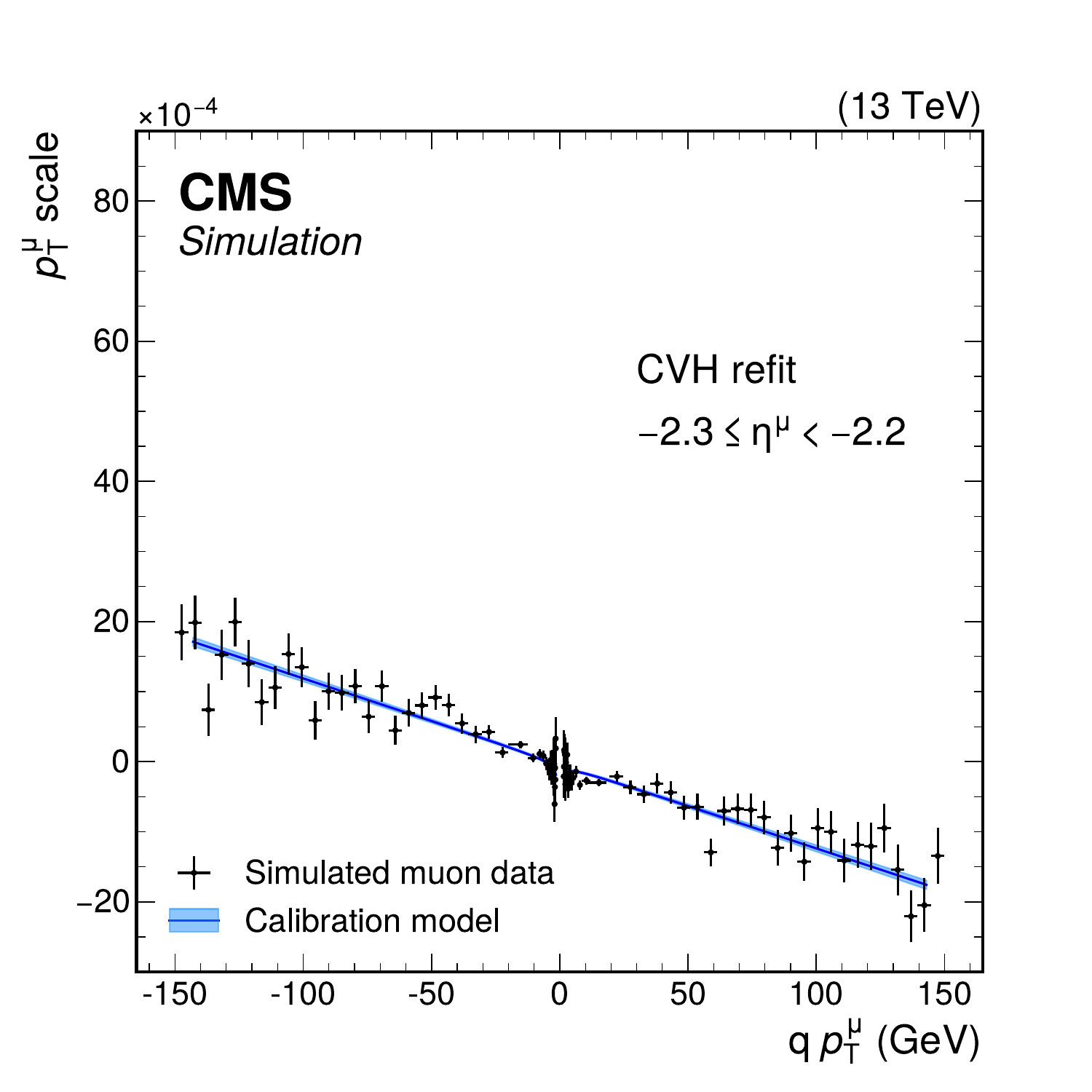}\quad
\includegraphics[width=0.45\textwidth]{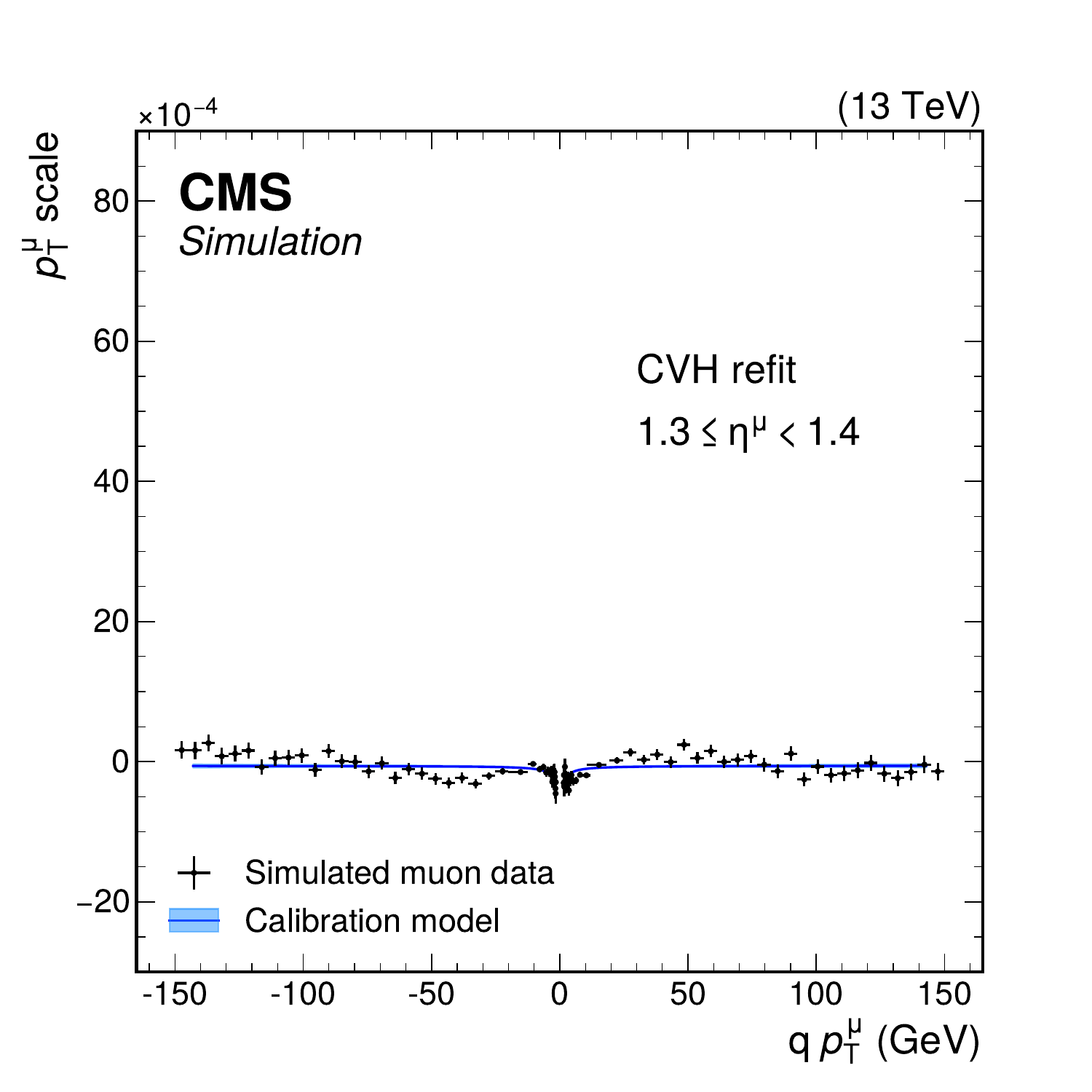}\\
\includegraphics[width=0.45\textwidth]{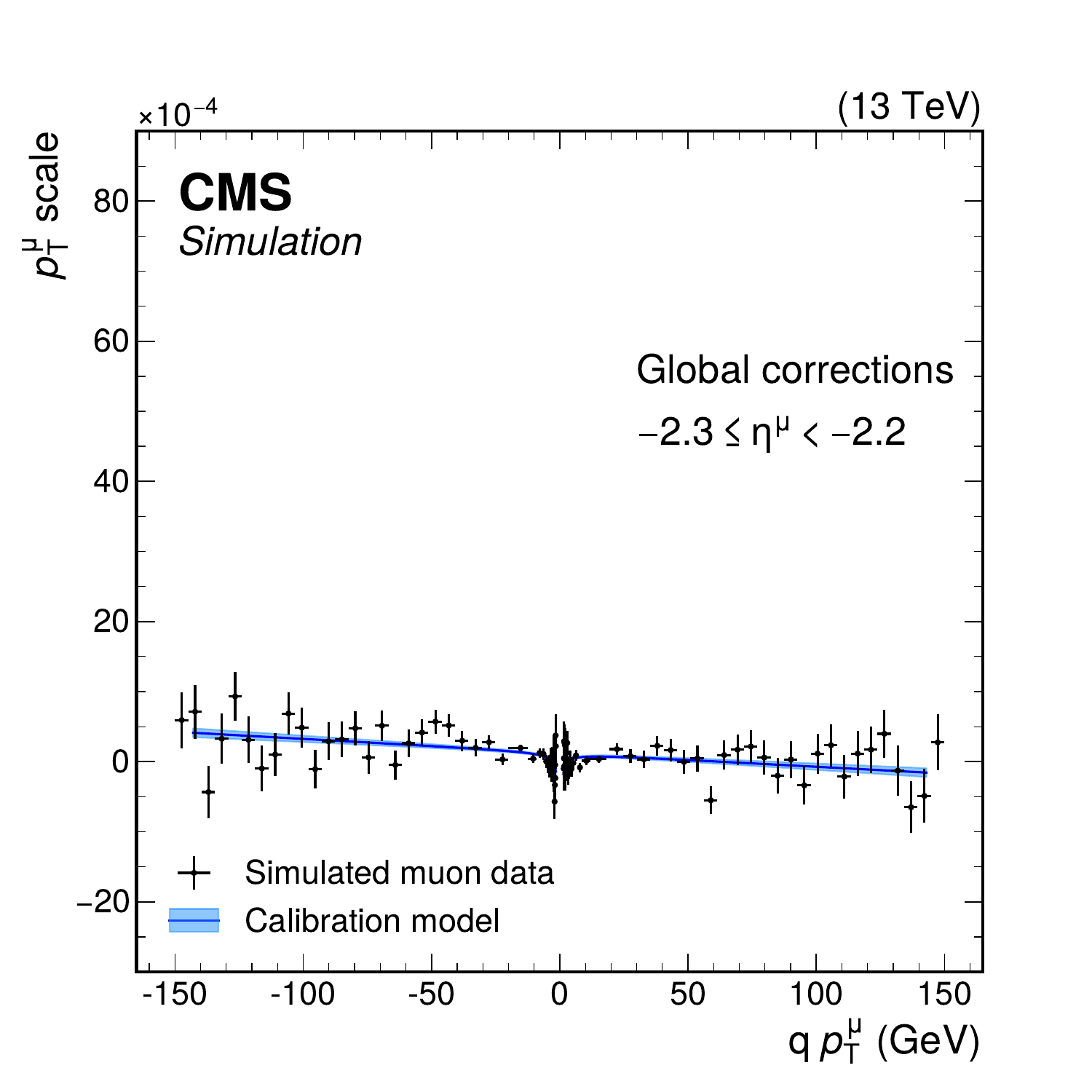}\quad
\includegraphics[width=0.45\textwidth]{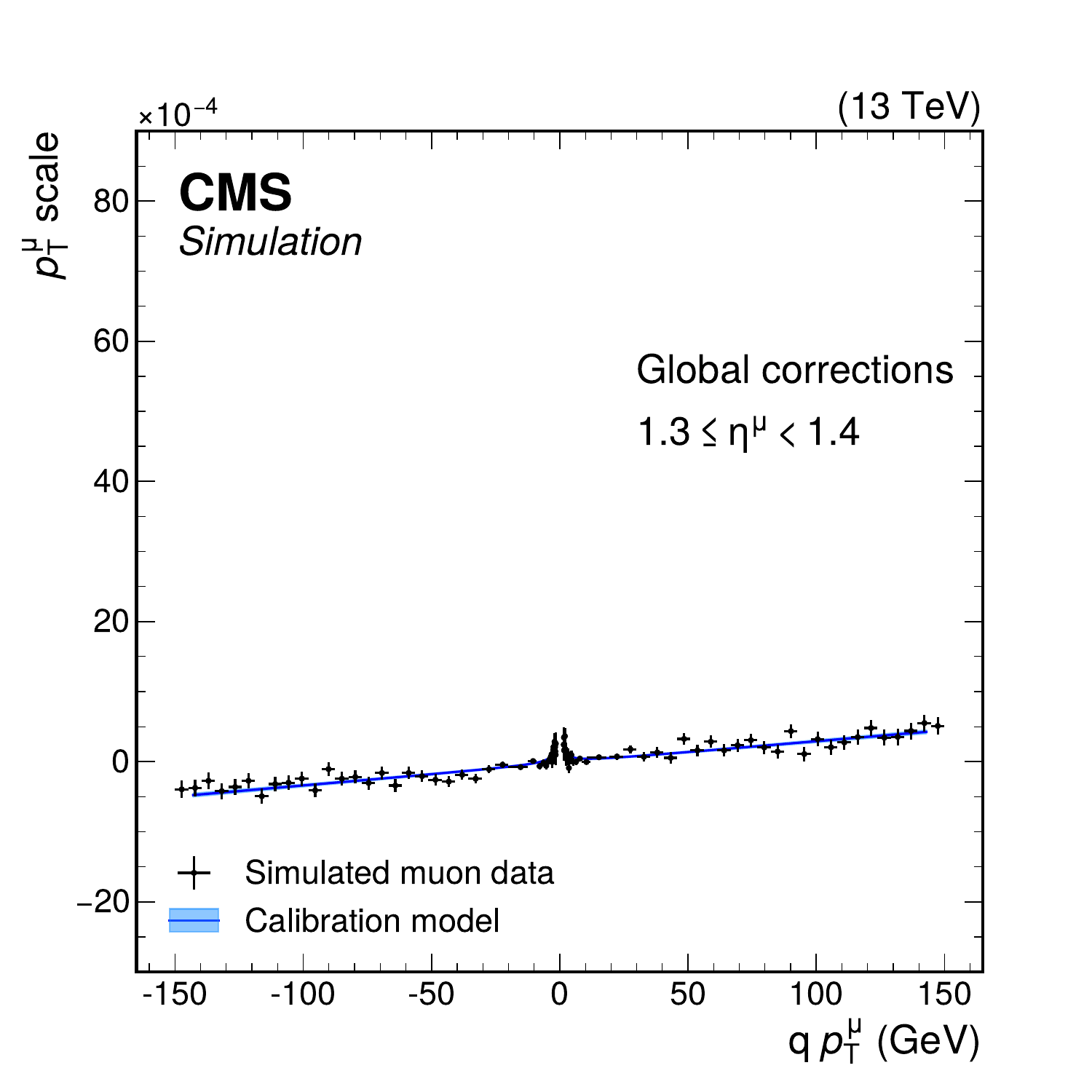}\\
\caption{Muon momentum scale bias evaluated in simulated events as a function of \ptmu times the muon charge $q$. 
The black dots represent simulated data, while the solid line is a fit of the calibration model. 
The bias is shown after the Kalman filter track fit (top), the CVH refit (middle), and the generalized global corrections applied on top of the CVH refit (bottom). 
The comparison is performed in two \etamu bins in the forward (left) and central (right) regions of the tracker.}
\label{fig:calib_model_scalept}
\end{figure}

\begin{figure}[hbtp]
\centering
\includegraphics[width=0.45\textwidth]{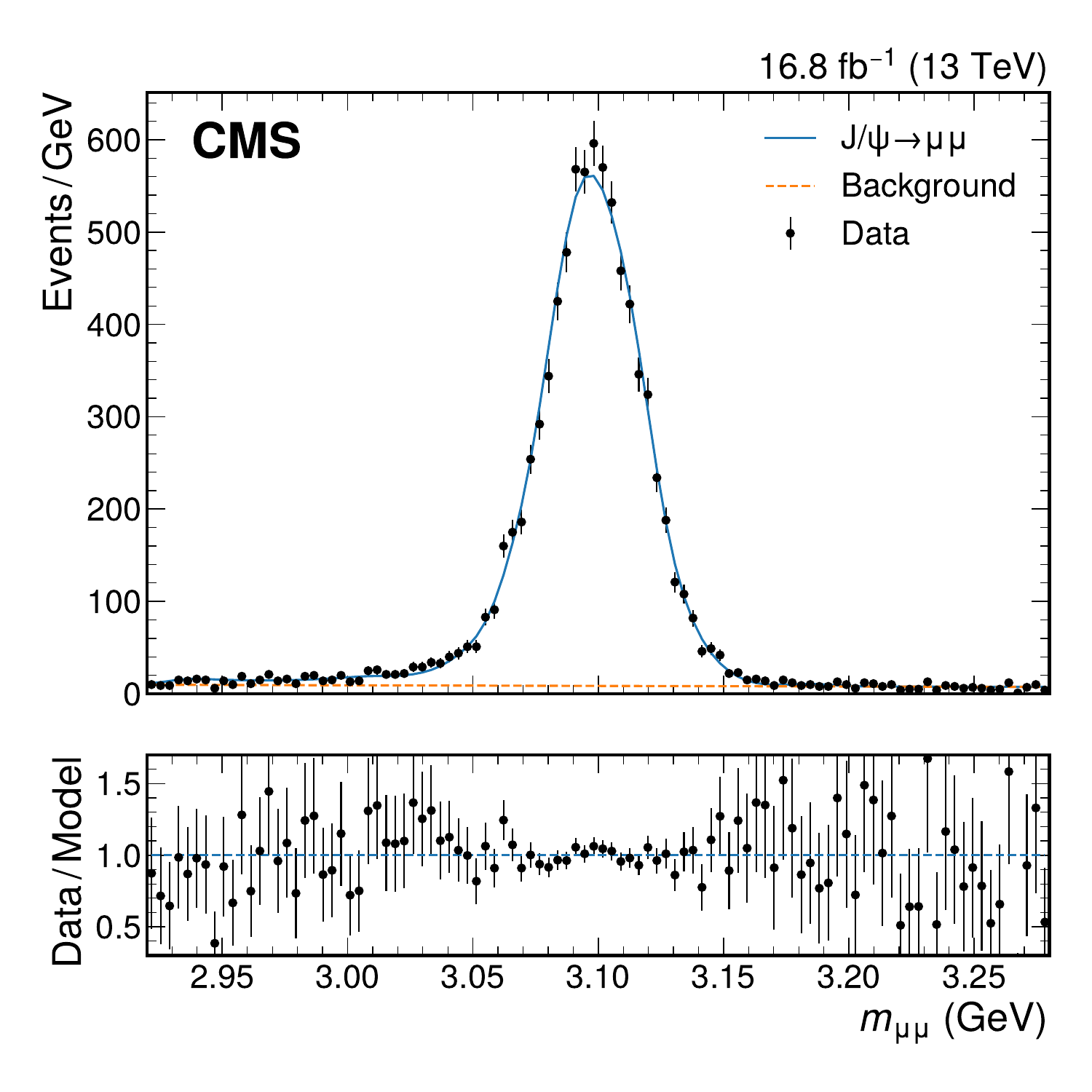}\quad
\includegraphics[width=0.45\textwidth]{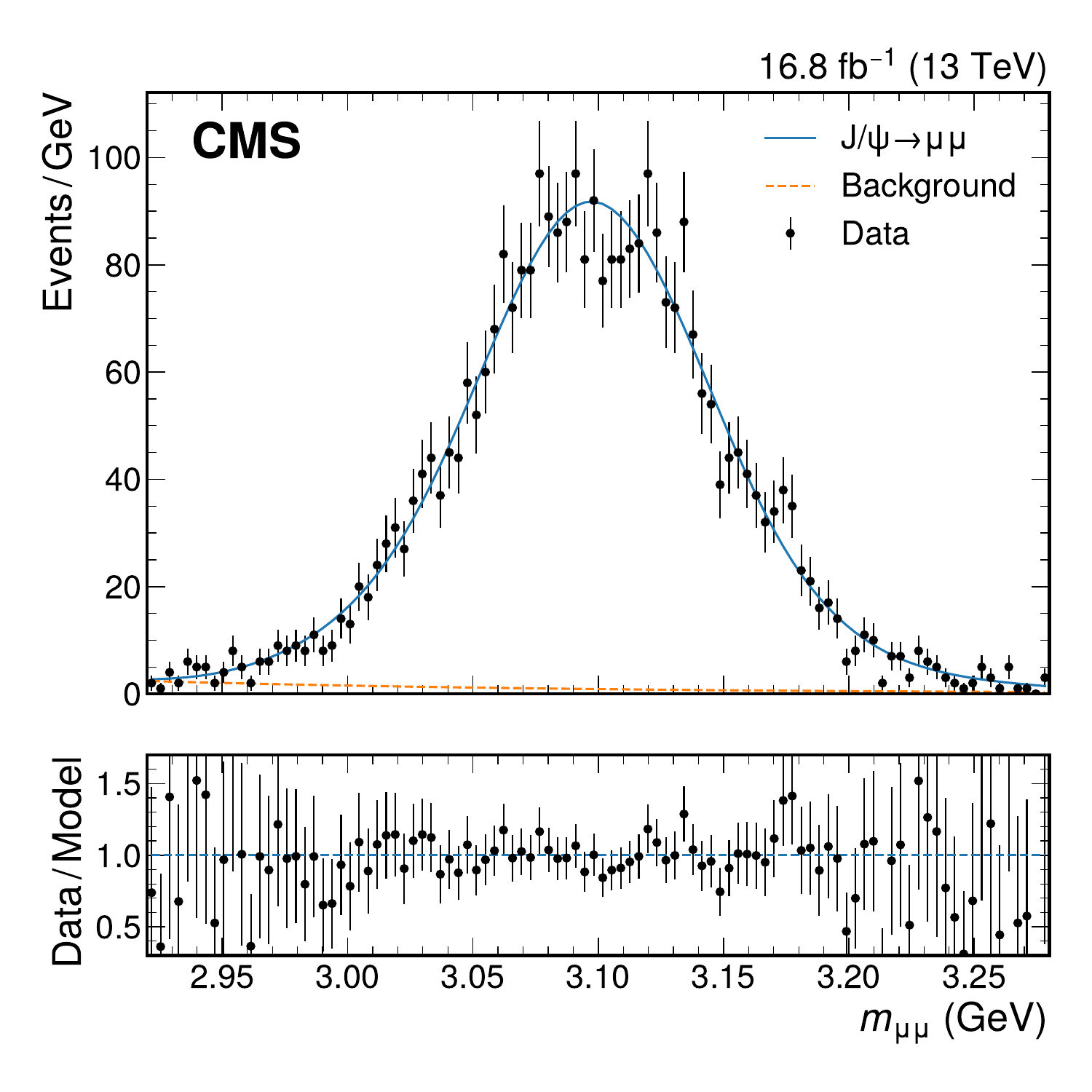}\\
\caption{Dimuon invariant mass distributions in \jmm decays reconstructed in data (black points) in two representative \etamu bins in the central (left) and forward (right) regions of the tracker. 
The blue line represents a fit to the distribution. 
The small background component is shown as a dashed orange line.}
\label{fig:jpsiMass_fits}
\end{figure}

\begin{figure}[hbtp]
\centering
\includegraphics[width=0.6\textwidth]{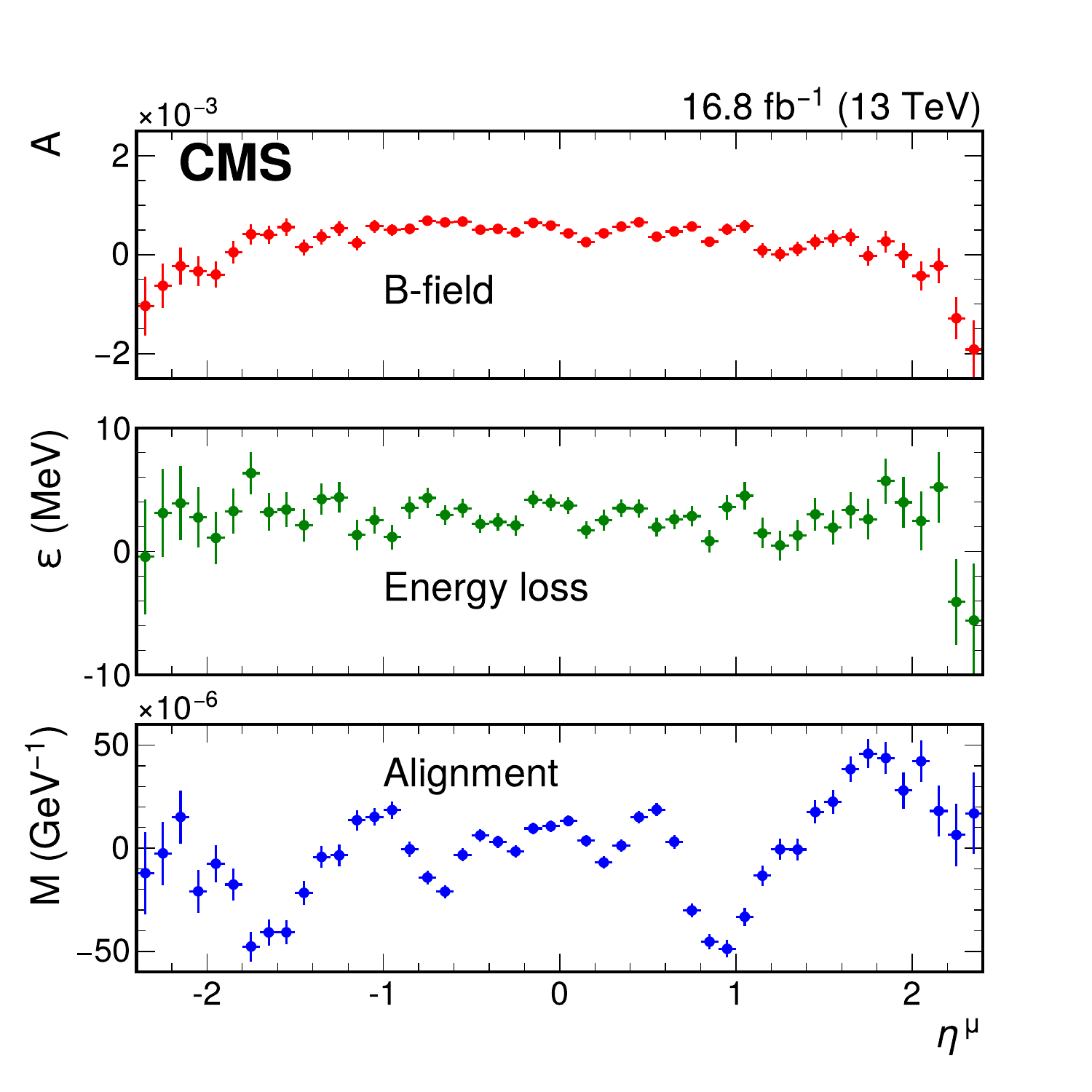}\\
\caption{Parameters of the calibration model as functions of \etamu, extracted from \jmm events.}
\label{fig:calib_corr_factors}
\end{figure}

\begin{table*}[!ht]
\centering
\topcaption{Goodness-of-fit test statistic for different PDF sets 
when fitting simultaneously the \etamu distributions for selected \PWp (\PWm) events 
and the \ymm distribution for \zmm events. 
Both the saturated likelihood ratios, which are expected to follow a
$\chi^2$ distribution with ndf degrees of freedom if the model
is an accurate representation of the data, and the associated $p$-value are shown.
The fit is performed in the nominal configuration with all uncertainties (left column), 
nominal configuration without PDF and \alphaS uncertainties (middle column), 
and nominal configuration without theory uncertainties (right column).}
\label{tab:pdf_fits_yll_etal_q}
\begin{tabular}{lcccccc}
\multirow{2}{*}{PDF set} & \multicolumn{2}{c}{Nominal fit} 
& \multicolumn{2}{c}{Without PDF$+$\alphaS unc.} & \multicolumn{2}{c}{Without theory unc.} \\
                & $\chi^2$/ndf & $p$-val.\ (\%) 
                & $\chi^2$/ndf & $p$-val.\ (\%) & $\chi^2$/ndf & $p$-val.\ (\%) \\ \hline
CT18Z           & 100.7/116 & 84  & 125.3/116  & 26 & 103.8/116 & 78 \\
CT18            & 100.7/116 & 84  & 153.2/116  & 1.0 & 105.7/116 & 74 \\
PDF4LHC21       & ~97.7/116  & 89  & 105.5/116  & 75 & 104.1/116 & 78 \\
MSHT20          & ~97.0/116  & 90  & 107.4/116  & 70 & ~98.8/116  & 87 \\
MSHT20aN3LO     & ~99.0/116  & 87  & 122.8/116  & 31 & 101.9/116 & 82 \\
NNPDF3.1        & ~99.1/116  & 87  & 105.5/116  & 75 & 115.0/116 & 51 \\
NNPDF4.0        & ~99.7/116  & 86  & 104.3/116  & 77 & 116.7/116 & 46 \\
\end{tabular}
\end{table*}

\begin{table*}[th!]
\centering
\topcaption{Number of nuisance parameters for the main groups of systematic uncertainties, 
for the \wlike \mz and \mw fits. 
The number of parameters is displayed only once when it is the same for both fits, 
while ``\NA'' means that this source is not relevant. 
For completeness, subgroups of parameters are also reported as indented labels for a few groups.} 
\label{tab:nuisParamFit}
\cmsTable{
\begin{tabular}{l c c}
Systematic uncertainties & \wlike \mz  & \mw  \\
\hline
Muon efficiency &  3127 & 3658 \\
\quad Muon eff.\ veto &  \NA & 531 \\
\quad Muon eff.\ syst. &  \multicolumn{2}{c}{343} \\	
\quad Muon eff.\ stat. &  \multicolumn{2}{c}{2784}  \\
Nonprompt-muon background   &  \NA & 387 \\
Prompt-muon background  &  2 & 3 \\
Muon momentum scale & \multicolumn{2}{c}{314} \\
L1 prefiring       & \multicolumn{2}{c}{14} \\
Integrated luminosity     &  \multicolumn{2}{c}{1}\\ 
\hline
PDF (CT18Z)      & \multicolumn{2}{c}{60} \\
Angular coefficients &  177 & 353\\
\quad \PW \MiNNLO \muF, \muR  &  \NA & 176\\
\quad \PZ \MiNNLO \muF, \muR  &  \multicolumn{2}{c}{176}\\
\quad \PYTHIA shower \kt & \multicolumn{2}{c}{1}\\
\ptv modeling   &  22 & 32\\
\quad Nonperturbative      &  4 & 10\\
\quad Perturbative          &  4 & 8\\
\quad Theory nuisance parameters   &  \multicolumn{2}{c}{10}\\
\quad \PQc, \PQb quark mass    &  \multicolumn{2}{c}{4} \\
Higher-order EW  &  6 & 7 \\
\PZ boson width   &  \multicolumn{2}{c}{1}\\
\PZ boson mass    &  \NA & 1 \\
\PW boson width   &  \NA & 1 \\
$\sin^2\theta_{W}$    &  \multicolumn{2}{c}{1}\\
\hline
Total         &  3725 & 4833 \\
\end{tabular}
}
\end{table*}

\begin{table*}[pth!]
\centering
\topcaption{Uncertainties in the \wlike \mz and \mw measurements, 
comparing the mass difference between charges and the nominal charge combination,
using nominal (upper) and global (lower) impacts.}
\label{tab:impacts_WlikeZandW_chargeDiff}
\cmsTable{
\begin{tabular}{lcccc}
\multirow{2}{*}{Source of uncertainty} & \multicolumn{4}{c}{Nominal impact (\MeVns)} \\
                           & in $m_{\PZ^{+}} - m_{\PZ^{-}}$  & in \mz & in $m_{\PW^{+}} - m_{\PW^{-}}$  & in \mw \\ \hline
 Muon momentum calibration        & 23.1    & 5.6  & 21.6    &   4.8 \\
 Muon reco.\ efficiency           & 7.1     & 3.8  & 7.2     &   3.0 \\
 \PW and \PZ angular coeffs.      & 14.5    & 4.9  & 18.7    &   3.3 \\
 Higher-order EW                  & 0.2     & 2.2  & 1.5     &   2.0 \\
 \ptv modeling                    & 0.6     & 1.7  & 7.4     &   2.0 \\
 PDF                              & 0.9     & 2.4  & 11.8    &   4.4 \\
 Nonprompt-muon background             & \NA  & \NA  & 7.5  &   3.2 \\
 Integrated luminosity            & $<$0.1  & 0.3  & 0.1     &   0.1 \\
 MC sample size                   & 4.9     & 2.5  & 3.0     &   1.5 \\ \hline
 Data sample size                 & 13.9    & 6.9  & 4.7     &   2.4 \\
 Total uncertainty                & 32.5    & 13.5 & 30.3    &   9.9 \\   
\end{tabular}
\vglue4mm
\begin{tabular}{lcccc}
\multirow{2}{*}{Source of uncertainty} & \multicolumn{4}{c}{Global impact (\MeVns)} \\
                           & in $m_{\PZ^{+}} - m_{\PZ^{-}}$  
                           & in \mz & in $m_{\PWp} - m_{\PWm}$  & in \mw \\ \hline
 Muon momentum scale                    & 21.2   &   5.3 & 20.0    &   4.4 \\
 Muon reco.\ efficiency                 & 6.5    &   3.0 & 5.8     &   2.3 \\
 \PW and \PZ angular coeffs.            & 13.9   &   4.5 & 13.7    &   3.0 \\
 Higher-order EW                        & 0.2    &   2.2 & 1.5     &   1.9 \\
 \ptv modeling                          & 0.4    &   1.0 & 2.7     &   0.8 \\
 PDF                                    & 0.7    &   1.9 & 4.2     &   2.8 \\
 Nonprompt-muon background                   & \NA & \NA & 4.8     &   1.7 \\
 Integrated luminosity                  & $<$0.1 &   0.2 & 0.1     &   0.1 \\
 MC sample size                         & 6.4    &   3.6 & 8.4     &   3.8 \\ \hline
 Data sample size                       & 18.1   &   10.1 & 13.4   &   6.0 \\
 Total uncertainty                      & 32.5   &   13.5 & 30.3   &   9.9 \\   
\end{tabular}
}
\end{table*}}
\cleardoublepage \section{The CMS Collaboration \label{app:collab}}\begin{sloppypar}\hyphenpenalty=5000\widowpenalty=500\clubpenalty=5000
\cmsinstitute{Yerevan Physics Institute, Yerevan, Armenia}
{\tolerance=6000
V.~Chekhovsky, A.~Hayrapetyan, V.~Makarenko\cmsorcid{0000-0002-8406-8605}, A.~Tumasyan\cmsAuthorMark{1}\cmsorcid{0009-0000-0684-6742}
\par}
\cmsinstitute{Institut f\"{u}r Hochenergiephysik, Vienna, Austria}
{\tolerance=6000
W.~Adam\cmsorcid{0000-0001-9099-4341}, J.W.~Andrejkovic, L.~Benato\cmsorcid{0000-0001-5135-7489}, T.~Bergauer\cmsorcid{0000-0002-5786-0293}, S.~Chatterjee\cmsorcid{0000-0003-2660-0349}, K.~Damanakis\cmsorcid{0000-0001-5389-2872}, M.~Dragicevic\cmsorcid{0000-0003-1967-6783}, P.S.~Hussain\cmsorcid{0000-0002-4825-5278}, M.~Jeitler\cmsAuthorMark{2}\cmsorcid{0000-0002-5141-9560}, N.~Krammer\cmsorcid{0000-0002-0548-0985}, A.~Li\cmsorcid{0000-0002-4547-116X}, D.~Liko\cmsorcid{0000-0002-3380-473X}, I.~Mikulec\cmsorcid{0000-0003-0385-2746}, J.~Schieck\cmsAuthorMark{2}\cmsorcid{0000-0002-1058-8093}, R.~Sch\"{o}fbeck\cmsAuthorMark{2}\cmsorcid{0000-0002-2332-8784}, D.~Schwarz\cmsorcid{0000-0002-3821-7331}, M.~Sonawane\cmsorcid{0000-0003-0510-7010}, W.~Waltenberger\cmsorcid{0000-0002-6215-7228}, C.-E.~Wulz\cmsAuthorMark{2}\cmsorcid{0000-0001-9226-5812}
% ...
\par}
\vskip\cmsinstskip
\dag:~Deceased\\
$^{1}$Also at Yerevan State University, Yerevan, Armenia\\
$^{2}$Also at TU Wien, Vienna, Austria\\
$^{3}$Also at Ghent University, Ghent, Belgium\\
$^{4}$Also at Universidade do Estado do Rio de Janeiro, Rio de Janeiro, Brazil\\
$^{5}$Also at FACAMP - Faculdades de Campinas, Sao Paulo, Brazil\\
$^{6}$Also at Universidade Estadual de Campinas, Campinas, Brazil\\
$^{7}$Also at Federal University of Rio Grande do Sul, Porto Alegre, Brazil\\
$^{8}$Also at University of Chinese Academy of Sciences, Beijing, China\\
$^{9}$Also at China Center of Advanced Science and Technology, Beijing, China\\
$^{10}$Also at University of Chinese Academy of Sciences, Beijing, China\\
$^{11}$Also at China Spallation Neutron Source, Guangdong, China\\
$^{12}$Now at Henan Normal University, Xinxiang, China\\
$^{13}$Also at University of Shanghai for Science and Technology, Shanghai, China\\
$^{14}$Now at The University of Iowa, Iowa City, Iowa, USA\\
$^{15}$Also at an institute formerly covered by a cooperation agreement with CERN\\
$^{16}$Also at Zewail City of Science and Technology, Zewail, Egypt\\
$^{17}$Also at British University in Egypt, Cairo, Egypt\\
$^{18}$Now at Ain Shams University, Cairo, Egypt\\
$^{19}$Also at Purdue University, West Lafayette, Indiana, USA\\
$^{20}$Also at Universit\'{e} de Haute Alsace, Mulhouse, France\\
$^{21}$Also at Istinye University, Istanbul, Turkey\\
$^{22}$Also at an international laboratory covered by a cooperation agreement with CERN\\
$^{23}$Also at The University of the State of Amazonas, Manaus, Brazil\\
$^{24}$Also at University of Hamburg, Hamburg, Germany\\
$^{25}$Also at RWTH Aachen University, III. Physikalisches Institut A, Aachen, Germany\\
$^{26}$Also at Bergische University Wuppertal (BUW), Wuppertal, Germany\\
$^{27}$Also at Brandenburg University of Technology, Cottbus, Germany\\
$^{28}$Also at Forschungszentrum J\"{u}lich, Juelich, Germany\\
$^{29}$Also at CERN, European Organization for Nuclear Research, Geneva, Switzerland\\
$^{30}$Also at HUN-REN ATOMKI - Institute of Nuclear Research, Debrecen, Hungary\\
$^{31}$Now at Universitatea Babes-Bolyai - Facultatea de Fizica, Cluj-Napoca, Romania\\
$^{32}$Also at MTA-ELTE Lend\"{u}let CMS Particle and Nuclear Physics Group, E\"{o}tv\"{o}s Lor\'{a}nd University, Budapest, Hungary\\
$^{33}$Also at HUN-REN Wigner Research Centre for Physics, Budapest, Hungary\\
$^{34}$Also at Physics Department, Faculty of Science, Assiut University, Assiut, Egypt\\
$^{35}$Also at Punjab Agricultural University, Ludhiana, India\\
$^{36}$Also at University of Visva-Bharati, Santiniketan, India\\
$^{37}$Also at Indian Institute of Science (IISc), Bangalore, India\\
$^{38}$Also at Amity University Uttar Pradesh, Noida, India\\
$^{39}$Also at IIT Bhubaneswar, Bhubaneswar, India\\
$^{40}$Also at Institute of Physics, Bhubaneswar, India\\
$^{41}$Also at University of Hyderabad, Hyderabad, India\\
$^{42}$Also at Deutsches Elektronen-Synchrotron, Hamburg, Germany\\
$^{43}$Also at Isfahan University of Technology, Isfahan, Iran\\
$^{44}$Also at Sharif University of Technology, Tehran, Iran\\
$^{45}$Also at Department of Physics, University of Science and Technology of Mazandaran, Behshahr, Iran\\
$^{46}$Also at Department of Physics, Faculty of Science, Arak University, ARAK, Iran\\
$^{47}$Also at Helwan University, Cairo, Egypt\\
$^{48}$Also at Italian National Agency for New Technologies, Energy and Sustainable Economic Development, Bologna, Italy\\
$^{49}$Also at Centro Siciliano di Fisica Nucleare e di Struttura Della Materia, Catania, Italy\\
$^{50}$Also at Universit\`{a} degli Studi Guglielmo Marconi, Roma, Italy\\
$^{51}$Also at Scuola Superiore Meridionale, Universit\`{a} di Napoli 'Federico II', Napoli, Italy\\
$^{52}$Also at Fermi National Accelerator Laboratory, Batavia, Illinois, USA\\
$^{53}$Also at Lulea University of Technology, Lulea, Sweden\\
$^{54}$Also at Consiglio Nazionale delle Ricerche - Istituto Officina dei Materiali, Perugia, Italy\\
$^{55}$Also at Department of Applied Physics, Faculty of Science and Technology, Universiti Kebangsaan Malaysia, Bangi, Malaysia\\
$^{56}$Also at Consejo Nacional de Ciencia y Tecnolog\'{i}a, Mexico City, Mexico\\
$^{57}$Also at Trincomalee Campus, Eastern University, Sri Lanka, Nilaveli, Sri Lanka\\
$^{58}$Also at Saegis Campus, Nugegoda, Sri Lanka\\
$^{59}$Also at National and Kapodistrian University of Athens, Athens, Greece\\
$^{60}$Also at Ecole Polytechnique F\'{e}d\'{e}rale Lausanne, Lausanne, Switzerland\\
$^{61}$Also at Universit\"{a}t Z\"{u}rich, Zurich, Switzerland\\
$^{62}$Also at Stefan Meyer Institute for Subatomic Physics, Vienna, Austria\\
$^{63}$Also at Laboratoire d'Annecy-le-Vieux de Physique des Particules, IN2P3-CNRS, Annecy-le-Vieux, France\\
$^{64}$Also at Near East University, Research Center of Experimental Health Science, Mersin, Turkey\\
$^{65}$Also at Konya Technical University, Konya, Turkey\\
$^{66}$Also at Izmir Bakircay University, Izmir, Turkey\\
$^{67}$Also at Adiyaman University, Adiyaman, Turkey\\
$^{68}$Also at Bozok Universitetesi Rekt\"{o}rl\"{u}g\"{u}, Yozgat, Turkey\\
$^{69}$Also at Marmara University, Istanbul, Turkey\\
$^{70}$Also at Milli Savunma University, Istanbul, Turkey\\
$^{71}$Also at Kafkas University, Kars, Turkey\\
$^{72}$Now at Istanbul Okan University, Istanbul, Turkey\\
$^{73}$Also at Hacettepe University, Ankara, Turkey\\
$^{74}$Also at Erzincan Binali Yildirim University, Erzincan, Turkey\\
$^{75}$Also at Istanbul University -  Cerrahpasa, Faculty of Engineering, Istanbul, Turkey\\
$^{76}$Also at Yildiz Technical University, Istanbul, Turkey\\
$^{77}$Also at School of Physics and Astronomy, University of Southampton, Southampton, United Kingdom\\
$^{78}$Also at IPPP Durham University, Durham, United Kingdom\\
$^{79}$Also at Monash University, Faculty of Science, Clayton, Australia\\
$^{80}$Also at Universit\`{a} di Torino, Torino, Italy\\
$^{81}$Also at Bethel University, St. Paul, Minnesota, USA\\
$^{82}$Also at Karamano\u {g}lu Mehmetbey University, Karaman, Turkey\\
$^{83}$Also at California Institute of Technology, Pasadena, California, USA\\
$^{84}$Also at United States Naval Academy, Annapolis, Maryland, USA\\
$^{85}$Also at Bingol University, Bingol, Turkey\\
$^{86}$Also at Georgian Technical University, Tbilisi, Georgia\\
$^{87}$Also at Sinop University, Sinop, Turkey\\
$^{88}$Also at Erciyes University, Kayseri, Turkey\\
$^{89}$Also at Horia Hulubei National Institute of Physics and Nuclear Engineering (IFIN-HH), Bucharest, Romania\\
$^{90}$Now at another institute formerly covered by a cooperation agreement with CERN\\
$^{91}$Also at Texas A\&M University at Qatar, Doha, Qatar\\
$^{92}$Also at Kyungpook National University, Daegu, Korea\\
$^{93}$Also at Yerevan Physics Institute, Yerevan, Armenia\\
$^{94}$Also at another international laboratory covered by a cooperation agreement with CERN\\
$^{95}$Also at Imperial College, London, United Kingdom\\
$^{96}$Also at Institute of Nuclear Physics of the Uzbekistan Academy of Sciences, Tashkent, Uzbekistan\\
$^{97}$Also at another institute formerly covered by a cooperation agreement with CERN\\
\end{sloppypar}
%%% END EDITABLE REGION %%%
% skeleton_end
\end{document}